\documentclass[12pt]{article}
\usepackage{amsmath,amssymb}
\usepackage{jheppub}

\begin{document}

\preprint{Imperial-TP-2023-CH-02}

\title{Covariant Action for Self-Dual $p$-Form 
  Gauge Fields 
  in General Spacetimes}
\author{C.M. Hull}

\affiliation{The Blackett Laboratory, Imperial College London, Prince Consort Road, London SW7 2AZ, United Kingdom}

\emailAdd{c.hull@imperial.ac.uk}

\abstract{ Sen's action for a $p$-form gauge field with self-dual field strength coupled to a spacetime metric $g$ involves an explicit Minkowski metric and the presence of this raises questions as to whether the action is coordinate independent and whether it can be used on a general spacetime manifold.
A natural generalisation of Sen's action is presented in which the Minkowski metric is replaced by a second metric $\bar g$ on spacetime. The theory is  covariant and can be formulated on any spacetime.
 The theory describes  a physical sector, consisting of the chiral $p$-form gauge field coupled to the dynamical metric $g$, plus a shadow sector consisting of a second chiral $p$-form and the second metric $\bar g$.
 The fields in this shadow sector     only couple to each other and have no interactions with the physical sector, so that they decouple from the physical sector.
The resulting theory is covariant and can be formulated on any spacetime. 
Explicit  expressions are found  for the interactions and
extensions to include interactions with other physical fields
or higher-derivative field equations are given. 
A spacetime with two metrics has some interesting geometry and some of this is explored here and used in the construction of the interactions.
The action has two diffeomorphism-like symmetries, one acting only on the physical sector 
and one acting only on the shadow sector, with the spacetime  diffeomorphism symmetry arising as the diagonal subgroup. This allows a further generalisation in which $\bar g$ is not  a tensor field but is instead a gauge field whose transition functions involve  the usual coordinate transformation together with a shadow sector gauge transformation.
}

%\keywords{}
                              
\maketitle

\section{Introduction}

In  \cite{Sen:2015nph,Sen:2019qit}, Sen proposed a remarkable action for self-dual antisymmetric tensor 
 gauge fields, which was 
inspired by string field theory. This approach was further discussed in 
\cite{Andriolo:2020ykk,Vanichchapongjaroen:2020wza, Andrianopoli:2022bzr,Chakrabarti:2022jcb,Barbagallo:2022kbt,Chakrabarti:2020dhv,Andriolo:2021gen,Evnin:2022kqn}. 
Sen's approach is covariant and quadratic in the fields, facilitating quantum calculations, and it  also generalises to allow interactions.
The construction of an action for an antisymmetric tensor gauge field with
self-dual field strength is a problem that has attracted a great deal of
attention, and many approaches have been used, including those in 
 \cite{Siegel:1983es}-\cite{Saemann:2019dsl}.
% \cite{Siegel:1983es,Floreanini:1987as,Pasti:1995tn,Perry:1996mk,Belov:2006jd, Mkrtchyan:2019opf,Townsend:2019koy}
  %\cite{Evnin:2022kqn,Arvanitakis:2022bnr}
  %\cite{Imbimbo:1987yt,Henneaux:1988gg,McClain:1990sx,Verlinde:1995mz,Dolan:1998qk,Henningson:1999dm, Chen:2013gca}.
For a recent review, see \cite{Evnin:2022kqn}. There
are many treatments that give the same classical theory but the corresponding
quantum theories are not all equivalent.  An important objective is to find an
action whose quantization meets the requirements discussed in e.g.\ \cite{Witten:1996hc}.

Sen's theory  \cite{Sen:2015nph,Sen:2019qit} is a theory for a $q - 1$-form gauge field
$A$ for odd $q$, $q = 2 n$+1, in $d = 2 q=4n+2$ dimensions. The field strength
\[ F = d A   \]
is a $q = d / 2$ form that is self-dual, $\text{}$
\[ F = \ast F \]
Here $\ast$ denotes the  Hodge dual constructed using the spacetime metric $g_{\mu
\nu}$, so that
\begin{equation}
  \label{duop}
   (\ast F)_{\mu_1 \ldots . \mu_q} = \frac{1}{q!} \sqrt{| g |}
  \epsilon_{\mu_1 \ldots . \mu_q \nu_1 \ldots \nu_q} g^{\nu_1 \rho_1} \ldots
  g^{\nu_q \rho_q} F_{\rho_1 \ldots \rho_q}
\end{equation}
with $\epsilon_{\mu_1 \ldots . \mu_q \nu_1 \ldots \nu_q}$ the alternating
symbol taking values $0, 1$ or $- 1$ (see Appendix A). The theory also gives a second $q -
1$-form gauge field $C$ with field strength
\[ G = d C \]
which satisfies a duality relation
\[ G = \ast_{\eta} G \]
where $\ast_{\eta}$ is the duality operation constructed using the Minkowski
metric $\eta_{\mu \nu}$ instead of $g_{\mu \nu}$. The gauge field $C$ is free and does not
couple to the spacetime metric $g_{\mu \nu}$ or to any of the other fields in
the theory, so it decouples from the theory.
Then Sen's action is a theory of two fields, $A, C$, with $A$ the
desired self-dual gauge field and $C$ an auxiliary free  field that decouples from
the rest of the theory and so can be regarded as non-physical.\footnote{Here the term {\it auxiliary fields} is used to refer to   extra fields that do not couple to the spacetime metric $g$ or any of the physical fields  so that they decouple from the physical theory. Note that here the auxiliary field $C$ is propagating, whereas in supersymmetry the term auxiliary  field is  often reserved for non-propagating fields in a  supermultiplet.}
This can be extended to include other physical fields which interact with $g$ and $A$ but not with the auxiliary field $C$. 
For example, a coupling to other gauge fields can involve a Chern-Simons-type term $\Omega$ with $F=dA+\Omega$.
Adding a kinetic term for $g$   makes it a dynamical gravitational field.
The Hamiltonian was shown in \cite{Sen:2019qit}
 to be positive for the physical fields including $g$ and $A$ but negative for the auxiliary field $C$.

The degrees of freedom in Sen's action are a $q - 1$-form gauge field $P$ and a
$q$-form-field $Q$ that is self-dual with respect to $\eta$, $Q = \ast_{\eta}
Q$. The gauge field strengths $F, G$ are then constructed from these. Sen's
construction uses a vielbein $e_{\mu} ^a$ and chooses a local Lorentz gauge in
which the vielbein is a symmetric matrix. This gauge blurs the distinction
between coordinate indices $\mu, \nu, \ldots$ and tangent space indices $a, b,
\ldots$ and requires every diffeomorphism to be accompanied by a compensating
local Lorentz transformation, so that  the transformations under diffeomorphisms are
non-standard. In  \cite{Andriolo:2020ykk} a treatment of Sen's theory was presented that did not
involve choosing this gauge. In both formulations there is a symmetry under
which $F, g_{\mu \nu}$ and the other physical fields transform pretty much as
would be expected under diffeomorphisms while $G$ and $\eta_{\mu \nu}$ are
{\it invariant} under these transformations.

This theory raises a number of issues. The first is the status of the
Minkowski metric $\eta$ in the construction. In the formulation of  \cite{Andriolo:2020ykk}, it is a
metric on the spacetime, i.e.\ the spacetime is a manifold equipped with two
metrics, $\eta_{\mu \nu}$ and $g_{\mu \nu}$. This is highly restrictive: most
manifolds do not admit a flat Minkowski metric, so the theory could not be
formulated on a general spacetime. 

Another possibility is that the Minkowski metric involved in the\,
construction is the tangent space metric $\eta_{a b} = e_a {}^{\mu} e_b {}^{\nu}
g_{\mu \nu}$. Then $G$ could be taken to have components $G_{a_1 \ldots a_q} =
\frac{1}{q!} \epsilon_{a_1 \ldots a_q b_1 \ldots b_q} G^{b_1 \ldots b_q}$ and
similarly for $Q$. However, the meaning of $d G = 0$ and $G = d C$ would then
need some care as the 1-forms $e^a = e_{\mu} ^a d x^{\mu}, {e'}^a = {e'}_{\mu}
^a d x^{\mu} $ in different  charts $U, U'$ would be related by a
local Lorentz transformation ${e' } ^a = \Lambda^a {}_b e ^b$ in the overlap $U
\cap U'$. If in the coordinate patch $U$ one had $G_{a_1 \ldots a_q} =
        \partial_{[a_1 } C_{a_2 \ldots  a_q]}$ with $\partial_a =
e_a ^{\mu} \partial_{\mu}$,  in the
coordinate patch $U'$ the field strength $G'_{a_1 \ldots a_q}$ would not take this form. In general, to find an expression valid in all
coordinate patches one would be led to the covariant expression $G_{a_1 \ldots
a_q} = D_{[a_1 } C_{a_2 \ldots  a_q]}$ where $D$ is the
covariant derivative involving a spin connection $\omega_{\mu} {}^a {}_b$.
However, with this covariantization, the gauge field $C$ couples to the gravitational field through the spin
connection and so globally there would be an issue as to whether $C$ really
decoupled from the physical fields.

Another issue is that of whether Sen's action is independent of the choice of
coordinates. Under a change of coordinates, all fields should change.\footnote{E.g.\
under $x \rightarrow x'$ a scalar field transforms $\varphi \rightarrow
\varphi'$ where $\varphi' (x') = \varphi (x)$.} On the other hand, the
symmetry referred to as a diffeomorphism in  \cite{Sen:2015nph,Sen:2019qit,Andriolo:2020ykk} leaves the $q - 1$ form field
$C$ and the metric $\eta_{\mu \nu}$ invariant, so that there is a problem in
interpreting these as passive coordinate transformations in which the coordinates change.
Of course, the key thing is that the physical fields $A, g_{\mu \nu}$ transform
appropriately under diffeomorphisms, and they do in Sen's theory \cite{Sen:2015nph,Sen:2019qit}. However,
this raises questions as to the geometric status of the field $C$ and the
metric $\eta_{\mu \nu}$.

In this paper, an action is proposed that generalises that of Sen and which
resolves the above problems. The action has two Lorentzian signature metrics
on spacetime,  the dynamical gravitational field $g_{\mu \nu}$ and a
second metric $\bar{g}_{\mu \nu}$ which can be regarded as a background or
non-dynamical field. Then, in addition to the Hodge dual $\ast$ defined in
(\ref{duop}), there is a second Hodge dual  $\bar{\ast}$ defined by replacing $g_{\mu
\nu}$ with $\bar{g}_{\mu \nu}$ in (\ref{duop}). The new action again gives a
$q$-form field strength $F = d A  $ that is self-dual, $F = \ast F$,
and again gives another field strength $G = d C$, but now $G$ satisfies the
self-duality condition with respect to the metric $\bar g$, 
$G = \bar{\ast} G$.
The other physical fields in the theory couple to the metric $g_{\mu \nu}$ and to $A$
but do not couple to $C$ or $\bar{g}_{\mu \nu}$. The field $C$ is a free field
coupling only to $\bar{g}_{\mu \nu}$ and so decouples from the physical
theory, while the extra metric $\bar{g}_{\mu \nu}$ only couples to $C$. In
this sense $(C, \bar{g}_{\mu \nu})$ is a shadow (auxiliary) sector decoupling from the
physical sector
consisting of $g,A$ and any other physical fields.
 The dynamical fields appearing in the action are again a $q - 1$-form gauge field
$P$ and a $q$-form-field $Q$, but now $Q$ is self-dual with respect to
$\bar{g}_{\mu \nu}$, $Q = \bar{\ast} Q$.

Sen's theory is regained on setting $\bar{g}_{\mu \nu} = \eta_{\mu \nu}$.
However, much is gained by replacing $\eta_{\mu \nu}$ with $\bar{g}_{\mu
\nu}$. First, this allows the theory to be formulated on any manifold, not
just on manifolds admitting a Minkowski metric. For a given manifold,
$\bar{g}_{\mu \nu}$ can be taken to be any metric on that space. Second, the
theory is formulated geometrically and is invariant under standard
diffeomorphisms, with $g_{\mu \nu}$, $\bar{g}_{\mu \nu}$, $F$, $G$ all
transforming as tensors. 

In addition to the diffeomorphism symmetry there are further
symmetries that generalise the ones found in \cite{Sen:2015nph,Sen:2019qit,Andriolo:2020ykk}. A theory with two gravitons $g,\bar g$
that don't interact with each other typically has two gauge symmetries, one for
which $g$ is the gauge field 
under which
 \[ \delta g_{\mu \nu} = 2 \partial_{(\mu } \zeta_{ \nu)} +
   \ldots, \quad \delta \bar{g}_{\mu \nu} = 0 \]
with a vector field gauge parameter
$\zeta^{\mu}$ 
and one for which $\bar{g}$ is the gauge field
under which
\[ \delta g_{\mu \nu} =0, \quad \delta \bar{g}_{\mu \nu} = 2 \partial_{(\mu }
   \chi_{  \nu)} + \ldots \]
with a vector field gauge parameter
$\chi^{\mu}$.
The theory presented here  indeed has two such symmetries. Only $C$ and $\bar{g}_{\mu \nu}$
transform under the symmetry with parameter $\chi$, with $g_{\mu \nu}$ and $A$
and the other physical fields all invariant. On the other hand, $C$ and
$\bar{g}_{\mu \nu}$ are invariant under the symmetry with parameter $\zeta$
while $g_{\mu \nu}$ and $A$ and the other physical fields all transform. The
diagonal subgroup in which $\chi^{\mu} = \zeta^{\mu}$ is the diffeomorphism
symmetry, while in the case in which $\bar{g}_{\mu \nu} = \eta_{\mu \nu}$
the symmetry with parameter $\zeta$ is the symmetry discussed
in \cite{Andriolo:2020ykk}.
The two types of gauge symmetry lead to two separately conserved
\lq energy-momentum' tensors $T_{\mu \nu}$ and $\bar T_{\mu \nu}$ and hence two types of momentum, angular momentum and mass-energy.

For most of this paper, the two fields $g,\bar g$ will be taken to be tensor fields  so that they can both be regarded as  metrics. However, in principle, all symmetries of a theory can 
be used in the transition functions -- for example, for a gauge theory the transition  functions are coordinate transformations with parameter $\xi^\mu$ combined with gauge transformations. Here, both the $\chi$ and $\zeta $ transformations could be used in transition functions along with the coordinate transformations with parameter $\xi$, so that the $g$ and $\bar g$ would be gauge fields, not tensor fields. To obtain a conventional physical sector, $g$ should be a tensor field so that it is a metric and gives the standard theory of gravity. Then the transition functions could consist of coordinate transformations with parameter $\xi$ combined with $\chi $ transformations, as the $\chi $ transformations leave the metric $g$ and physical fields invariant and only act on  the shadow sector. This would give transition functions for $\bar g$   of the form 
\begin{equation}
\bar g'_{\mu\nu}=\bar g_{\mu\nu}+2 \partial_{(\mu  } \xi_{  \nu)} +2 \partial_{(\mu  } \zeta_{  \nu)} +\ldots
\end{equation}
 Remarkably, this allows choosing $\zeta=-\xi$ so that $\bar g'_{\mu\nu}=\bar g_{\mu\nu}$ and the components of $\bar g$ are unchanged between patches.
 In the case that
$\bar g$ is not a tensor field but is a gauge field so that the change in its components between patches resulting from the change of coordinates is cancelled by a symmetric tensor gauge transformation. In particular, this would allow taking  the components of $\bar g$ to be $\bar g_{\mu\nu}=\eta_{\mu\nu}
 $ in every coordinate patch provided that $\eta_{\mu\nu} $ is regarded as the components of a gauge field, not a tensor field. Thus there are two ways of promoting Sen's action to one that can be used on any spacetime: one is to replace $\eta_{\mu \nu}$ with a second metric tensor $\bar g_{\mu\nu}$, resulting in a conventionally covariant theory, and the other is to take the $\eta_{\mu\nu} $ appearing in Sen's action as a 
 gauge field in which the coordinate transformations in the transition functions are cancelled by $\chi$ gauge transformations. This paper will mostly focus on the case in which both $g$ and $\bar g$ are tensor fields.

The existence of two metrics on a spacetime leads to some interesting geometry and in particular to     a tensor that  interpolates between the two metric structures is introduced. This is used to construct a map between tensors that are self-dual with respect to $\bar g$ and those that are self-dual with respect to $  g$. This, together with the enriched gauge symmetry discussed above,  plays an interesting role in the theory investigated here.

An important part of Sen's theory was the construction of a term in
the action that is bilinear in $Q$ but has complicated non-linear dependence
on the metric $g$. This is generalised here to a term depending on both metrics.
The analysis here provides a geometric understanding of its construction and
it is given explicitly. An important part of the construction is the inversion of a certain operator (called $N$ here) and here it will be shown that an inverse exists.
The theory presented here is covariant and can  be formulated on any spacetime.

The plan of the paper is as follows. In section 2 Sen's action and its interactions are reviewed.
In section 3 the action with two metrics $g,\bar g$ is introduced. In section 4 some aspects of local and global geometry with two metrics are discussed and an \lq interpolating structure' is introduced which is a tensor that plays an important role and which is used to relate forms that are $g$-self-dual to ones that are $\bar g$ self-dual.  In section 5
the interpolating structure is used to construct the interaction term that is needed in the two-metric action. This requires  an  operator $N$  defined in that section to be invertible. The action is shown to be globally well-defined. 
In section 6  two kinds of diffeomorphism-like symmetries are found, with one acting on the physical sector and one acting on the shadow sector. In section 7 the finite form of the symmetries is discussed and the global structure that arises with transition functions between coordinate patches involving all gauge transformations is considered.
In section 8 the theory is analysed  in  a suitable basis for $q$-forms.
In section 9, all the operators introduced in section 5 are given explicitly and it is shown that the operator $N$ is invertible. This gives an explicit construction of the interaction term in the action. In section 10,  Chern-Simons are included in the field strength $F$ and in section 11 non-linear higher derivative generalisations are considered.
In section 12 the case of two dimensions is considered explicitly and the results of previous sections checked.
Section  13 provides some final comments.

\section{Sen's Action} \label{sec2}

\subsection{Sen's Action in Minkowski Space}
\label{sec21}

Consider first Sen's  theory  formulated  in Minkowski space in $d = 2 q$ dimensions  (with $q=2n+1$ for some integer $n$) in terms of a $q - 1$-form
$P$ and a $q$-form $Q$ that is self-dual, $Q = \ast Q$, with action
\begin{equation}
\label{senact}
  S = \int \left( \frac{1}{2} d P \wedge \ast d P - 2 Q \wedge d P \right)
\end{equation}
Here $\ast$ is the Hodge dual in Minkowski space given by (\ref{duop}) with Minkowski metric $g=\eta$.
The field equations are
\begin{equation}
  d \left( \frac{1}{2} \ast d P + Q \right) = 0
\end{equation}
\begin{equation}
  d P = \ast d P
\end{equation}
Defining
\begin{equation}
  G \equiv \frac{1}{2}  (dP + \ast d P) + Q, \quad F \equiv Q
\end{equation}
which are self-dual,
\begin{equation}
  \ast F = F, \quad \ast G = G,
\end{equation}
the field equations give
\begin{equation}
  \label{dGF} dG = 0, \qquad dF = 0
\end{equation}
This implies there are local $q - 1$-form potentials $A, C$ with
\begin{equation}
  \label{pots} F = d A, \quad   G = d C
\end{equation}
This gives the theory of the desired chiral $q - 1$-form $A$ with self-dual field strength $F$  together with an
auxiliary chiral $q - 1$-form $C$.

\subsection{Non-Linear Form of Sen's Action in Minkowski Space}
\label{sec22}

Sen's quadratic action in Minkowski Space has a    generalisation to give non-linear field equations, e.g.\ a Born-Infeld type generalisation of the quadratic theory \cite{Sen:2015nph,Sen:2019qit,Evnin:2022kqn}.
The action is
\begin{equation}
  S = \int \left( \frac{1}{2} d P \wedge \ast  d P - 2 Q \wedge d P - {\mathcal F}(Q)  \right)
\end{equation}
where ${\mathcal F}(Q) $ is a top form constructed from $Q$.
The variation of the interaction term 
\begin{equation}
\delta \int {\mathcal F}(Q) = 2 \int \delta Q \wedge {\mathcal R}(Q)
\end{equation}
defines a  $q$-form ${\mathcal R}(Q)$ which is anti-self-dual:
$$\ast {\mathcal R}(Q)=-{\mathcal R}(Q)
$$
as the wedge product of two self-dual forms vanishes.
Then the field equations are
\begin{equation}
  \label{dqa} d \left( \frac{1}{2} \ast  d P + Q \right) = 0
\end{equation}
\begin{equation}
  \label{mqa} {\mathcal R}(Q) + d P = \ast   ({\mathcal R}(Q) + d P)
\end{equation}
Defining
\begin{equation}
  G \equiv \frac{1}{2}  (d P + \ast d P) + Q
\end{equation}
so that
\begin{equation}
  \ast G = G
\end{equation}
it follows that equation (\ref{dqa}) gives
\begin{equation}
  dG = 0
\end{equation}
so that there is a potential $C$ with
$$G=dC$$

Taking the exterior derivative of (\ref{mqa}) and using ${\mathcal R}(Q) = - \ast {\mathcal R}(Q)$ gives
\[ d [2 {\mathcal R} (Q) - \ast d P] = 0 \]
Using (\ref{dqa}) and defining
\begin{equation}
\label{FFis}
  F \equiv Q + {\mathcal R}(Q)
\end{equation}
one obtains
\begin{equation}
  d F = 0
\end{equation}
so that locally  there is a potential $A$ with
$$
F=d A$$
Defining
\begin{equation}
F_\pm =\frac 1 2 (F+\ast F)
\end{equation}
then (\ref{FFis}) gives
$$F_+=Q
$$
and 
$$
F_-={\mathcal R}(F_+)
$$
As discussed in \cite{Evnin:2022kqn}, this is a consistent non-linear generalisation of the self-duality constraint $F_-=0$ and halves the number of degrees of freedom. Note that, in this formulation, ${\mathcal R}$ is not completely arbitrary but must come from the variation of some ${\mathcal F}$.

\subsection{Sen's Action in Minkowski Space with Chern-Simons
Terms}\label{Sencs}

Following \cite{Sen:2015nph,Sen:2019qit,Andriolo:2020ykk},    a coupling to matter can be introduced through a Chern-Simons term. This
involves a gauge-invariant $q + 1$-form $\Upsilon$ locally given by $\Upsilon
= d \Omega$ for a  $q$-form $\Omega$, where $\Omega$ is typically a Chern-Simons term constructed from other gauge fields plus a gauge-invariant term such as a fermion bilinear; the term $\Omega$ will be referred to as a  Chern-Simons term here. The  field
strength $F = d A + \Omega$  is self-dual, $F = \ast F$ and satisfies $d
F = \Upsilon$. 
This implies that $\ast d \ast F= \ast \Upsilon$ so that $\ast \Upsilon$ can be viewed as a source for $A$.
The Chern-Simons term 
transforms as a total derivative under the corresponding gauge transformations,
$\delta \Omega = d \Lambda$ for some $\Lambda$. The field strength $F= d A + \Omega$ is
invariant under this if $\delta A = - \Lambda$.

The action  proposed for such a self-dual gauge field with Chern-Simons
interactions is \cite{Sen:2015nph,Sen:2019qit,Andriolo:2020ykk}
\begin{equation}
  \label{actom} S = \int \left( \frac{1}{2} d P \wedge \ast d P - 2 Q \wedge d
  P + 2 Q \wedge \Omega - \frac{1}{2} \Omega \wedge \ast \Omega \right)
\end{equation}
This can be generalised to the non-linear case as in the previous subsection, but in this subsection only the quadratic action will be considered.\footnote{The term $-\frac{1}{2} \Omega \wedge \ast \Omega$ in this action  (and its generalisations appearing later in this paper) does not affect the equations of motion for $P$ and $Q$ but is included here so that on-shell  the action agrees with the usual action for a non-self-dual gauge field whose equations of motion are supplemented by the self-duality constraint; see \cite{Andriolo:2020ykk} for further discussion.}
The field equations for $P, Q$ are now
\begin{equation}
  \label{adpist} d \left( \frac{1}{2} \ast d P + Q \right) = 0
\end{equation}

\begin{equation}
  \label{pcon} d P - \Omega = \ast (d P - \Omega)
\end{equation}
Defining as before
\begin{equation}
  \label{giss} G \equiv \frac{1}{2}  (dP + \ast d P) + Q
\end{equation}
gives, using (\ref{adpist}),
\begin{equation}
  \label{gfie} \ast G = G, \quad d G = 0
\end{equation}
Taking the exterior derivative of (\ref{pcon}) and using (\ref{adpist}) to
eliminate $P$ gives
\begin{equation}
  d (Q - \Omega_-) = 0
  \label{erty}
\end{equation}
where
\begin{equation}
  \label{omplu} \Omega_{\pm} = \frac{1}{2}  (1 \pm \ast) \Omega
\end{equation}
Defining
\begin{equation}
  \label{fiss} F \equiv Q + \Omega_+
\end{equation}
so that $F$ is  self-dual,
\begin{equation}
  \label{fdu} \ast F = F
\end{equation}
the field equation (\ref{erty}) implies
\begin{equation}
  \label{fom} d F = d \Omega
\end{equation}
 there are then $q-1$-form potentials $A, C$ with
\begin{equation}
  F = d A + \Omega, \quad G = d C
\end{equation}

The action (\ref{actom}) has a symmetry $$\delta P = d \alpha$$ where $\alpha$ is a
$q - 2$ form. Under the transformations
\begin{equation}
  \label{symom} \delta P = \Lambda, \quad \delta \Omega = d \Lambda, \quad
  \delta Q = - \frac{1}{2}  (1 + \ast) d \Lambda
\end{equation}
the field strengths (\ref{giss}),(\ref{fiss}) are invariant, $\delta F =
\delta G = 0$. This is then  a symmetry of the field equations
(\ref{adpist}),(\ref{pcon}), and hence is a symmetry of
(\ref{gfie}),(\ref{fdu}),(\ref{fom}). The variation of the action under these
transformations is

\begin{equation}
  \label{delss} \delta S = - \int \Lambda \wedge d \Omega
\end{equation}
Thus the action is invariant only under transformations for which this is
satisfied \cite{Sen:2019qit,Andriolo:2020ykk}. Then the transformations (\ref{symom}) are in general only a symmetry of the equations of motion, not of the action. An alternative action which has this as a symmetry will be discussed in section \ref{secCoup}.

\subsubsection{IIB Supergravity}

For IIB supergravity in d=10 dimensions, $A$ is a 4-form. The supergravity has a NS-NS
2-form gauge field $B$ with field strength $H = d B$ and  gauge transformation
$\delta B = d \lambda$ together with a RR 2-form gauge field ${\cal C}^{(2)}$ with
gauge transformations $\delta {\cal C}^{(2)} = d \lambda'$ and a gauge-invariant RR
3-form field strength ${\cal G}
= d {\cal C}^{(2)}$. Then $\Upsilon =H\wedge
{\cal G}$ (plus terms involving fermion fields)
and the Chern-Simons 5-form can be taken to be
\begin{equation}
  \label{omis} \Omega = B \wedge {\cal G}
\end{equation}
(plus terms involving fermion fields)
transforming as
\begin{equation}
  \delta \Omega = d (\lambda \wedge {\cal G})
\end{equation}
so that one can take
\begin{equation}
  \label{liss} \Lambda = \lambda \wedge {\cal G}
\end{equation}
Then (\ref{delss}) gives
\begin{equation}
  \delta S = - \int \lambda \wedge {\cal G} \wedge d B \wedge {\cal G}
\end{equation}
which vanishes identically as ${\cal G} \wedge {\cal G}= 0$. In this case
(\ref{symom}) with (\ref{omis}),(\ref{liss}) is a symmetry of the action
(\ref{actom}), as is
\begin{equation}
  \delta P = d \alpha
\end{equation}
for any $3$-form $\alpha $.

Consider now the case $d = 6$ with $F$ a self-dual 3-form with $\Omega$ the
Chern-Simons 3-form for a Yang-Mills  field with Lie algebra-valued connection
$\mathcal{A}$ transforming as $\delta \mathcal{A}  = D \lambda$ and field
strength $\mathcal{F}  = d\mathcal{A}  +\mathcal{A} \wedge \mathcal{A}$.
Then
\begin{equation}
\Upsilon =  \rm {tr} (\mathcal{F} \wedge \mathcal{F})  , \quad
   \Omega =  \rm{tr} \left( \mathcal{A} \wedge d\mathcal{A}+ \frac{2}{3}
   \mathcal{A} \wedge \mathcal{A} \wedge \mathcal{A} \right)
\end{equation}
so that
\begin{equation}
\label{acsom} \delta \Omega = d \Lambda (\mathcal{A}, \lambda), \quad
   \Lambda (\mathcal{A}, \lambda) = \rm{tr} (\lambda d\mathcal{A}) 
\end{equation}
for a 2-form $\Lambda (\mathcal{A}, \lambda)$ (which is the 2-form giving the
consistent gauge anomaly). Now $F = d A + \Omega$ satisfies $d F = \Upsilon$
and for $F$ to be a well-defined 3-form requires, as usual, the integrability
condition
\begin{equation}
  \label{chern} \int \rm{tr} (\mathcal{F} \wedge \mathcal{F}) = 0
\end{equation}
restricting to gauge bundles with vanishing 2nd Chern class. The theory for
such a self-dual $F = d A + \Omega$ has the field equation
\[ F = \ast F \]
with gauge symmetry
\[ \delta A = - \Lambda (\mathcal{A}, \lambda), \quad \delta \mathcal{A}^a = D
   \lambda^a \]
with no constraint on the gauge parameter $\lambda .$
Now the Sen action for such an $F$ that is self-dual gives the constraint
(\ref{delss}) which in this case is
\[ \int [\Lambda (\mathcal{A}, \lambda) \wedge \rm{tr} (\mathcal{F}
   \wedge \mathcal{F})] = \int [\rm{tr} (\lambda d\mathcal{A}) \wedge
   \rm{tr} (\mathcal{F} \wedge \mathcal{F})] = 0 \]
so that the action (\ref{actom}) does not have a gauge invariance
(\ref{symom}),(\ref{acsom}) for general parameters $\lambda $, but these are symmetries of the equations of motion.

\subsection{Sen's Action with Spacetime Metric $g_{\mu \nu}$}

Sen's theory (as formulated in \cite{Andriolo:2020ykk}) involves a spacetime metric $g_{\mu \nu}$ with Hodge duality operator $\ast$
and a Minkoswki metric $\eta_{\mu \nu}$ with Hodge  dual operator $\ast_{\eta}$ and the
degrees of freedom are a $q - 1$-form field $P$, together with a $q$-form field $Q$ that
is self-dual with respect to $\eta, Q = \ast_{\eta} Q$. The action is
\begin{equation}
  S = \int \left( \frac{1}{2} d P \wedge \ast_{\eta} d P - 2 Q \wedge d P - Q
  \wedge M (Q) \right)
  \label{sensss}
\end{equation}
where $M (Q)$ is a linear map taking a self-dual $q$-form field $Q$ to a
$q$-form $M (Q)$ that is anti-self-dual, $M (Q) = - \ast_{\eta} M (Q)$ \cite{Andriolo:2020ykk}.
Moreover, $M (Q)$   has the symmetry property $Q \wedge M
(\delta Q) = \delta Q \wedge M (Q)$ \cite{Andriolo:2020ykk}. 
This is then a special case of the interacting theory considered in subsection \ref{sec22}
 with 
${\mathcal F}(Q)=Q\wedge M(Q)$ and so the analysis is similar to that case. However, here $M(Q)$ is carefully chosen with a particular dependence on the metric $g$ so as to give the required self-duality conditions.

The field equations are then
\begin{equation}
  \label{dq} d \left( \frac{1}{2} \ast_{\eta} d P + Q \right) = 0
\end{equation}
\begin{equation}
  \label{mq} M (Q) + d P = \ast_{\eta}  (M (Q) + d P)
\end{equation}
Defining
\begin{equation}
  G \equiv \frac{1}{2}  (d P + \ast_{\eta} d P) + Q
\end{equation}
so that
\begin{equation}
  \ast_{\eta} G = G
\end{equation}
equation (\ref{dq}) gives
\begin{equation}
  dG = 0
\end{equation}
Taking the exterior derivative of (\ref{mq}) and using $M (Q) = - \ast_{\eta}
M (Q)$ gives
\[ d [2 M (Q) - \ast_{\eta} d P] = 0 \]
Using (\ref{dq}) and defining
\begin{equation}
  F \equiv Q + M (Q)
\end{equation}
gives
\begin{equation}
  d F = 0
\end{equation}
The metric $g$ only enters through $M (Q)$. 

The remarkable achievement of
\cite{Sen:2015nph,Sen:2019qit} was to find a choice of function $M (Q)$ such that $F \equiv Q + M (Q)$
is self-dual with respect to the metric $g$,
\begin{equation}
  \ast F = F
\end{equation}
 for any $Q$ that is $\eta$-self-dual, $Q =
\ast_{\eta} Q$. The construction of $M (Q) $ will be discussed in 
section \ref{sec5}. Defining
\begin{equation}
F_\pm=\frac 1 2 (1\pm \bar \ast F)
\end{equation}
one has $F_+=Q$ so that $F \equiv Q + M (Q)$ can be written as
\begin{equation}
F_-=M(F_+)
\end{equation}

\section{The Action with two metrics}
\label{twometrics}

In this section, Sen's theory will be generalised to one in which the
Minkowski metric $\eta_{\mu \nu}$ is replaced by an arbitrary metric
$\bar{g}_{\mu \nu}$ on the spacetime. The result is a theory with two metrics on
the spacetime: a ``dynamical'' metric $g_{\mu \nu}$ which couples to all the
physical fields and carries the gravitational degerees of freedom, and a
second metric $\bar{g}_{\mu \nu}$ which doesn't couple to the physical fields
and only couples to the auxiliary field $C$. The metric $\bar{g}_{\mu \nu}$
can be regarded as a ``background'' metric or as an auxiliary field. For each of the two metrics there
is a corresponding Hodge dual operation (see Appendix A). The
Hodge dual for the ``dynamical'' metric $g_{\mu \nu}$ will be denoted here by $\ast$ and the Hodge
dual for the ``auxiliary'' metric $\bar{g}_{\mu \nu}$ by $\bar{\ast}$. Forms
of rank $q$ that are $g$-self-dual satisfying $\ast X = X$ and $q$-forms that
are $\bar{g} $-self-dual satisfying $Y = \bar{\ast} Y$ both play a role in
what follows.

The degrees of freedom of the theory are a $q-1$-form field $P$ and a $q$-form
field $Q$ which is $\bar{g} $-self-dual, $Q = \bar{\ast} Q$. The action is
\begin{equation}
  \label{act}
   S = \int \left( \frac{1}{2} d P \wedge \bar{\ast} d P - 2 Q
  \wedge d P - Q \wedge M (Q) \right)
\end{equation}
so that Sen's theory is recovered when $\overline{g} = \eta$. Here $M$ is a
linear map on $q$-forms $Q$ which can be written in components as
\begin{equation}
  \label{Mcomps} M (Q)_{\mu_1 \ldots \mu_q} = \frac{1}{q !} M_{\mu_1
  \ldots \mu_q}^{\nu_1 {\ldots \nu_q} } Q_{\nu_1 \ldots \nu_q}
\end{equation}
for some coefficients $M_{\mu_1 \ldots \mu_q}^{\nu_1 {\ldots \nu_q} } (x)$.
The  coefficients $M_{\mu_1 \ldots \mu_q}^{\nu_1 {\ldots \nu_q} } (x)$ depend on the metrics $g,\bar g$ and will be determined in later 
sections.

Following \cite{Andriolo:2020ykk},  $M$  can be taken to be symmetric in the sense that
\begin{equation}
  \label{symm} R \wedge M (Q) = Q \wedge M (R)
\end{equation}
for any two \ $q$-forms $Q, R$ which are self-dual, $Q = \bar{\ast} Q$, $R =
\bar{\ast} R$, and $M$  can be taken to be  $\bar g$-anti-self-dual,
\begin{equation}
  \label{mqsd} \bar{\ast} M (Q) = - M (Q)
\end{equation}
There is no loss of generality with these assumptions as any antisymmetric
part or self-dual part of $M (Q)$ drops out of the action. 
$M(Q)$ will be constructed  in section
\ref{sec5}
 and it will be seen that it is indeed anti-self-dual and symmetric.
Note that the metric $g$ only enters through $M (Q)$. 

The analysis of the field equations  is similar to that of the last section, but with $\eta$ replaced
by $\bar{g} .$ The field equations are (using (\ref{mqsd}) and the  symmetry
and linearity of $M$)
\begin{equation}
\label{feq1}
  d \left( \frac{1}{2}  \bar{\ast} d P + Q \right) = 0
\end{equation}
\begin{equation}
\label{feq2}
  M + d P = \bar{\ast}  (M + d P)
\end{equation}
Defining
\begin{equation}
\label{Giss}
  G \equiv \frac{1}{2}  (d P + \bar{\ast} d P) + Q
\end{equation}
which is $\bar g$-self-dual \begin{equation}
  \bar{\ast} G = G
\end{equation}
together with
\begin{equation}
  F \equiv Q + M (Q)
\end{equation}
the field equations (\ref{feq1}),(\ref{feq2}) imply
\begin{equation}
  dG = 0, \qquad dF = 0
\end{equation}
There are then $q-1$-form potentials
$A, C$ with $F = d A$, $G = d C$. 

The key point is that $M(Q)$ can be chosen so that $F$
is $g$-self-dual
\begin{equation}
  \ast F = F
\end{equation}
This will be shown in detail in section \ref{sec5}.
With such a choice of $M(Q)$, this is then a theory of the desired $q -
1$-form $A$ with self-dual field strength $\ast F = F$ and an auxiliary $q - 1
$-form $C$ whose field strength is self-dual with respect to the background
metric $\bar{\ast} G = G$. It is important that the auxiliary field $C$ does
not couple to the physical metric $g_{\mu \nu}$ and
 the physical  field $A$ does
not couple to the  auxiliary metric $\bar g_{\mu \nu}$.

Choosing $\bar g$ to be the same as $g$ simplifies the story and the term involving $M$ is then not needed as both $F$ and $G$ are self-dual with respect to $g$.
This just gives the  theory of subsection 
\ref{sec21}
minimally coupled to the metric $g$. However, this is problematic if the metric $g$ is dynamical as, with $g=\bar g$, the auxiliary field  $C$  couples to the gravitational field and  has negative energy.

\section{Bi-Metric Geometry}

\subsection{Interpolating Structure}

Here spacetime is a manifold equipped with two Lorentzian-signature metrics $g_{\mu
\nu}$, $\bar{g}_{\mu \nu}$. Spacetime will be assumed to be orientable with  the volume forms for
$\bar{g}, g$ given by 
$V_{\bar{g}} = \bar{\ast} 1$, $V_g = \ast 1$ specifying the same orientation.
 It will be useful to introduce an \lq interpolating structure' which is a tensor
$f_{\mu}  {}^{\nu}$  that  satisfies
\begin{equation}
  \label{fra} 
  g_{\mu \nu} = f_{\mu}  {}^{\rho} f_{\nu} {} ^{\sigma}  \bar{g}_{\rho
  \sigma}
 \end{equation}
This $f$ would be the vielbein if $\bar{g}_{\mu \nu} = \eta_{\mu \nu}$.
Such an $f$ can always be found locally; see subsection \ref{sec43}. 
A solution  $f_{\mu}  {}^{\nu}$ 
of (\ref{fra})
 that is a
  tensor field on spacetime will be referred to as a {\it global interpolating structure}.
 More generally, for an open cover of spacetime by charts $U_\alpha$, a local solution consists of a tensor field
 $f_{(\alpha)}$ in each chart $U_\alpha$ satisfying (\ref{fra}), but which need not 
 fit together to give a tensor field on the entire spacetime.
 Such a solution will be referred to as a {\it local interpolating structure}.
 
 Much of the paper will discuss the theory on $U_\alpha$ for a local interpolating structure $f_{(\alpha)}$.
 If there is a global interpolating structure, the local discussion then  extends immediately to the whole of spacetime.  In the local case, it will be necessary to carefully check that the theory  patches together to give a well-defined action on the whole spacetime; this will be done in subsection \ref{globM}
 and section \ref{sec7}.

Consider then a local interpolating structure $f_{\mu}  {}^{\nu}$ which is a tensor field on some chart (coordinate patch) $U$
 with  coordinates $x^\mu$ on $U$ chosen so that there are no coordinate singularities and in particular the components of the metric in $U$  satisfy $\det (g_{\mu \nu}) \ne 0$, $\det (\bar g_{\mu \nu}) \ne 0$. Then $\det (f_{\mu}  {}^{\nu})\ne 0$
 and $f_{\mu}  {}^{\nu}(x)$ is an invertible matrix at each point $x$ in $U$, with inverse
 $(f^{-
1})_{\mu}{}^{ \nu}$, satisfying $(f^{- 1})_{\mu} {}^{\nu} f_{\nu} {} ^{\rho}
= \delta_{\mu}  {}^{\rho}$.
 Such an $f$
defines an invertible map on $r$-forms $X$ on $U$ by
\begin{equation}
  \Phi : X \rightarrow \Phi (X)
\end{equation}
where
\begin{equation}
\label{phiis}
  \Phi (X)_{\mu_1 \ldots \mu_r} = f_{\mu_1} {}^{\alpha_1} \ldots f_{\mu_r}{}
  ^{\alpha_r} X_{\alpha_1 \ldots \alpha_r}
\end{equation}
and satisfies
\[ \Phi (X \wedge Y) = \Phi (X) \wedge \Phi (Y) \]
This map converts between the two Hodge duals for the two metrics (see appendix A):
\begin{equation}
  \ast \Phi (X) = \Phi (\bar{\ast} X)
  \label{phistar}
\end{equation}
which can be written as an operator relation
\begin{equation}
  \ast \Phi = \Phi \bar{\ast}
\end{equation}
This map has the important property that it maps $\bar{g}$-self-dual forms to
$g$-self-dual forms: for any $\bar{g}$-self-dual $q$-form $X = \bar{\ast} X$,
the $q$-form
\[ X' = \Phi (X) \]
is $g$-self-dual:
\[ X' = \ast X' \]
The inverse map $\Phi^{- 1}$ has a similar definition involving $(f^{-
1})_{\mu}{}^{ \nu}$.

For $r$-forms $X, Y$ one can define the inner products $(X, Y)_{\bar{g}}$,
$(X, Y)_g$ by
\[ V_{\bar{g}} (X, Y)_{\bar{g}} = X \wedge \bar{\ast} Y \]
\[ V_g (X, Y)_g = X \wedge \ast Y \]
Then
\[ (\Phi (X), \Phi (Y))_g = (X, Y)_{\bar{g}} \]

\subsection{Orthogonal Transformations}

The structure equation (\ref{fra}) can be written in matrix notation as
\begin{equation}
  g = f \bar{g} f^t
  \label{matf}
\end{equation}
and is invariant under
\begin{equation}
  \label{abc} f \rightarrow L f \bar L
\end{equation}
for transformations $L _{\mu}{}^{\nu},
\bar L _{\mu}{}^{ \nu}$ satisfying the orthogonality conditions
\begin{equation}
  L g L^t = g, \qquad \bar L \bar{g} \bar L^t = \bar{g}
\end{equation}
Note that an $L _{\mu}{}^{\nu}$ satisfying $L g L^t = g$ is mapped  through the interpolating structure to an $\bar L '$ given by
\begin{equation}
\bar L '= f^{-1}Lf
   \label{asdafd}
\end{equation}
that satisfies $\bar L' \bar{g}({ \bar L'})^t = \bar{g}$.
Then the transformation $f \rightarrow L f$ is equivalent to $f \rightarrow  f \bar L'$ 
with $\bar L'$ given by (\ref{asdafd})
so that the general transformation between solutions can be taken to  be
\begin{equation}
  \label{abcd} f \rightarrow  f \bar L
\end{equation}

Let
\begin{equation}
 \hat f _{\mu \nu }\equiv f_\mu{}^\rho \bar g _{\rho \nu}
 \end{equation}
The transformations (\ref{abc}) can usually be used to choose an interpolating structure such that $ \hat f _{\mu \nu }$ is symmetric on $U$, 
\begin{equation}
\hat f _{\mu \nu }= \hat f _{ \nu \mu}
\label{fsym}
\end{equation}
If $ \hat f _{\mu \nu }$ 
  is a symmetric tensor, the condition (\ref{matf}) can be written as the matrix equation
\begin{equation}
g = \hat f  (\hat f  \bar g ^{-1} )^t  = \hat f \bar g ^{-1} \hat f = ff\bar g
\end{equation}
Then
\begin{equation}
g \bar g ^{-1}=f^2
\end{equation}
and so $f$ is a matrix square root of $g \bar g ^{-1}$. Such a symmetric \lq gauge choice' will not be assumed here.

\subsection{Vielbeins}\label{sec43}

Let $e_{\mu}{}^a$ be a vielbein for the metric $g$ and  $\bar{e}_{\mu}{}^a$ be  a vielbein for the metric $\bar{g}$ satisfying
\begin{equation}
  e \eta e^t = g, \quad \bar{e} \eta \bar{e}^t = \bar{g}
\end{equation}
Then (\ref{fra}) is solved by taking the interpolating structure  to be
\begin{equation}
  f_{\mu} {} ^{\nu} = e_{\mu} {} ^a  \bar{e}_a  {}^{\nu}
  \label{fviel}
\end{equation}
where $\bar{e}_a  {}^{\mu}$ is the inverse of $\bar{e}_{\mu}  {}^a$. More
generally, a relative Lorentz transformation $\Theta_a  {}^b$ can be included
to give
\begin{equation}
  f_{\mu}  {}^{\nu} = e_{\mu} {} ^a  \Theta_a {} ^b  \bar{e}_b  {}^{\nu}
  \label{fvielL}
\end{equation}
with $\Theta \eta \Theta^t = \eta$; this also satisfies (\ref{fra}).
One could also include transformations
of the form (\ref{abc}) to give a general solution
\begin{equation}
  f_{\mu}  {}^{\nu} =L_{\mu}{}^\rho e_{\rho} {} ^a \Theta _a {} ^b  \bar{e}_b  {}^{\tau}
  \bar L_{\tau}{}^\nu
  \label{fvielLa}
\end{equation}
However, this can be rewritten in the form  (\ref{fvielL})
as
\begin{equation}
  f_{\mu}  {}^{\nu} = e_{\mu} {} ^a \hat \Theta_a {} ^b  \bar{e}_b  {}^{\nu},
  \qquad
  \hat \Theta \equiv (e^{-1} Le )\Theta
 (\bar e ^{-1} \bar L \bar e) \label{afvielLab}
\end{equation}
so that the general solution can be written in the form  (\ref{fvielL}).
Any  two solutions $f,f'$ of (\ref{fra}) then take the form
\begin{equation}
  f_{\mu}  {}^{\nu} = e_{\mu} {} ^a  \Theta_a {} ^b  \bar{e}_b  {}^{\nu}, \qquad f'_{\mu}  {}^{\nu} = e_{\mu} {} ^a  \Theta'_a {} ^b  \bar{e}_b  {}^{\nu}
  \label{fvielLl}
\end{equation}
for some local Lorentz transformations $\Theta_a {} ^b,\Theta'_a {} ^b$.
This can be rewritten as
\begin{equation}
  f'_{\mu}  {}^{\nu} =  f_{\mu}  {}^{\rho} 
  \bar L
  _\rho{}^\nu  \label{fvielLlsdf}
\end{equation}
where
\begin{equation}
  \bar L _{\mu}  {}^{\nu} =  \bar e_{\mu} {} ^a 
  ({\Theta} ^{-1})_a{}^b{\Theta '}_b{}^c \bar e _c{}^\nu
   \label{fvielLlssdsf}
\end{equation}
which is   of the form (\ref{abcd})  with $\bar L \bar{g} \bar L^t = \bar{g}$.

Defining frame components
\[ \bar{P}_{a_1 \ldots a_{q - 1}} = \bar{e}_{a_1}{} ^{\mu_1} \ldots
   \bar{e}_{a_{q - 1}} {}^{\mu_{q - 1}} P_{\mu_1 \ldots \mu_{q - 1}} \]
\[ \overline{Q }_{a_1 \ldots a_q} = \bar{e}_{a_1} {}^{\mu_1} \ldots
   \bar{e}_{a_q} {}^{\mu_q} Q_{\mu_1 \ldots \mu_q} \]
and similarly for other forms, 
 $d P$ has frame components
\begin{equation}
  \bar{D}_{[a_1  } \overline{ P}_{a_2 \ldots a_q  ]}
\end{equation}
where 
$$\bar{D}_{a_1} \overline{ P}_{a_2 \ldots a_q} = \bar{e}_{a_1}{} ^{\nu}
(\partial_{\nu} \overline{ P}_{a_2 \ldots a_q} - \bar{\omega}_{\nu} {}^b 
_{[  a_2} \overline{   P}_{| b | a_2 \ldots a_q  ]})$$
 with
$\bar{\omega}_{\mu}  {}^a{} _b$ the (torsion-free) spin-connection for the
vielbein $\bar{e}_{\mu} {}^a$. The action (\ref{act}) can then be written as
\begin{eqnarray}
  S  =  \frac{1}{q!} \int d^{2 q} x \, \bar{e}& \Biggl( &\frac{1}{2}
  \bar{D}_{[a_1 } \overline{ P}_{a_2 \ldots a_q ]} \bar{D}
  ^{[a_1 } \overline{ P}  ^{a_2 \ldots a_q ]} 
   \nonumber   \\
  &  &  - 2 \overline{Q }^{a_1 \ldots a_q} \bar{D}_{[a_1
  } \overline{ P}_{a_2 \ldots a_q ]} + \overline{Q }^{a_1
  \ldots a_q} \bar{M} (\bar{Q})_{a_1 \ldots a_q}\Biggr) 
  \label{act3}
\end{eqnarray}
where $\bar{e} = \det (\bar{e}_{\mu} {}^a)$.

\subsection{Transition Functions for Local Interpolating Structures}

In each patch $U_{\alpha}$ there are basis 1-forms $e_{(\alpha) } {}^a= e
_{{(\alpha) }  \mu}{} ^a d x^{\mu}$ and $\bar{e}_{{(\alpha) } } {}^a= \bar{e} _{{(\alpha) }  \mu} {}^a d x^{\mu}$
with vielbeins $e _{{(\alpha) }  \mu} {}^a$ and
  $\bar{e} _{{(\alpha) }  \mu} {}^a$. 
Then \begin{equation}
  f_{(\alpha) \mu} {}^{\nu} = e _{{(\alpha) }  \mu} {}^a \bar{e} _{{(\alpha) }  a}
  {}^{\nu}
\end{equation}
is a
 local
interpolating structure on $U_\alpha$.
This is a tensor on 
$U_{\alpha}$, with
$\delta  f_{(\alpha) \mu} {}^{\nu} ={\cal L}_\xi f_{(\alpha) \mu} {}^{\nu}  $
under an infinitesimal change of coordinates $x^\mu\to x^\mu-\xi^\mu$.
In overlaps $U_{\alpha} \cap U_{\beta}$, the frames are related by local
Lorentz transformations
\begin{equation}
  e_{{(\beta) } } {}^a= e_{{(\alpha) } }{}^b (\Lambda_{(\alpha \beta)})_b {}^a,
  \quad \bar{e}_{{(\beta) } } {}^a= \bar{e}_{{(\alpha) } }{}^b
  (\bar{\Lambda}_{(\alpha \beta)})_b {}^a
\end{equation}
In general 
the  chart $U_{\alpha}$ has coordinates $x_{(\alpha)}^{\mu}$
and
the  chart $U_{\beta}$ has coordinates $x_{(\beta)}^{\mu}$. In the following, all expressions on $U_{\alpha} \cap U_{\beta}$ will be given in the   $x_{(\alpha)}^{\mu}$ coordinate system so as to avoid factors of 
$\partial x_{(\alpha)}^{\mu}
/\partial
x_{(\beta)}^{\nu}$. More general expressions including such factors will be given in the following subsection.

In matrix notation,
\begin{equation}
\label{floc1}
  f_{(\alpha)} = e_{(\alpha)} (\bar{e}_{(\alpha)})^{- 1}
\end{equation}
so that in $U_{\alpha} \cap U_{\beta}$ 
\begin{equation}
  f_{(\beta)} = e_{(\beta)} \bar{e}_{(\beta)}^{- 1} = e_{(\alpha)}
  \Lambda_{(\alpha \beta)} \bar{\Lambda}_{(\alpha \beta)}^{- 1}
  \bar{e}_{(\alpha)}^{- 1}
\end{equation}
which can be rewritten as
\begin{equation}
  f_{(\beta)} = f_{(\alpha)} \bar{L}_{(\alpha \beta)}
  \label{flis}
\end{equation}
where
\begin{equation}
  \bar{L}_{(\alpha \beta)} = \bar{e}_{(\alpha)} \Lambda_{(\alpha \beta)}
  \bar{\Lambda}_{(\alpha \beta)}^{- 1} \bar{e}_{(\alpha)}^{- 1}
\end{equation}

Choosing instead the more general ansatz (\ref{fvielL}) in each patch
\begin{equation}
\label{floc2}
  f_{(\alpha)} = e_{(\alpha)} \Theta_{(\alpha)} (\bar{e}_{(\alpha)})^{- 1}
\end{equation}
for some $(\Theta_{(\alpha)})_a {}^b$ 
again gives a tensor on 
$U_{\alpha}$, with
$\delta  f_{(\alpha) \mu} {}^{\nu} ={\cal L}_\xi f_{(\alpha) \mu} {}^{\nu}  $.
Then, on the overlap
$U_{\alpha} \cap U_{\beta}$, 
\begin{equation}
  f_{(\beta)} = e_{(\beta)} \Theta_{(\beta)} \bar{e}_{(\beta)}^{- 1} =
  e_{(\alpha)} \Lambda_{(\alpha \beta)} \Theta_{(\beta)}
  \bar{\Lambda}_{(\alpha \beta)}^{- 1} \bar{e}_{(\alpha)}^{- 1}
\end{equation}
which can be written as (\ref{flis}) with
\begin{equation}
  \bar{L}_{(\alpha \beta)} = \bar{e}_{(\alpha)} \Theta_{(\alpha)}^{- 1}
  \Lambda_{(\alpha \beta)} \Theta_{(\beta)} \bar{\Lambda}_{(\alpha \beta)}^{-
  1} \bar{e}_{(\alpha)}^{- 1}
\end{equation}
Then $f$ will be a global interpolating structure with $f_{(\alpha)} =
f_{(\beta)}$ in $U_{\alpha} \cap U_{\beta}$ provided the $\Theta$'s satisfy the patching condition 
\begin{equation}
  \Theta_{(\alpha)}  = \Lambda_{(\alpha \beta)} \Theta_{(\beta)}
  \bar{\Lambda}_{(\alpha \beta)}^{- 1}
\end{equation}
in the overlap  $U_{\alpha} \cap U_{\beta}$.
If  the $\Theta$'s don't satisfy this, then the  $f_{(\alpha)}$ constitute a local interpolating structure.

\subsection{Local and Global Interpolating Structures}
\label{sec44}

It has been seen that any two solutions of (\ref{fra}) are related by $f'=f\bar L $ for some $\bar L$ satisfying 
\begin{equation}\bar L \bar g \bar L^t
=\bar g
\label{LgL}
\end{equation}
For a spacetime with open cover   $U_\alpha$, a local interpolating structure is given by a tensor field $f_{(\alpha)}$ 
on each $U_\alpha$ such that in overlaps $U_\alpha
\cap U_\beta$
\begin{equation}
f_{(\alpha)}=f_{(\beta)}\bar L_{(\alpha \beta)}
\end{equation}
where $\bar L_{(\alpha \beta)}$ satisfies
(\ref{LgL})
and in triple overlaps $U_\alpha
\cap U_\beta\cap U_\gamma$ satisfies the usual consistency condition $\bar L_{(\alpha \beta)}\bar L_{( \beta \gamma)}\bar L_{(\gamma\alpha  )}=1$. For the purposes of this paper, a local interpolating structure will be shown to be sufficient for the construction of a well-defined theory, but if there were a global interpolating structure  then this would simplify the discussion.

In more detail,  each coordinate chart $U_{\alpha}$ has coordinates $x_{(\alpha)}^{\mu}$ and the local
interpolating structure $f_{(\alpha)}$ on $U_{\alpha}$ has  components $f_{(\alpha) \mu}{}
^{\nu}$ in the  $x_{(\alpha)}^{\mu}$  coordinate system. In an overlap $U_{\alpha} \cap U_{\beta}$ the coordinates are related by
a smooth function $x_{(\alpha)} (x_{(\beta)})$. For a {\it global} interpolating
structure, $f$ is a tensor with the usual relation between components in the
different coordinate charts, so that in $U_{\alpha} \cap U_{\beta}$
\begin{equation}
  {f _{(\alpha) \mu}}^{\nu} (x_{(\alpha)}) = (A_{(\alpha \beta)}^{- 1}) 
  _{\mu} {}^{\rho} {f _{(\beta) \rho}}^{\sigma} (x_{(\beta)}) A_{(\alpha \beta)
  \sigma} {} ^{\nu}
\end{equation}
where
\begin{equation}
\label{aiss}
  A_{(\alpha \beta) \mu} {} ^{\nu} = \frac{\partial x_{(\alpha)}^{\nu}}{\partial
  x_{(\beta)}^{\mu}}
\end{equation}
This can be written in matrix notation as
\begin{equation}
  f _{(\alpha)} {= A_{(\alpha \beta)}^{- 1}}  f _{(\beta)} A_{(\alpha \beta)} 
\end{equation}
and the components of the metric in each coordinate system are related by
\begin{equation}
 g_{(\beta)} = A_{(\alpha \beta)}  g _{(  \alpha)} A_{(\alpha \beta)} ^t
\end{equation}

For a {\it local} interpolating structure, this is twisted by a transformation (\ref{abcd}) to
give
\begin{equation}
  f _{(\alpha)} {= A_{(\alpha \beta)}^{- 1}}  f _{(\beta)} \bar{L}_{(\alpha
  \beta)}  A_{(\alpha \beta)} 
\end{equation}
where $\bar{L}_{(\alpha \beta)} $ satisfies
\begin{equation}\bar L _{(\alpha \beta)}\bar g_{(  \beta)} \bar L_{(\alpha \beta)}^t
=\bar g_{(  \beta)}
\label{LgLb}
\end{equation}
 In $U_{\alpha} \cap
U_{\beta}$, one can choose one set of coordinates, say $x_{(\alpha)}$, and
consider the components of all tensors on $U_{\alpha} \cap
U_{\beta}$ in this coordinate system in order to
to avoid  $A_{(\alpha \beta)} $ contributions.
 The
components of $f_{(\alpha)}, f_{(\beta)}$  in the $x_{(\alpha)}$
coordinate system are then related by
\begin{equation}
  {f _{(\alpha) \mu}}^{\nu} {= f _{(\beta) \mu}}^{\sigma} \bar{L}_{(\alpha
  \beta) \sigma} {} ^{\nu}
\end{equation}
Then the condition for $f$ to be a global interpolating structure is $\bar{L}_{(\alpha
  \beta) }=1$ so that $f_{(\alpha)}= f_{(\beta)}$ on the overlap.
In the following, coordinates will often be chosen in this way to avoid  $A_{(\alpha \beta)} $ contributions.

Instead of requiring the spacetime to have two metrics $g,\bar g$
and constructing $f$ from them, one could instead require the spacetime with dynamical metric $g$ to also have a globally defined tensor field $f_{\mu}  {}^{\nu}(x)$ satisfying
$\det (f_{\mu}  {}^{\nu}(x))\ne 0$  and which is non-trivial, $f_{\mu}  {}^{\nu}\ne \delta_\mu{}^\nu$.
 A second metric $\bar g$ could then be constructed from $g,f$:
\begin{equation}
  \label{fram} \bar g_{\mu \nu} =( f^{-1})_{\mu}  {}^{\rho} ( f^{-1})_{\nu} {} ^{\sigma}  {g}_{\rho
  \sigma}
 \end{equation}
Then $\bar g$ will be a globally defined tensor field and the signature of this metric is the same as that of $g$ (as a result of Sylvester's Theorem). Locally this is equivalent to postulating two metrics and finding $f$, but postulating instead the existence of $f$ would avoid the question of whether the interpolating structure is local or global. 
In this way, for a given metric $g$, there is always a metric $\bar g$ such that there is a global interpolating structure $f$. For the theory of chiral forms moving in  a fixed spacetime, there is no problem in choosing  
  the second auxiliary metric $\bar g$ in this way. However, for a theory with dynamical gravity, this would mean that $\bar g$ would not be independent of the spacetime metric $g$ and this would be an obstacle to  the decoupling of the shadow (auxiliary) sector from the physical sector.

\section{Construction of Self-Dual Forms}
\label{sec5}

\subsection{Construction of $M (Q)$}

In the first part of this section a local discussion will be given in a coordinate patch $U$ in which there is a local interpolating structure $f$.
 First $M(Q)$ will be constructed on $U$ and in the next subsection it will be shown that $M(Q)$ is independent of the choice of local interpolating structure. Finally, in subsection \ref{globM}
 it will be shown that this local construction of $M(Q)$ in each patch 
 leads to an $M(Q)$ that is
  globally well-defined.

It will be useful to introduce projectors
\begin{equation}
  \bar{\Pi}_{\pm} = \frac{1}{2}  (1 \pm \bar{\ast}), \quad \Pi_{\pm} =
  \frac{1}{2}  (1 \pm \ast)
\end{equation}
The goal of this section is, given a $\bar{g}$-self-dual $q$-form $Q$ with $Q =
\bar{\ast} Q$, to construct a $g$-self-dual $q$-form $F$ with
\[ \ast F = F \]
whose $\bar{g}$-self-dual part is $Q, $ i.e.\ $\bar{\Pi}_+ F = Q$. Then this
determines $M = \bar{\Pi}_- F$ as a function of $Q$, giving the $M (Q)$
required for the action (\ref{act}).

For a given $g$-self-dual $F$, acting with $\Phi^{- 1}$ on $F$ gives a $q$-form
\begin{equation}
  \alpha \equiv \Phi^{- 1} (F)
\end{equation}
which is $\bar{g}$-self-dual:
\begin{equation}
  \bar{\ast} \alpha = \alpha
\end{equation}
Decomposing $F$ using the projectors $\frac{1}{2}  (1 \pm \bar{\ast})$ then
gives
\begin{equation}
  F = Q + M
\end{equation}
with
\begin{equation}
  Q \equiv \bar{\Pi}_+ F, \qquad M \equiv \bar{\Pi}_- F
\end{equation}
The aim is now to construct an $F$ satisfying $\ast F = F$ from $Q \equiv
\frac{1}{2}  (1 + \bar{\ast}) F.$ It has been seen that $F$ is given by $\Phi
(\alpha)$ where $\alpha = \bar{\ast} \alpha$. The first step is to find $\alpha$ as a function of
$Q$ and then $F = \Phi (\alpha (Q))$ gives $F$ in terms of $Q$, and in particular
gives $M (\alpha (Q)) .$

The  relation
\begin{equation}
  F = Q + M = \Phi (\alpha)
\end{equation}
can be decomposed using the projectors ${\bar {\Pi}}_{{\pm}}$ to give
\begin{equation}
  F_+=Q = N (\alpha)
\end{equation}
\begin{equation}
  F_-=M = K (\alpha)
\end{equation}
with
\begin{equation}
F_\pm
\equiv
{\bar {\Pi}}_{{\pm}}F
\end{equation}
where the maps $N, K$ are
\begin{equation}
\label{wertwed1}
  N \equiv \bar{\Pi}_+ \Phi \bar{\Pi}_+
\end{equation}
\begin{equation}
\label{wertwed2}
  K \equiv \bar{\Pi}_- \Phi \bar{\Pi}_+
\end{equation}
An operator  $\tilde{N}$ satisfying
\begin{equation}
  \label{genin} \tilde{N} N = \bar{\Pi}_+
\end{equation}
will be referred to here as 
a generalised inverse of $N$.
Such an inverse   can be
constructed as a formal power series, following \cite{Sen:2015nph,Sen:2019qit}, at least when $g$ is
sufficiently close to $\bar{g}$; this will be done in Appendix B. In what follows in this section, the existence of such a generalised
inverse will be assumed. Later, in section \ref{sec8}, the existence  of  a generalised inverse $\tilde{N}$ will be established.

Then
\begin{equation}
  \alpha = \tilde{N}  (Q)
\end{equation}
so that finally one can write $M$ as a linear function $M (Q)$ of $Q$
\begin{equation}
\label{wertwed3}
  M (Q) = K \tilde{N}  (Q)
\end{equation}
This is the desired function $M (Q)$ and, by construction, satisfies
$\bar{\ast} M = - M$.

As $Q = \bar{\Pi}_+ F$ and $F = Q + M$, it follows that any self-dual $F =
\ast F$ is determined by $\bar{\Pi}_+ F$:
\begin{equation}
  F = \bar{\Pi}_+ F + M (\bar{\Pi}_+ F)
\end{equation}
so that  $$F_-=M(F_+)$$
For any $\bar{g}$-self-dual $q$-form $Q$ with $Q = \bar{\ast} Q$, the
construction above ensures that $Q + M (Q)$ is $g$-self-dual
\begin{equation}
  Q + M = \ast (Q + M)
\end{equation}
Following  \cite{Andriolo:2020ykk}, it is useful to introduce a map $\Psi = 1 + M$ from $q$-forms to
$q$-forms
\begin{equation}
  \label{siis} \Psi (Y) = Y + M (Y)
\end{equation}
For any $q$-form $Y$
\begin{equation}
  \Pi_- \Psi (\bar{\Pi}_+ Y) = 0
\end{equation}
so that $\Psi$ takes a $\bar{g}$-self-dual form $X $ (with
$X = \bar{\ast} X$) to a $g$-self-dual one $\Psi (X)$ (with $\Psi (X) = \ast
\Psi (X)$). This map is nilpotent, $\Psi^2 = 0$, and invertible  \cite{Andriolo:2020ykk}. For any
$g$-self-dual form $F = \ast F$, there is then a $\overline{g}$-self-dual form
$Q \equiv \Psi^{- 1} (F)$ and
\begin{equation}
  \bar{\Pi}_+ F = \bar{\Pi}_+ \Psi (Q) = \bar{\Pi}_+ (Q + M (H)) = Q
\end{equation}
so that $F = \Psi (Q)$ implies that $F$ satisfies  \cite{Andriolo:2020ykk}
\begin{equation}
  \label{psiiden} F = \Psi (F_+)
\end{equation}

It remains to check that $M$ satisfies the symmetry property (\ref{symm}). For
any two $\bar{g}$-self-dual $q$-forms $Q, R$ (with $Q = \bar{\ast} Q, R =
\bar{\ast} R$),
\begin{equation}
  \label{qris} Q \wedge R = 0, \quad M (Q) \wedge M (R) = 0
\end{equation}
Moreover, $\Psi (Q)$ and $\Psi (R)$ are $g$-self-dual $\Psi (Q) = \ast \Psi
(Q)$, $\Psi (R) = \ast \Psi (R)$ so $\Psi (Q) \wedge \Psi (R) = 0$, i.e.\
\[ (Q + M (Q)) \wedge (R + M (R)) = 0 \]
Expanding this and using (\ref{qris}) gives the symmetry property
(\ref{symm}).

The above construction can be written in matrix form by choosing a basis for
$q$-forms; this is done in section \ref{sec7a}  and this reproduces the analysis of  \cite{Andriolo:2020ykk}
in the case in which $\bar{g} = \eta $. Alternatively, explicit perturbative
expressions can be found following  \cite{Sen:2015nph,Sen:2019qit}; this is done in Appendix B.

\subsection{Independence of $M(Q)$ on choice of Interpolating Structure}

As was seen in section 4, any two local interpolating structures $f,f'$  in a given  coordinate patch
are related by
\begin{equation}
{f'}_\mu{}^\nu= f_\mu {}^\rho \bar L_\rho {}^\nu
\end{equation}
where $\bar L$ satisfies the $\bar g$-orthogonality condition  (\ref{LgL}) so that $\det ( \bar L)=\pm 1$. Here attention   will be restricted to structures $f'$ for which  $\det ( \bar L)=  1$.
Then as well as the map $\Phi$ given in terms of $f$ by (\ref{phiis}) there is a map $\Phi'$ given by replacing $f$ with $f'$ in (\ref{phiis}). The maps are related by
\begin{equation}
\Phi'=\Phi \Delta
\end{equation}
where $\Delta$ is the map on $r$-forms $X$ given by
\begin{equation}
\label{deliss}
  \Delta (X)_{\mu_1 \ldots \mu_r} = \bar L_{\mu_1} {}^{\alpha_1} \ldots \bar L_{\mu_r}{}
  ^{\alpha_r} X_{\alpha_1 \ldots \alpha_r}
\end{equation}
As $\det ( \bar L)=  1$, $\Delta$ preserves the tensor $\bar \varepsilon _{\alpha_1 \ldots \alpha_d}$ 
and so 
\begin{equation}
 \bar \ast \Delta (X) = \Delta (\bar{\ast} X)
\end{equation}
and as a result $\Delta$ maps $\bar g$-self-dual forms to $\bar g$-self-dual forms and
$\Phi '$ satisfies (\ref{phistar}) so that it maps $\bar g$-self-dual forms to $ g$-self-dual forms.

Consider now the arguments of the previous subsection, but with $\Phi$ replaced with $\Phi'$.
For  a $g$-self-dual $F$, acting with ${\Phi'}^{- 1}$ on $F$ gives a $q$-form
\begin{equation}
  \alpha' \equiv \Phi'^{- 1} (F)
\end{equation}
which is $\bar{g}$-self-dual:
\begin{equation}
  \bar{\ast} \alpha' = \alpha'
\end{equation}
Decomposing $F$ using  $\bar{\Pi}_\pm $ then
gives
$  F = Q + M
$
with
$
  Q \equiv \bar{\Pi}_+ F$, $ M \equiv \bar{\Pi}_- F
$ as before.
Then
\begin{equation}
   Q = N' (\alpha')
,\qquad 
  M = K '(\alpha')
\end{equation}
where  $ \alpha'=\Delta^{-1}\alpha$ and
\begin{equation}
  N '\equiv \bar{\Pi}_+ \Phi '\bar{\Pi}_+
,\qquad 
  K '\equiv \bar{\Pi}_- \Phi ' \bar{\Pi}_+
\end{equation}
However $\Phi'=\Phi \Delta$ and
$\Delta$ commutes with the projectors $\bar{\Pi}_\pm$ so that 
\begin{equation}
N '= \bar{\Pi}_+ \Phi \bar{\Pi}_+\Delta=N\Delta
,\qquad 
  K '= \bar{\Pi}_- \Phi  \bar{\Pi}_+\Delta=K\Delta
\end{equation}
and
\begin{equation}
   Q = N(\Delta \alpha' )
,\qquad 
  M = K (\Delta \alpha')
\end{equation}
Eliminating $\Delta \alpha'=\alpha$ then gives
\begin{equation}
  M (Q) = K \tilde{N}  (Q)
\end{equation}
as before.
Hence changing from $f$ to $f'$ leaves $M(Q)$ unchanged and the action is independent of the choice of interpolating structure.

\subsection{Global construction of $M(Q)$ from a local interpolating structure}
\label{globM}

As seen in section 4, a  local interpolating structure is given by a tensor field $f_{(\alpha)}$ 
on each chart $U_\alpha$. Then, in $U_\alpha$, $f_{(\alpha)}$  defines  a map $\Phi_{(\alpha)}$ through (\ref{phiis}) and this in turn gives $N_{(\alpha)}$, $K_{(\alpha)}$ and $M_{(\alpha)}$   through (\ref{wertwed1}),(\ref{wertwed2}),(\ref{wertwed3}).
In overlaps $U_\alpha
\cap U_\beta$
\begin{equation}
f_{(\alpha)}=f_{(\beta)}\bar L_{(\alpha \beta)}
\end{equation}
where $\bar L_{(\alpha \beta)}$ satisfies
(\ref{LgL}) (and as explained in subsection \ref{sec44}, coordinates are chosen to avoid factors of (\ref{aiss})).
Then $\bar L_{(\alpha \beta)}$ defines a map $\Delta_{(\alpha \beta)}$ through (\ref{deliss}). The  last subsection 
showed how $N,K,M$ change under $f\to f\bar L$ and this then
 gives  the patching conditions
 \begin{equation}
N _{(\alpha )}= N_{( \beta)}\Delta_{(\alpha \beta)}
,\qquad 
  K_{(\alpha )} =K_{( \beta)}\Delta_{(\alpha \beta)}
\end{equation}
and
\begin{equation}
M _{(\alpha )}= M_{( \beta)}
\end{equation}
so that $M$ is a globally defined tensorial operator and in particular $M(Q)$ is a globally defined $q$-form.
Thus the action (\ref{act}) is well-defined globally, even if the interpolating structure is only defined locally.

\section{Diffeomorphism invariance and other symmetries}
\label{symsec}

\subsection{Diffeomorphism Invariance}

No gauge has been fixed in the discussion here and, in particular, 
 the 
symmetric \lq gauge choice' (\ref{fsym}) (which for $\bar{g} = \eta$
played a key role in Sen's analysis) has not been assumed. This is important:
without fixing a gauge, everything is tensorial and diffeomorphism invariance
is manifest. In particular, $g, \bar{g},Q,dP$  are  tensors and transform
under an infinitesimal diffeomorphism as
\begin{equation}
  \delta g =\mathcal{L}_{\xi} g, \quad \delta \bar{g} =\mathcal{L}_{\xi} 
  \bar{g}  \quad
  \delta Q =\mathcal{L}_{\xi} Q, \quad
  \delta d P =\mathcal{L}_{\xi} d P
   \end{equation}
where $\mathcal{L}_{\xi}$ is the Lie derivative with respect to the
infinitesimal vector field $\xi^{\mu} $. The field $P$ is a $q-1$ form gauge field with gauge symmetry $\delta P=d\lambda$.
If $f$ is a global interpolating structure, it is a tensor field 
transforming as 
\begin{equation}
\label{delff}
\delta f =\mathcal{L}_{\xi} f
\end{equation}
The maps $\Phi, \ast, \bar{\ast}$ then map
tensors  to tensors, and this implies that $N, K$ are also tensorial, i.e.\  they map tensors to tensors.
If $f$ is a local interpolating structure, then $f_{(\alpha)}$ is a tensor field on the patch $U_\alpha$, again transforming under an infinitesimal coordinate transformation on $ U_\alpha$ as
\begin{equation}
\label{asdelff}
\delta f =\mathcal{L}_{\xi} f
\end{equation}
This is consistent with the defining equation (\ref{fra}) and with the ans\" atze for $f_{(\alpha)}$ in (\ref{floc1}) or (\ref{floc2}).
The map $\Phi_{(\alpha)}$ then maps
tensors on $U_\alpha$   to tensors  on $U_\alpha$, and this implies that $N_{(\alpha)}, K_{(\alpha)}$ are also tensorial, i.e.\ they map tensors on $U_\alpha$ to tensors on $U_\alpha$.

For a $q$-form $\alpha$ with components $\alpha_{\mu_1 \ldots \mu_q}$ the linear map $N
(\alpha)$ is given by
\begin{equation}
  N (\alpha)_{\mu_1 \ldots \mu_q} = \frac{1}{q!} N_{\mu_1 \ldots \mu_q} {}^{\nu_1
  \ldots \nu_q} \alpha_{\nu_1 \ldots \nu_q}
\end{equation}
with $N_{\mu_1 \ldots \mu_q} {}^{\nu_1 \ldots \nu_q}$ constructed from $g,
\bar{g}$ as described in section \ref{sec5}
 and satisfies
\begin{equation}
  N = \bar{\Pi}_+ N \bar{\Pi}_+
\end{equation}
As the map $N$ is tensorial, the $N_{\mu_1 \ldots \mu_q} {}^{\nu_1 \ldots
\nu_q}$ transform as the components of a tensor. The generalised inverse
$\tilde{N}$ satisfies
\begin{equation}
  \tilde{N} = \bar{\Pi}_+ \tilde{N} \bar{\Pi}_+
\end{equation}
and has components $\tilde{N}_{\mu_1 \ldots \mu_q} {}^{\nu_1 \ldots \nu_q}$
\begin{equation}
  \tilde{N} (Q)_{\mu_1 \ldots \mu_q} = \frac{1}{q!} \tilde{N}_{\mu_1 \ldots
  \mu_q}{} ^{\nu_1 \ldots \nu_q} Q_{\nu_1 \ldots \nu_q}
\end{equation}
satisfying
\begin{equation}
  \label{ntilcon} \frac{1}{q!} \tilde{N}_{\mu_1 \ldots \mu_q} {}^{\nu_1 \ldots
  \nu_q} N_{\nu_1 \ldots \nu_q} {} ^{\rho_1 \ldots \rho_q} =
  (\bar{\Pi}_+)_{\mu_1 \ldots \mu_q} {}  ^{\rho_1 \ldots \rho_q}
\end{equation}
As the components $N_{\nu_1 \ldots \nu_q} {} ^{\rho_1 \ldots \rho_q},
(\bar{\Pi}_+)_{\mu_1 \ldots \mu_q}  {} ^{\rho_1 \ldots \rho_q}$ transform
tensorially, the components $\tilde{N}_{\mu_1 \ldots \mu_q} {}^{\nu_1 \ldots
\nu_q}$ found by solving (\ref{ntilcon}) do also, and the map $\tilde{N} $ is
then tensorial. This is in accord with the explicit expression (\ref{nti}) for
$\tilde{N}$ that can be expanded as a power series (\ref{npow}) with each term
in the expansion a tensorial map.

Given that $K, \tilde{N}$ are tensorial, the map $M$ is also tensorial so that $M (Q)$
transforms under coordinate transformations as a $q$-form, with the coefficients $M_{\mu_1 \ldots
\mu_q}^{\nu_1 {\ldots \nu_q} } (x)$ appearing in the expression (\ref{Mcomps})
for $M (Q)$ transforming as the components of a tensor. As a result, the action
(\ref{act}) is constructed covariantly and so, for global interpolating structures, is
  invariant under 
diffeomorphisms.
For local interpolating structures  the lagrangian transforms covariantly under infinitesimal coordinate transformations on the patch $U_{\alpha}$.
The interpolating structure only appears in the lagrangian through $M(Q)$ and in subsection \ref{globM}
it was shown that $M(Q)$ is a  globally well-defined $q$-form, so that the action (\ref{act}) is well-defined.

\subsection{Symmetries of free field actions}
\label{sec62}

The field equations from the action (\ref{act}) are
\[ \ast F = F, \qquad \bar{\ast} G = G \]
\[ d F = 0 \qquad d G = 0 \]
so that there are local $p$-form potentials $A, C$:
\[ G = d C  , \qquad F = d A \]
An alternative formulation giving these field equations is to take the action
\begin{equation}
  \label{salt} S_{\rm{alt}} = S_g - S_s
\end{equation}
with
\begin{equation}
  \label{salty} S_g = \frac{1}{2}  \int d A \wedge \ast d A, \qquad S_s =
  \frac{1}{2}  \int d C \wedge \bar{\ast} d C
\end{equation}
\[ \  \]
where the field equations are to be supplemented by the constraints
\begin{equation}
  \ast F = F, \qquad \bar{\ast} G = G
\end{equation}
Under a passive diffeomorphism -- a change of coordinates with $\delta x^{\mu}
= \xi^{\mu}$ -- the action is invariant with all fields transforming
tensorially. In the usual way, this can also be viewed as an active
transformation in which the coordinates are unchanged ($\delta x^{\mu} = 0$)
but the fields transform.

For active transformations with fields transforming, but points unchanged,
there are more possibilities. As $g, \bar{g}$ are independent gauge fields,
one might expect there to be two gauge symmetries with parameters $\chi^{\mu}$
and $\zeta^{\mu}$ with
\[ \delta g_{\mu \nu} = 2 \partial_{(\mu  } \zeta_{  \nu)} +
   \ldots, \quad \delta \bar{g}_{\mu \nu} = 2 \partial_{(\mu  }
   \chi_{  \nu)} + \ldots \]
Indeed, one can consider ``diffeomorphisms'' with parameter $\chi^{\mu}$
acting  {\it{only}} on the fields in $S_s $ or ``diffeomorphisms'' with
parameter $\zeta^{\mu}$ acting  {\it{only}} on the fields in $S_g$. The
action $S_{\rm{alt}} $ is in fact invariant under
\begin{equation}
  \delta C =\mathcal{L}_{\chi} C  , \quad \delta \bar{g}
  =\mathcal{L}_{\chi}  \bar{g}, \quad \delta A = 0  , \quad \delta g =
  0
\end{equation}
giving
\begin{equation}
  \delta G = d i_{\chi} G =\mathcal{L}_{\chi} G, \quad \delta F = 0
\end{equation}
and under
\begin{equation}
  \delta C = 0, \quad \delta \bar{g} = 0, \quad \delta A =\mathcal{L}_{\zeta}
  b, \quad \delta g =\mathcal{L}_{\zeta} g
\end{equation}
giving
\begin{equation}
  \delta F = d i_{\zeta} F =\mathcal{L}_{\zeta} F, \quad \delta G = 0
\end{equation}
for vector fields $\chi^{\mu}, \zeta^{\mu}$. The active transformation with
$\chi^{\mu} = \zeta^{\mu}$ is the active diffeomorphism considered above, with
$\xi^{\mu} = \chi^{\mu} = \zeta^{\mu}$.

The theory  is also invariant under  the gauge transformations
\begin{equation}
  \delta A = d \beta, \quad \delta C = d \gamma
\end{equation}
Combining $\chi^{\mu}, \zeta^{\mu}$ transformations with suitable gauge
transformations gives
\begin{equation}
  \delta C = i_{\chi} G, \quad \delta A = i_{\zeta} F
\end{equation}
The field equations for the   action (\ref{act}) are the same as the
ones for this alternative action, so the field equations for (\ref{act}) might be expected to have both
$\chi^{\mu}, \zeta^{\mu}$ symmetries; this will be checked in the following subsections.

In general, a coordinate-independent action (invariant under the passive
transformations) with interactions between $g$ and $\bar{g}$ will break the
$\chi^{\mu}, \zeta^{\mu}$ symmetries down to the diagonal subgroup with
$\chi^{\mu} = \zeta^{\mu}$. The quantity $M (Q)$ depends on both $g$ and
$\bar{g}$ so the action (\ref{act}) might be expected to be invariant under only the
diagonal subgroup. However, Sen's action (with $\bar{g} = \eta$) is invariant
under a symmetry under which $\bar{g} = \eta$ and $G$ are invariant \cite{Sen:2015nph,Sen:2019qit,Andriolo:2020ykk}
and it will now be shown that this generalises to the action with
general $\bar{g}$.

\subsection{$\zeta$-Symmetry}\label{zeet}

\subsubsection{Transformations}
\label{Transformations}

Motivated by the $\zeta$-symmetry of the action (\ref{salt}),(\ref{salty}) and
the results of \cite{Sen:2015nph,Sen:2019qit,Andriolo:2020ykk}, a symmetry of the action (\ref{act}) will now be sought
which has a gauge parameter given by an infinitesimal vector field $\zeta^{\mu}$ under
which
\begin{equation}
  \label{cdiff} \delta G = 0, \quad \delta \bar{g} = 0, \quad \delta g
  =\mathcal{L}_{\zeta} g
\end{equation}
The invariance of the field strength $G$, $\delta G = 0,$ is achieved by
choosing
\begin{equation}
  \label{Qtrans} \delta Q = - \frac{1}{2}  (1 + \bar{\ast}) d \delta P
\end{equation}
It remains to find $\delta P$.

The variation of the action (\ref{act}) under (\ref{cdiff}),(\ref{Qtrans})
gives (see Appendix C)
\begin{equation}
  \delta S = \label{varS3} \int  \left\{ - 2 F \wedge \left( d \delta P -
  \frac{1}{2} [\delta M] (Q) \right) \right\}
\end{equation}
where $\delta M$ is the change in $M$ due to the change in $g$:
\begin{equation}
  [\delta M] (Q) = \delta (M (Q))- M (\delta Q)
\end{equation}
 Then
invariance requires $\delta P$ to satisfy
\begin{equation}
\Pi_- \left( d \delta P - \frac{1}{2} [\delta M] (Q) \right) = 0 
\end{equation}

\subsubsection{Variation of M}\label{mvary}

To proceed, it is necessary to calculate $[\delta M] (Q) $. Under the
transformations $\delta \bar{g} = 0, \delta g =\mathcal{L}_{\zeta} g$ with
parameter a vector field $\zeta^{\mu}$ (combined with an orthogonal
transformation (\ref{abc}); see appendix C)
\begin{equation}
  \delta f_{\mu}{}^{ \nu} = \zeta_{\mu}{}^{\rho}
  f_{\rho}{}^{ \nu}
\end{equation}
where
\begin{equation}
  \zeta_{\mu}{}^{ \nu} = (\nabla_{\mu} \zeta^{\nu})
\end{equation}
and $\nabla_{\mu}$ is the Levi-Civita connection for the metric $g$.

Define the map $R_{\zeta}$ on $r$-forms $X$ by
\begin{equation}
  \label{ris} R_{\zeta} (X)_{\mu_1 \ldots \mu_r} = r (\nabla_{[ 
  \mu_1} \zeta^{\rho}) X_{| \rho | \mu_2 \ldots   \mu_r]}
\end{equation}
Then
\begin{equation}
  \delta (\Phi (X)) = \Phi (\delta X) + R_{\zeta} (\Phi (X))
\end{equation}
This map $R_{\zeta}$ on $r$-forms $X$ (\ref{ris}) can be rewritten (see
Appendix C) as
\begin{equation}
  \label{riisl} R_{\zeta} (X) =\mathcal{L}_{\zeta} {X + \nabla_{\zeta}}  X
\end{equation}
where ${\nabla_{\zeta}}  X = \zeta^{\rho} \nabla_{\rho} X$.

Consider now the variation of
\begin{equation}
  F = \Phi (\alpha)
\end{equation}
giving
\begin{equation}
  \label{delf} \delta F = R_{\zeta} (F) + \Phi (\delta \alpha)
\end{equation}
Note that $\Phi (\delta \alpha)$ is self-dual with respect to $g$, $\ast \Phi
(\delta \alpha) = \Phi (\delta \alpha)$, but $R_{\zeta} (F)$ is not $g$-self-dual so
that $\delta F$ is not self-dual with respect to $g$. Instead, by
construction, $F + \delta F$ is (to linearised order) self-dual with respect
to $g + \delta g$: to lowest order in $\delta g$, $F + \delta F = \ast' (F +
\delta F)$ where $\ast'$ is the Hodge dual for $g + \delta g$.

Decomposing $\delta F$ (\ref{delf}) using the projectors $\bar{\Pi}_{\pm}$ (which
are invariant since $\delta \bar{g} = 0$) gives
\begin{equation}
  \delta Q = \bar{\Pi}_+ R_{\zeta} (F) + N \delta \alpha
\end{equation}
and
\begin{equation}
  \delta (M (Q)) = \bar{\Pi}_- R_{\zeta} (F) + K \delta \alpha
\end{equation}
Then
\begin{equation}
  \delta \alpha = \tilde{N} \Delta Q
\end{equation}
where
\begin{equation}
  \Delta Q \equiv \delta Q - \bar{\Pi}_+ R_{\zeta} (F)
\end{equation}
giving
\begin{equation}
  \delta (M (Q)) = \bar{\Pi}_- R_{\zeta} (F) + K \tilde{N} \Delta Q
\end{equation}
so that
\begin{equation}
  \delta (M (Q)) = \bar{\Pi}_- R_{\zeta} (F) + M (\Delta Q)
\end{equation}
and as a result
\begin{equation}
  \delta (M (Q)) = \bar{\Pi}_- R_{\zeta} (F) + M (\delta Q) - M (\bar{\Pi}_+
  R_{\zeta} (F))
\end{equation}
Then
\begin{equation}
  (\delta M) (Q) = \delta [M (Q)]- M (\delta Q)
\end{equation}
is given by
\begin{equation}
  (\delta M) (Q) = \bar{\Pi}_- R_{\zeta} (F) - \{ \bar{\Pi}_+ R_{\zeta} (F) +
  M (\bar{\Pi}_+ R_{\zeta} (F)) \} + \bar{\Pi}_+ R_{\zeta} (F)
\end{equation}
i.e.\
\begin{equation}
  \label{delm} (\delta M) (Q) = R_{\zeta} (F) - \{ \Xi + M (\Xi) \} =
  R_{\zeta} (F) - \Psi (\Xi)
\end{equation}
where
\begin{equation}
  \label{xis} \Xi \equiv \bar{\Pi}_+ R_{\zeta} (F)
\end{equation}

\subsubsection{Variation of the Action}

From subsection \ref{Transformations} the variation of the action is
\begin{equation}
  \delta S = \label{varS3a} \int  \left\{ - 2 F \wedge \left( d \delta P -
  \frac{1}{2} (\delta M) (Q) \right) \right\}
\end{equation}
and from subsection \ref{mvary}
\begin{equation}
  \label{delmis} (\delta M) (Q) = R_{\zeta} (F) - \{ \Xi + M (\Xi) \}
\end{equation}
where
\begin{equation}
  \Xi \equiv \bar{\Pi}_+ R_{\zeta} (F)
\end{equation}
Substituting the expression for $\delta M$ (\ref{delmis}) in (\ref{varS3})
gives (see Appendix C)
\begin{equation}
  \delta S = \label{varS3aasdf} \int  \{ - 2 F \wedge (d \delta P - d i_{\zeta}
  F) \}
\end{equation}
and so the action is invariant if 
\begin{equation}
  \delta P = i_{\zeta} F
\end{equation}
(More generally, it is invariant if $\delta P = i_{\zeta} F + d \alpha$  for any
$\alpha$.)

Next consider  the variation of $F = Q + M (Q)$, so that
\[ \delta F = \delta Q + M (\delta Q) + (\delta M) (Q) \]
gives, using (\ref{siis}),(\ref{delm}),(\ref{xis}) (see Appendix C)
\begin{equation}
  \delta F =\mathcal{L}_{\zeta} F - \Psi (\bar{\Pi}_+ i_{\zeta} d F)
\end{equation}
The field equations imply $d F = 0$, so that on-shell
\begin{equation}
  \delta F \approx \mathcal{L}_{\zeta} F
\end{equation}
Thus the transformation of $F$ is a combination of a diffeomorphism with
parameter $\zeta$ and an on-shell trivial symmetry, i.e.\ a transformation that
vanishes on-shell. This is in agreement (up to an on-shell trivial symmetry) with
what was found in subsection \ref{sec62}.

To summarise, the action is invariant under the $\zeta$-transformations
\[ \delta G = 0, \quad \delta \bar{g} = 0, \quad \delta g =\mathcal{L}_{\zeta}
   g \]
\begin{equation}
  \label{Qtransabc} \delta Q = - \frac{1}{2}  (1 + \bar{\ast}) d \delta P, \qquad \delta P = i_{\zeta} F
\end{equation}
Then
\[ \delta G = 0 \]
while
\[ \delta F \approx d i_{\zeta} F =\mathcal{L}_{\zeta} F \]
up to terms involving $d F$ which vanish on-shell using the field equation $d
F = 0$, so that on-shell this agrees with the $\zeta$ symmetry of the action
$S_{ \rm{alt}}$ given above.

\subsection{The $\chi$-symmetry}

Combining the $\zeta$-transformations of the previous subsection with a  diffeomorphism with parameter $\xi = - \zeta$
gives a transformation with parameter $\chi \equiv
\xi = - \zeta 
$ given by
\[  \delta \bar{g} =\mathcal{L}_{\chi}  \bar{g}, \quad \quad \delta g = 0
\]
\[
  \delta G = d i_{\chi} G =\mathcal{L}_{\chi} G, \quad \delta F =  \Psi
   (\bar{\Pi}_+ i_{\chi} d F) \]
\begin{equation}
  \label{chitrans}\delta P= \mathcal{L}_{\chi}P-i_{\chi} F, \qquad \delta Q=\mathcal{L}_{\chi}Q
+ \frac{1}{2}  (1 + \bar{\ast}) d i_{\chi} F
\end{equation}
As a result,  $\delta F = 0$ on-shell, i.e.\ up to terms involving $d F$. 
Then on-shell $\delta F \approx 0$, $\delta G = \mathcal{L}_{\chi} G$,
which agrees  with the $\chi$ symmetry of the action $S_{ \rm{alt}}$
given in subsection \ref{sec62}.

The theory then has two gauge  symmetries, one with parameter $\zeta$ and one with parameter
$\chi$, with the diffeomorphism symmetry  arising as the diagonal subgroup with $\zeta=\chi$.

\section{Transition functions and Global Structure}
\label{sec7}

\subsection{Finite Gauge Transformations}

In this subsection, finite forms of the gauge transformations will be found to facilitate the discussion of 
of patching conditions.
A coordinate transformation $x\to x'(x)$ induces changes of tensor fields $T(x)\to T'(x)$. For example a scalar field $\Phi(x)$ transforms to a field $\Phi'(x)$ with
\begin{equation}
\Phi'(x')=\Phi(x)
\label{scal}
\end{equation}
and a metric $g(x)$ transforms to a metric $g'(x)$ with
\begin{equation}
  g'_{\mu \nu} (x') = g_{\rho \sigma} (x) \frac{\partial x^{\rho}}{{\partial
  x'}^{\mu}}  \frac{\partial x^{\sigma}}{{\partial x'}^{\nu}} \label{transf}
\end{equation}
These can be viewed as passive or active transformations. In a passive transformation  there is a change of coordinates so that the coordinates $x(p)$ of a point $p$ are changed to $x'(p)$, while in an active transformation a diffeomorphism takes a point $p$ to a point $p'$ and so takes the coordinates
$x(p)$  of $p$ to the coordinates $x'=x(p')$ of $p'$ (assuming for simplicity that $p,p'$ are in the same coordinate patch).

A  vector field $\xi^\mu(x)$ induces a  coordinate transformation with 
$$x'=\exp (-\xi^\nu \partial _\nu)x^\mu$$
so that ${x'}^\mu=x^\mu -\xi ^\mu +O(\xi^2)$ and for any function $f$
$$f(x')=\exp (-\xi^\mu \partial _\mu)f(x)$$
(The minus sign is chosen for later convenience.) As an active transformation, this results from the diffeomorphism given by flowing along the integral curves of $-\xi$ by a unit parameter distance.
In terms of local coordinates, the integral curves are curves $x^\mu(t)$ that are solutions of
$$
\frac {dx^\mu(t)}{dt}=-\xi^\mu (x(t))
$$
For a point $x_0$, take the integral curve with $x(0)=x_0$. Then the diffeomorphism takes $x(0)=x_0$ to $x(1)$. 

For  a tensor field $T(x)$, this transformation takes the components of $T(x)$ to the components of $T'(x)$ which will be denoted $T'=\sigma(\xi)_*T$.
For example, for a scalar field $\Phi$, $\Phi'=\sigma(\xi)_*\Phi\,   $ is given  (using (\ref{scal})) by
\begin{equation}
[\sigma(\xi)_*\Phi ](x) =\exp (\xi^\mu \partial _\mu)\Phi (x)=\Phi(x) +\xi^\mu \partial _\mu\Phi (x)+O(\xi^2)
\end{equation}
For a metric with  (\ref{transf}),
\begin{equation}
[\sigma(\xi)_*\, g_{\mu\nu} ](x) =g_{\mu\nu} (x)+[{\cal L}_\xi g_{\mu\nu}] (x)
+O(\xi^2)
\end{equation}
so that
\begin{equation}
g_{\mu\nu} \to g'_{\mu\nu} =\sigma(\xi)_*g_{\mu\nu}
\end{equation}
is the finite form of the gauge transformation given for infinitesimal $\xi $ by
\begin{equation}
g_{\mu\nu} \to g'_{\mu\nu} =g_{\mu\nu} +{\cal L}_\xi g_{\mu\nu}
\end{equation}
Consider now the $\zeta$ and $\chi$ transformations. The infinitesimal $\zeta$   transformations
given by
\begin{equation}
\delta G = 0, \quad \delta \bar{g} = 0, \quad \delta g =\mathcal{L}_{\zeta}
   g, \quad
   \delta F \approx  \mathcal{L}_{\zeta} F 
\end{equation}
(where $\approx$ indicates equality on-shell) have a finite form given by 
\begin{equation}
G' = G, \quad   \bar{g}' =  \bar{g}, \quad  g' =\sigma(\zeta)_*g
   , \quad
    F' \approx  \sigma(\zeta)_*F 
\end{equation}
for a finite vector field $\zeta ^\mu$, while the
infinitesimal $\chi$   transformations
given by
\begin{equation}
\delta \bar{g} =\mathcal{L}_{\chi}  \bar{g}, \quad  \delta g = 0, \quad
 \delta G = \mathcal{L}_{\chi} G, \quad
 \delta F \approx 0
\end{equation}
have a finite form given by 
\begin{equation}
 \bar{g} '= \sigma(\chi)_*
  \bar{g}, \quad   g' = g, \quad
  G '=\sigma(\chi)_* G, \quad
   F' \approx F
\end{equation}
for a finite vector field $\chi ^\mu$.

\subsection{Transition functions}

Spacetime is a manifold covered with open sets $U_{\alpha}$ each of which is
diffeomorphic to an open set $\hat{U}_{\alpha} \subseteq \mathbb{R}^d$ and (for
each $\alpha$) the diffeomorphism takes a point $p \in U_{\alpha}$ to the
coordinates of $p$, $x_{(\alpha)}^{\mu} (p)$. For a point in an overlap $p \in
U_{\alpha} \cap U_{\beta}$, either of the sets of coordinates $x_{(\alpha)}^{\mu} (p)$
and $x_{(\beta)}^{\mu} (p)$ can be used and these are related by a
transition function $\varphi_{\alpha \beta} : \hat{U}_{\beta} \rightarrow
\hat{U}_{\alpha}$
\begin{equation}
  x_{(\alpha)}  = \varphi_{\alpha \beta} (x_{(\beta)} )
\end{equation}
If the transition function is induced by a vector field $\xi_{\alpha \beta}$
defined on $U_{\alpha} \cap U_{\beta}$ so that
\begin{equation}
  {x^{\nu}_{(\alpha)}}  = \exp \left( \xi_{\alpha \beta}^{\mu} (x_{(\beta)}) 
  \frac{\partial}{\partial x_{(\beta)}^{\mu}} \right) x^{\nu}_{(\beta)}
\end{equation}
then a tensor $T (p)$ for $p \in U_{\alpha} \cap U_{\beta}$  has
components $T_{(\alpha)}$ in the $x_{(\alpha)}^{\mu}$ coordinate system and
components $T_{(\beta)}$ in the $x_{(\beta)}^{\mu}$ coordinate system that are
related by
\[ T_{(\alpha)} = \sigma (\xi_{\alpha \beta})_{\ast} T_{(\beta)} \]
For the metric tensor $g$,
\begin{equation}
  g_{(\alpha)} = \sigma (\xi_{\alpha \beta})_{\ast} g_{(\beta)}
  \label{mtran}
\end{equation}
can be written explicitly as
\begin{equation}
  g_{(\alpha) \mu \nu} (x_{(\alpha)}) = g_{(\beta) \rho \sigma} (x_{(\beta)})
  \frac{\partial x^{\rho}_{(\beta)}}{\partial x^{\mu}_{(\alpha)}}
  \frac{\partial x^{\sigma}_{(\beta)}}{\partial x^{\nu}_{(\alpha)}}
  \label{mtran2}
\end{equation}
giving the standard relation between the components in the two coordinate
systems.

\subsection{Gauge Fields and Tensor Fields}

For a gauge theory, the transition functions can also involve a gauge
transformation. For example, for a Yang-Mills gauge field $A,$ the components
$A_{(\alpha)}$ and $A_{(\beta)}$ in the two patches will be related by a
coordinate transformation combined with a gauge transformation $h_{(\alpha
\beta)}$ which is a map from $U_{\alpha} \cap U_{\beta}$ to the gauge group
$H$. Explicitly,
\begin{equation}
  A_{(\alpha) \mu} (x_{(\alpha)}) = \frac{\partial x^{\nu}_{(\beta)}}{\partial
  x^{\mu}_{(\alpha)}} \left( h_{(\alpha \beta)}^{- 1} \left[
  \frac{\partial}{\partial x_{(\beta)}^{\nu}} + A_{(\beta) \nu} (x_{(\beta)})
  \right] h_{(\alpha \beta)} \right)
\end{equation}
where $h_{(\alpha \beta)} (x_{(\beta)}) \in H$.
The transition functions have to satisfy the usual consistency conditions in triple overlaps $U_{\alpha} \cap U_{\beta}\cap U_{\gamma}$. This means that $A$ is not a tensor field but has  more general transition functions, which are those for a bundle connection.

For the theory considered in this paper, there are two fields $g, \bar{g}$ which so far
have been taken to be tensor fields. This means they have 
  transition functions 
  \begin{equation}
  g_{(\alpha)} = \sigma (\xi_{\alpha \beta})_{\ast} g_{(\beta)}, \qquad \bar g_{(\alpha)} = \sigma (\xi_{\alpha \beta})_{\ast} \bar g_{(\beta)}
  \label{mettran}
\end{equation}  
in $U_{\alpha} \cap U_{\beta}$,
so that both can be regarded as
metrics on spacetime. This ensures diffeomorphism invariance in the usual way.

The theory considered here  has $\zeta$ and $\chi$ gauge symmetries for which $g,
\bar{g}$ can be regarded as gauge fields so that in principle a more general
structure is possible in which the transition functions consist of coordinate
transformations combined with $\zeta$ and $\chi$ transformations. 
The transition functions would then be
  \begin{equation}
  g_{(\alpha)} =\sigma (\zeta_{\alpha \beta})_{\ast}  \sigma (\xi_{\alpha \beta})_{\ast} g_{(\beta)}, \qquad \bar g_{(\alpha)} =\sigma (\chi_{\alpha \beta})_{\ast}  \sigma (\xi_{\alpha \beta})_{\ast} \bar g_{(\beta)}
  \label{patg}
\end{equation}  
for vector fields $\zeta_{\alpha \beta},\chi_{\alpha \beta}$ on $U_{\alpha} \cap U_{\beta}$.
The other fields transforming would have patching relations involving $\zeta$ and $\chi$ transformations
so that
\begin{equation}
\label{patF}
  F_{(\alpha)} \approx
  \sigma (\zeta_{\alpha \beta})_{\ast}  \sigma (\xi_{\alpha \beta})_{\ast} F_{(\beta)}, \qquad  G_{(\alpha)} =\sigma (\chi_{\alpha \beta})_{\ast}  \sigma (\xi_{\alpha \beta})_{\ast} G_{(\beta)}
\end{equation}  
(Note that one could instead choose other forms of combining the gauge and coordinate transformations, such as  e.g.\ $g_{(\alpha)} =\sigma (\xi_{\alpha \beta})_{\ast} \sigma (\zeta_{\alpha \beta})_{\ast}   g_{(\beta)}$ or $g_{(\alpha)} =\sigma (\zeta_{\alpha \beta}+\xi_{\alpha \beta})_{\ast}   g_{(\beta)}$, but these give similar results.)
In  triple overlaps $U_{\alpha} \cap U_{\beta}\cap U_{\gamma}$, these transition functions should satisfy
the consistency conditions
\begin{equation}
\sigma (\zeta_{\alpha \beta})_{\ast} \sigma (\zeta_{ \beta \gamma})_{\ast} \sigma (\zeta_{\gamma \alpha })_{\ast} =1, \quad
\sigma (\chi_{\alpha \beta})_{\ast} \sigma (\chi_{ \beta \gamma})_{\ast} \sigma (\chi_{\gamma \alpha })_{\ast} =1
\end{equation}
To linear order in the vector fields, these transition functions
are
\begin{equation}
  g_{(\alpha)} = g_{(\beta)} +\mathcal{L}_{\xi_{\alpha \beta}}g_{(\beta)}+\mathcal{L}_{\zeta_{\alpha \beta}}g_{(\beta)}+\dots  , 
  \qquad \bar g_{(\alpha)} =   \bar g_{(\beta)}+\mathcal{L}_{\xi_{\alpha \beta}}\bar g_{(\beta)}+\mathcal{L}_{\chi_{\alpha \beta}}\bar g_{(\beta)}
  +\dots
\end{equation}

If $g$ is to be
the physical spacetime metric,  it should be a tensor field with
components in different coordinate systems related by  (\ref{mtran}),(\ref{mtran2}). 
Then the transition functions should not include 
$\zeta$ transformations as these would act on $g$ but they could involve  $\chi$
transformations which leave $g$ invariant. This gives patching conditions (\ref{patg}),(\ref{patF}) with $\zeta_{\alpha \beta}=0$, giving for $g,\bar g$
\begin{equation}
  g_{(\alpha)} = \sigma (\xi_{\alpha \beta})_{\ast} g_{(\beta)}, \qquad \bar g_{(\alpha)} =\sigma (\chi_{\alpha \beta})_{\ast}  \sigma (\xi_{\alpha \beta})_{\ast} \bar g_{(\beta)}
  \label{patga}
\end{equation}  
Then the physical fields $g,F$ would be tensors while the non-physical fields $\bar g,G$ would have non-standard transition functions with the coordinate transformation twisted by a $\chi$ gauge transformation.

Requiring the transition functions to consist of $\xi$ and $\chi$ transformations but not $\zeta$ transformations in general restricts the gauge symmetries acting on the physical fields to be diffeomorphisms but still allows  both  $\xi$ and $\chi$ transformations
as gauge transformations acting on the shadow sector.

With $\bar{g}$ a tensor gauge field with transition functions including $\chi$ transformations, more general
configurations are possible  than would have been the case if $\bar g$ were  a tensor field.
In particular, choosing
$ \zeta_{\alpha \beta}=-\xi_{\alpha \beta}$
gives
$$ \bar g_{(\alpha)} =   \bar g_{(\beta)}$$
so, remarkably,
it is possible for $\bar{g}$
 to have the same conmponents
 in every patch $U_\alpha$. 
In particular, this allows taking $\bar{g}_{\mu \nu} = \eta_{\mu \nu}$ in every
patch $U_\alpha$, which is consistent with Sen's form of the action. That is, Sen's
form of the action is valid globally provided $\eta$ is not a tensor but is
instead a tensor gauge field of the type proposed here. Furthermore, taking
$\bar{e}_{\mu}^a = \delta_{\mu}^a$ gives $f_{\mu}{}^{\nu} = e_{\mu}^a
\delta_a^{\nu}$ which is not a tensor but instead has transition functions involving $\chi$ transformations.
The transition functions with 
$ \zeta_{\alpha \beta}=-\xi_{\alpha \beta}$
are precisely $\zeta$ transformations, so that the same transition functions would be obtained by setting
$ \chi_{\alpha \beta}=\xi_{\alpha \beta}=0$ in (\ref{patg}),(\ref{patF}) and then renaming $ \zeta_{\alpha \beta}\to\xi_{\alpha \beta}$.

In some circumstances, such as those discussed in section \ref{secCoup}, it is of
interest to consider regarding $\bar{g}$ as the spacetime metric and $g$ as
the auxiliary field. In those circumstances $\bar{g}$ should be a tensor field
but one could then consider taking $g$ to be a   gauge field. Then the
transition functions can involve diffeomorphisms together with $\zeta$
transformations but not $\chi$ transformations, so that the situation is
similar to that above but with the roles of $g$ and $\bar{g}$ interchanged, with the transition functions given by (\ref{patg}),(\ref{patF}) with $\chi_{\alpha \beta}=0$.

\section{Basis for $q$-forms}
\label{sec7a}

The (anti-) self-dual forms at a point $x$ form a vector space of dimension $N_q=(2q!)/2(q!)^2$.
Following  \cite{Andriolo:2020ykk}, it is useful to introduce a basis $\omega^A_+ (x)$
for the $\bar g$-self-dual $q$-forms at $x$ and a basis 
$ \omega_{- A}
(x)$ for the $\bar g$-anti-self-dual $q$-forms at $x$, with $A,B=1,\dots, N_q$.
\begin{equation}
  \bar{\ast} \omega^A_+ = \omega^A_+, \qquad \bar{\ast} \omega_{- A} = -
  \omega_{- A}
\end{equation}
Then acting with $\Phi$ 
 induces bases
\begin{equation}
  \varphi^A = \Phi (\omega^A_+), \qquad \varphi_A = \Phi (\omega_{- A})
\end{equation}
for the $  g$-self-dual and $  g$-anti-self-dual $q$-forms at $x$,
satisfying
\begin{equation}
  \ast \varphi^A = \varphi^A, \qquad \ast \varphi_A = - \varphi_A
\end{equation}
If the $\omega$ are orthornormal with respect to $\bar{g}$
\begin{equation}
  (\omega^A, \omega_P)_{\bar{g}} = \delta^A{} _P
\end{equation}
then the induced basis is orthonormal with respect to $g$
\begin{equation}
  (\varphi^A, \varphi_P)_g = \delta^A {}_P
\end{equation}
The bases will be chosen to be orthonormal in what follows.

Expressing $\varphi^A$ in the $\omega$ basis gives
\begin{equation}
  \varphi^A = N^A{}_{ P} \omega^P_+ + K^{AP} \omega_{- P}
\end{equation}
where
\begin{equation}
  N^A{}_P \omega^P_+ = \frac{1}{2}  (1 + \bar{\ast}) \varphi^A
  = \frac{1}{2}  (1 + \bar{\ast}) \Phi \omega^A_+ = N (\omega^A_+)
\end{equation}
\begin{equation}
  K^{AP} \omega_{- P} = \frac{1}{2}  (1 - \bar{\ast}) \varphi^A = \frac{1}{2} 
  (1 - \bar{\ast}) \Phi \omega^A_+ = K (\omega^A_+)
\end{equation}
with the operators
$N,K$ defined by (\ref{wertwed1}),(\ref{wertwed2}).
 
 The generalised inverse $\tilde N$
 would be  an inverse $(N^{- 1})^A{}_P = (\tilde{N})^A{}_P$ to the matrix
$N^A{}_P$, satisfying $N^A{}_B (N^{- 1})^B{}_C = \delta^A{}_C$. 
A key issue for the construction discussed in this paper is whether 
such an inverse exists.
 If $g = \bar{g}$, then
$\varphi^A = \omega^A$ and $N^{A }{}_B = \delta^{A }{}_B$ is invertible. It should
remain invertible provided $g$ and $\bar{g}$ are sufficiently close, and this can be seen in the perturbation expansion in Appendix B.\footnote{ Note that if $g$
and $\bar{g}$ were equal but the orientations were chosen to be opposite so that $\ast = -
\bar{\ast}$, then $N = 0$ and $K^{{AB}} = \delta^{{AB}}$. Such cases are avoided here by choosing the orientations defined by $\ast $ and $
\bar{\ast}$
to be the same.}

Consider now the equations
\begin{equation}
  F = Q + M = \Phi (\alpha)
\end{equation}
\begin{equation}
  Q = N (h), \quad M = K (\alpha)
\end{equation}
In bases, the forms have components
\begin{equation}
  h = h_A \omega^A_+, \quad Q = Q_A \omega^A_+, \quad M = M^A \omega_{- A}
\end{equation}
so that
\begin{equation}
  Q = Q_A \omega^A_+ = N (\alpha_A \omega^A_+) = \alpha_A N (\omega^A_+) = \alpha_A
  N^A{}_{P} \omega^P_+
\end{equation}
from which
\begin{equation}
  Q_P = \alpha_A N^A{}_{ P}
\end{equation}
and
\begin{equation}
  \alpha_P = Q_A (N^{- 1})^A{}_{P}
\end{equation}
Next,
\begin{equation}
  M = M^A \omega_{- A} = K (\alpha_A \omega^A_+) = \alpha_A K (\omega^A_+) = \alpha_A K^{AP}
  \omega_{- P}
\end{equation}
so
\begin{equation}
  M^P = U_A K^{AP} = Q_C (N^{- 1})^C{}_{ P} K^{AP}
\end{equation}
Define components $M^{A B}$ by
\begin{equation}
  M^A (Q) = M^{AB} Q_B
\end{equation}
so that
\begin{equation}
  M^{{AB}} = (N^{- 1})^B{}_{{} C} K^{A C}
\end{equation}
in agreement with \cite{Andriolo:2020ykk}. From (\ref{symm}), $M^{A B}$ is symmetric,
$M^{A B} = M^{B A}$.

%%%%
\section{Invertibility of $N$ and Explicit Expressions for $N,K,M$}
\label{sec8}

In this section it will be shown that $N$ is invertible and  explicit expressions
for $N, K$ and $M$ will be given. 
For the case in which $g$ and $\bar g$ are both tensor fields, this will be done by choosing a  frame
which is orthonormal for $\bar g$ and then using local Lorentz transformations (which are the frame rotations preserving this orthonormality) to bring the frame components of $g$ to a diagonal or near-diagonal form in which $N, K$ and $M$ are readily calculated. This is done in the following subsections.

Alternatively if $\bar g$ is a gauge field, the symmetries can be used to bring both metrics at a given point $p$ to the Minkowski metric so that $M(p)=0$. 
 To do this, one can first  choose normal coordinates $x^{\mu}$ 
so that   the coordinates of $p$ are $x^{\mu} (p) = 0$ and the metric $ {g}_{\mu \nu} (p) $ at $p$ is the Minkowski metric,
$ {g}_{\mu \nu} (p) = \eta_{\mu \nu}$, with
\begin{equation}
  {g}_{\mu \nu} (x) = \eta_{\mu \nu} - \frac{1}{6}  {R}_{\rho \mu \nu
  \sigma} x^{\rho} x^{\sigma} + O (x^3)
  \label{normal}
\end{equation}
For the case in which $g$ is a tensor field but $\bar g$ is a gauge field,  one can then use $\chi $ transformations to set $\bar g(p)
=\eta_{\mu \nu}$. Then, at $p$, $g(p)=\bar g(p)=\eta$ and  $f_\mu{}^\nu(p)=\delta_\mu {}^\nu$.
This then gives $N(p)=\bar \Pi_+$, $K(p)=0$ and hence $M(p)=0$. Then the generalised inverse exists and is given by  $\tilde N(p)=\bar \Pi_+$.
Thus at any point $p$ the symmetries can be used to bring the structure of the model to that of the Minkowski space theory discussed in subsection \ref{sec21}.

%%%%

\subsection{Frames and Standard Forms}

For the remainder of this section attention will be restricted to a  local patch $U$.
As in section \ref{sec43},  a vielbein $\bar{e}_{\mu}{}^a$ is introduced defining a frame  on $U$ that is orthonormal with
respect to $\bar{g}$
\begin{equation}
  \eta_{a b} = \bar{e}_a^{\mu} \bar{e}_b^{\nu}\,  \bar{g}_{\mu \nu}
\end{equation}
In this frame, the other metric $g$ has components
\begin{equation}
   G_{ab}=\bar e _a{}^\mu\bar e _b{}^\nu\, g_{\mu\nu}
   \end{equation}
Some linear algebra establishes that
  local Lorentz transformations $\bar{e}^a_{\mu} \rightarrow L^a{} _b
{\bar{e}^b_{\mu}} $   can be used to choose a frame in which the matrix $G_{a b}$ at a point $p$  is
brought to one of two standard forms $\bar{G}_{a b}$. The metric $\bar{G}_{a b}$  at $p$   can be 
brought to
 either the diagonal form
\begin{equation}
  \bar{G}_{a b}^{(D)} =  \rm{diag} (- \lambda_0^2, \lambda_1^2, \ldots,
  \lambda_{d - 1}^2)
  \label{diagl}
\end{equation}
with $0 < \lambda_1 ^2\leqslant \lambda_2^2  \leqslant \ldots \leqslant \lambda_{d
- 1}^2$ in a basis in which $\eta_{a b} =  \rm{diag} (- 1, 1, \ldots, 1)$, or to the form
\begin{equation}
  \overline{G }_{a b}^{(N)} = \left(\begin{array}{cc}
    0 & \lambda^2\\
    \lambda^2 & a
  \end{array}\right) \oplus  \rm{diag} (\lambda_1^2, \ldots, \lambda_{d -
  2}^2)
  \label{ndiagl}
\end{equation}
with $0 < \lambda_1^2 \leqslant \lambda_2^2 \leqslant \ldots \leqslant \lambda_{d
- 2}^2$ in a basis in which
\begin{equation}
  \eta = \left(\begin{array}{cc}
    0 & 1\\
    1 & 0
  \end{array}\right) \oplus  \rm{diag} (1, \ldots, 1)
\end{equation}
If $a = 0$ (\ref{ndiagl}) can be brought to the diagonal form (\ref{diagl}) by a change of basis.
The standard forms
(\ref{diagl}),(\ref{ndiagl})
 correspond to bringing $\eta ^{-1}G$ to either diagonal form
\begin{equation}
  \eta ^{-1} \bar{G} ^{(D)} =  \rm{diag} (\lambda_0^2, \lambda_1^2, \ldots,
  \lambda_{d - 1}^2)
\end{equation}
with eigenvalues $\lambda_0^2, \lambda_1^2, \ldots,
  \lambda_{d - 1}^2$
or to the Jordan normal form
\begin{equation}
 \eta ^{-1} \bar {G } ^{(N)} = \left(\begin{array}{cc}
     \lambda^2& a\\
0 &    \lambda^2 
  \end{array}\right) \oplus  \rm{diag} (\lambda_1^2, \ldots, \lambda_{d -
  2}^2)
\end{equation}
with eigenvalues $\lambda ^2 ,\lambda ^2 , \lambda_1^2, \ldots, \lambda_{d -
  2}^2$.
  Note that the eigenvalues $\lambda _a$ never vanish as a coordinate system has been chosen for each patch
   that has no coordinate singularities, so that in particular $\det g $ and $\det \bar g $ are non-zero.   
   
The patch $U$ is then $U=U_D\cup U_N$ where 
at each point in $U_D$ the metric can be brought to the diagonal form (\ref{diagl}) 
and at each point in $U_N$ the metric can be brought to the form (\ref{ndiagl}). 
 In the intersection $U=U_D\cap U_N$ the metric can be brought to the form (\ref{ndiagl}) with $a=0$. 
As the components of the metric $G_{ab}$ are smooth functions on $U$, the eigenvalues $\lambda_a$ and $a$ will  be smooth functions almost everywhere on $U$.
It is convenient to use local Lorentz transformations to choose a frame  so that $G=\bar G^{(D)}$ on $U_D$ and
  $G=\bar G^{(N)}$ on $U_N$.

An interpolating structure $f_\mu{}^\nu$ satisfying (\ref{fra}) has frame components $f_a{} ^b$ satisfying
\begin{equation}
  f_a {}^c f_b{} ^d \eta_{c d} = {G}_{a b}
  \label{eryery}
\end{equation}
so that  finding $f_a{} ^b$ involves taking a \lq square root' of $G_{ab}$ and in particular choosing a square root $\pm \lambda_i$ for each of the eigenvalues $\lambda _i^2$ of $G$.
Here the positive square root will be taken, so that all $\lambda _i$ will be taken to be positive. As it is assumed that there neither $g$ nor $\bar g$  have coordinate singularities  in $U$, the determinants  $\det  (g_{\mu\nu}) $ and
$\det  (\bar g_{\mu\nu}) $ are non-zero throughout $U$, the $\lambda _i$ are non-zero in $U$ so that if they are positive at one point, they are positive throughout $U$.

For  points in $U_D$ in which $G=\bar G^{(D)}$ with the  diagonal form (\ref{diagl}) one can take
\begin{equation}
  f_a{} ^b =  \rm{diag} (\lambda_0, \lambda_1, \ldots, \lambda_{d - 1})
  \label{diagf}
\end{equation}
with
$0 < \lambda_0 \leqslant \lambda_1 \leqslant \ldots \leqslant \lambda_{d
- 1}$ which satisfies (\ref{eryery}).
For  points in $U_N$ in which  $G=\bar G^{(N)}$ takes the  non-diagonal form (\ref{ndiagl}),  $f$ can be taken to be
\begin{equation}
f_a {}^b = \left(\begin{array}{cc}
     \lambda &  0\\
    a / 2 \lambda & \lambda
   \end{array}\right) \oplus  \rm{diag} (\lambda_1, \ldots, \lambda_{d - 2})
\end{equation}
with
$0 < \lambda_1 \leqslant \lambda_2 \leqslant \ldots \leqslant \lambda_{d
- 2}$ which
satisfies (\ref{eryery}). In either case,
\begin{equation}
\label{drte}
  f_{\mu} {}^{\nu} = \bar{e}_{\mu}{}^a f_a{} ^b \bar{e}_b{}^{\nu}
\end{equation}
satisfies (\ref{fra}) and
\[ e_{\mu}{}^a = f_{\mu} {}^{\nu} \bar{e}_{\nu}{}^a = \bar{e}_{\mu}{}^b f_b{} ^a \]
defines an orthonormal frame for $g$ on $U$:
\begin{equation}
  e_{\mu}{}^a e_{\nu}{}^b \eta_{a b} = g_{\mu \nu}
\end{equation}
Then the $f$ defined by (\ref{drte}) satisfies (\ref{fviel}).

The action of $\Phi$ on the vielbein is
\begin{equation}
\Phi (\bar{e}_{\mu}{}^a)= f_{\mu} {}^{\nu} \bar{e}_{\nu}{}^a
\end{equation}
which using (\ref{drte}) becomes
\begin{equation}
\Phi (\bar{e}_{\mu}{}^a)=  \bar{e}_{\mu}{}^b f_b{} ^a
\end{equation}
A basis for the $q$-forms at $p$ is
\begin{equation}
  E^{a_1 \ldots a_q} = \bar e^{a_1} \wedge \ldots \wedge \bar e^{a _q}
\end{equation}
where
$$ \bar{e}^a =
 \bar{e}_{\mu}{}^a dx^\mu$$
Then
\begin{equation}
 \Phi( E^{a_1 \ldots a_q} )= \bar e^{b_1} \wedge \ldots \wedge \bar e^{b _q} 
 f_{b_1}{}^{a_1}
 \ldots 
  f_{b_q}{}^{a_q}
\end{equation}

Diffeomorphisms can be used to simplify the form of the metrics.
For example, in two dimensions,  conformal gauge can be chosen for the metric $g_{\mu\nu}$
in which $g_{\mu\nu}=e^\phi \hat g_{\mu\nu}$ for some convenient background metric 
$\hat g_{\mu\nu}$. 
Then, if $\bar g_{\mu\nu}$ is taken to be $\hat g_{\mu\nu}$,  $G_{ab}$ is of the diagonal form (\ref{diagl}) with $\lambda _0^2=\lambda _1^2=e^\phi $ and only the diagonal form arises, so $U=U_D$.
Note that if $g_{\mu\nu}=e^\phi \bar g_{\mu\nu}$, then  $M=0$
with $N=1$ and $K=0$: no $M$ term is needed in this case as the Hodge duals $*$ and $\bar *$ are the same.
Choosing instead
 the Polyakov gauge, the metric  $ g_{\mu\nu}$ takes the form
\begin{equation}
 \left(\begin{array}{cc}
    0 & 1\\
    1 & g_{--}
  \end{array}\right)
\end{equation}
 in a  null coordinate system $x^\pm$ in which the Minkowski metric  is $ds^2=2dx^+dx^-$.
 Then
  taking  $\bar g_{\mu\nu}$ to be the Minkowski metric 
 with  vielbein $\bar e_\mu^a=\delta_\mu^a$,
$G_{ab}$ takes the non-diagonal form (\ref{ndiagl}) with $a=g_{--} $, $\lambda=1$.
Then in this gauge only the non-diagonal form arises, so $U=U_N$.
Thus in two dimensions, a coordinate choice can ensure that only the diagonal case  or only the non-diagonal case arise in $U$. The two-dimensional case is discussed in more detail in section \ref{2dims}.
In higher dimensions, a coordinate choice again simplifies the situation. However, for completeness both the diagonal and the non-diagonal cases will be considered below.

\subsection{A basis for self-dual forms}
\subsubsection{The Diagonal Case} \label{diagN}

For a $q$-form $\omega$ with components $\omega_{a_1\dots a_q}$ in the frame
defined by $\bar{e}_{\mu}^a$, the dual $\bar \ast \omega$ has components
$$\frac 1 {q!}\epsilon_{a_1\dots a_q}{}^{b_1\dots b_q}\omega_{b_1\dots b_q}$$
where $\epsilon_{a_1\dots a_d}$ is the alternating symbol (\ref{epon}) taking values $\pm1,0$. 
 
 Choosing a vielbein $\bar{e}_{\mu}^a=(\bar{e}_{\mu}^0,\bar{e}_{\mu}^i)$ where $i=1,\dots ,d-1$ with
 $$\bar{g}_{\mu \nu}  =-\bar{e}_{\mu}^0\bar{e}_{\nu}^0+\delta_{ij}\bar{e}_{\mu}^i\bar{e}_{\nu}^j$$
  a convenient basis for the $q$-forms at $p$ that are self-dual with respect to
$\bar{g}_{\mu \nu}  = \eta_{\mu \nu}$ is given by
\begin{equation}
  \omega^{i_1 \ldots i_{q - 1}} \equiv \bar{\Pi}_+ E^{0 i_1 \ldots i_{q - 1}}
\end{equation}
where $ E^{a_1 \ldots a_q} = \bar e^{a_1} \wedge \ldots \wedge \bar e^{a _q}$, while a basis for the anti-self-dual $q$-forms at $p$ is
\begin{equation}
  \omega_{i_1 \ldots i_{q - 1}} \equiv \bar{\Pi}_- E^{0 i_1 \ldots i_{q - 1}}
\end{equation}

\subsubsection{The Non-Diagonal Case}
\label{subs922}

Choosing a vielbein $\bar{e}_{\mu}^a=(\bar{e}_{\mu}{}^+,\bar{e}_{\mu}{}^-,\bar{e}_{\mu}{}^i,\bar{e}_{\mu}{}^{\times })$ where now $i=1,\dots ,d-3$ 
and $\bar{e}_{\mu}{}^{\times}=\bar{e}_{\mu}{}^{d-2}$
with
 $$\bar{g}_{\mu \nu}  =2\bar{e}_{(\mu}{}^+\bar{e}_{\nu)}{}^- 
 +\delta_{ij}\bar{e}_{\mu}{}^i\bar{e}_{\nu}{}^j\delta_{ij}+\bar{e}_{\mu}{}^{\times}\bar{e}_{\nu}{}^{\times}$$
then a convenient basis for the $q$-forms at $p$ that are self-dual with respect to
$\bar{g}_{\mu \nu}  = \eta_{\mu \nu}$ is given by
\begin{eqnarray}
  \omega^{+i_1 \ldots i_{q - 2}}& \equiv &\bar{\Pi}_+ E^{+ i_1 \ldots i_{q - 2} \times }, \qquad
  \omega^{-i_1 \ldots i_{q - 2}} \equiv \bar{\Pi}_+ E^{- i_1 \ldots i_{q - 2} \times }
\cr
\omega^{i_1 \ldots i_{q - 2}}& \equiv &\bar{\Pi}_+ E^{-+ i_1 \ldots i_{q - 2}  }, \qquad
  \omega^{-i_1 \ldots i_{q - 3}} \equiv \bar{\Pi}_+ E^{-+ i_1 \ldots i_{q - 3} \times }
\end{eqnarray}
 while a basis for the anti-self-dual $q$-forms at $p$ is
\begin{eqnarray}
  \omega_{+i_1 \ldots i_{q - 2}}& \equiv &\bar{\Pi}_- E^{+ i_1 \ldots i_{q - 2} \times }, \qquad
  \omega_{-i_1 \ldots i_{q - 2}} \equiv \bar{\Pi}_- E^{- i_1 \ldots i_{q - 2} \times }
\cr
\omega_{i_1 \ldots i_{q - 2}}& \equiv &\bar{\Pi}_- E^{-+ i_1 \ldots i_{q - 2}  }, \qquad
  \omega_{-i_1 \ldots i_{q - 3}} \equiv \bar{\Pi}_-E^{-+ i_1 \ldots i_{q - 3} \times }
\end{eqnarray}

\subsection{Explicit forms for matrices in the diagonal case}

Using the basis given in subsection \ref{diagN} it is
straightforward to calculate the action of $\Phi$ on the basis
forms and hence to find $N, \widetilde{N}, K, M$ explicitly at $p$. The cases of $d = 2$ and $d = 6$ will be consider  before moving on to the general case.

\subsubsection{Two dimensions}

In $d = 2$ dimensions, $x^{\mu} = (x^0, x^1)$, and $E^{a} = \bar e^a_\mu d x^{\mu}$. Any
form that is self-dual with respect to $\bar{g} $ is a multiple of
\begin{equation}
  \omega^+ = \bar{\Pi}_+ E^0 = \frac{1}{\sqrt 2} (\bar e^0 + \bar e^1)
\end{equation}
while any form that is anti-self-dual with respect to $\eta$ is a multiple of
\begin{equation}
  \omega^- = \bar{\Pi}_- E^0 = \frac{1}{\sqrt 2} (\bar e^0 - \bar e^1)
\end{equation}

For the diagonalisable case (\ref{diagl}),(\ref{diagf})
\begin{equation}
   f_{a}{}^{b} 
  =  \rm{Diag} \left( {\lambda^0} , \lambda^1 \right)
\end{equation}
so that 
\begin{equation}
  \Phi (E^0) = \lambda^0 E^0, \quad \Phi (E^1) = \lambda^1 E^1
\end{equation}
Then
\begin{equation}
  \sigma^+ = \Phi (\omega^+) = \frac{1}{\sqrt 2} (\lambda^0 \bar e^0 + \lambda^1\bar e^1)
\end{equation}
is self-dual with respect to $g  $ and
\[ \sigma^- = \Phi (\omega^-) = \frac{1}{\sqrt 2} (\lambda^0 \bar e^0 - \lambda^1\bar e^1) \]
is anti-self-dual with respect to $g$. Expressed in terms of
the basis $\omega^{\pm}$
\begin{equation}
  \sigma^+ = N \omega^+ + K \omega^-
\end{equation}
where
\begin{equation}
  N = \frac{1}{2} (\lambda^0 + \lambda^1), \quad K = \frac{1}{2}
  (\lambda^0 - \lambda^1)
\end{equation}
As $\lambda^0 > 0, \lambda^1 > 0,$ $N > 0$ and is an invertible $1 \times 1$
matrix with inverse
\begin{equation}
  \tilde{N} = N ^{- 1} = \frac{2}{\lambda^0 + \lambda^1}
\end{equation}
Then $M $ is the $1 \times 1$ matrix
\begin{equation}
  M = \tilde{N}  K = \frac{\lambda^0 - \lambda^1}{\lambda^0 +
  \lambda^1}
  \label{mlam}
\end{equation}

\subsubsection{Six dimensions}

In $d = 6$ dimensions,  $E^{abc} = \bar e^a 
 \wedge \bar e^b \wedge \bar e^c$. A basis for the $3$-forms
at $p$ that are self-dual with respect to $\bar{g}$ is given by
\begin{equation}
  \omega^{i  j} \equiv \bar{\Pi}_+ E^{0 i j}
\end{equation}
while a basis for the $\eta$-anti-self-dual $3$-forms at $p$ is
\begin{equation}
  \omega_{i j}^{} \equiv \bar{\Pi}_- E^{0 i j}
\end{equation}
For example,
\begin{equation}
  \omega^{1 2} = \frac{1}{2} (\bar e^0 \wedge \bar e^1 \wedge \bar e^2 + \bar e^3
  \wedge \bar e^4 \wedge \bar e^5)
\end{equation}
\begin{equation}
  \omega_{1 2}  = \frac{1}{2} (\bar e^0 \wedge \bar e^1 \wedge \bar e^2 - \bar e^3
  \wedge \bar e^4 \wedge \bar e^5)
\end{equation}
For the diagonal case
\begin{equation}
   f_{a}{}^{b}  
  =  \rm{Diag} \left( {\lambda^0} , \lambda^1, \ldots, \lambda^5 \right)
\end{equation}
so that, for each $a$,
\begin{equation}
  \Phi (\bar e^{a}) = \lambda^{a} \bar e^{a}
\end{equation}
with no summation over $a$. Then
\begin{equation}
  \sigma^{1 2} = \Phi (\omega^{1 2}) = \frac{1}{2} (\lambda^0 \lambda^1
  \lambda^2\,  \bar e^0 \wedge \bar e^1 \wedge \bar e^2 + \lambda^3 \lambda^4
  \lambda^5\,  \bar e^3 \wedge \bar e^4 \wedge \bar e^5)
\end{equation}
is self-dual with respect to $g$ and
\begin{equation}
  \sigma_{1 2}  = \Phi (\omega _{1 2}) = \frac{1}{2} (\lambda^0 \lambda^1
  \lambda^2 \, \bar e^0 \wedge \bar e^1 \wedge \bar e^2 - \lambda^3 \lambda^4
  \lambda^5 \, \bar e^3 \wedge \bar e^4 \wedge \bar e^5)
\end{equation}
is anti-self-dual with respect to $g$. Then expressing
$\sigma^{1 2}$ in the $\omega$ basis gives
\begin{equation}
  \sigma^{1 2} = N^{1 2}{}_{1 2} \omega^{1 2} + K^{1 2 \, 1 2} \omega_{1 2}
\end{equation}
where
\begin{equation}
  N^{1 2}{}_{1 2} = \frac{1}{2} (\lambda^0 \lambda^1 \lambda^2 + \lambda^3
  \lambda^4 \lambda^5), \quad K^{1 2 \, 1 2} = \frac{1}{2} (\lambda^0 \lambda^1
  \lambda^2 - \lambda^3 \lambda^4 \lambda^5)
\end{equation}
Note that $\frac{1}{2} (\lambda^0 \lambda^1 \lambda^2 + \lambda^3 \lambda^4
\lambda^5) > 0$ as $\lambda^{a} > 0$.

The general form of $N$ and $K$ is found similarly and is
\[ N^{i j}{}_{k l} = \frac{1}{2} \left( \lambda^0 \lambda^i \lambda^j +
   \frac{\Lambda}{\lambda^0 \lambda^i \lambda^j} \right) \delta^{i j}_{k l} \]

\[ K^{i j \, k l}  = \frac{1}{2} \left( \lambda^0 \lambda^i \lambda^j -
   \frac{\Lambda}{\lambda^0 \lambda^i \lambda^j} \right) \delta^{i j | k l}  \]
where there is no summation over repeated indices,
\[ \Lambda = \det (f_{\mu}^{\nu}) = \lambda^0 \lambda^1 \lambda^2 \lambda^3
   \lambda^4 \lambda^5 \]
and
\begin{equation}
  \delta^{i j}_{k l} = \delta^{[i }_k \delta^{ j]}_l,
  \quad \delta^{i j | k l}  = \delta^{i j}_{m n} \delta^{k m} \delta^{l n}
\end{equation}
Then $N $ is diagonal with eigenvalues
\begin{equation}
  \Lambda^{i j} = \frac{1}{2} \left( \lambda^0 \lambda^i \lambda^j +
  \frac{\Lambda}{\lambda^0 \lambda^i \lambda^j} \right)
\end{equation}
and $\Lambda^{i j} > 0$ as $\lambda^{a} > 0$. Then $N $ has a generalised
inverse
\begin{equation}
  \tilde{N}^{i j}{}_{k l} = \frac{1}{\Lambda^{i j}} \delta^{i j}_{k l}
\end{equation}
and $M$ is also diagonal and given by
\begin{equation}
  M^{i j\,  k l}  = \frac{\tilde{\Lambda}^{i j}}{\Lambda^{i j}} \delta^{i j | k l} 
\end{equation}
where
\begin{equation}
  \tilde{\Lambda}^{i j} = \frac{1}{2} \left( \lambda^0 \lambda^i \lambda^j -
  \frac{\Lambda}{\lambda^0 \lambda^i \lambda^j} \right)
\end{equation}

\subsubsection{General Dimensions }%$d = 4 n + 2$ }

The analysis in $d = 2 q= 4 n + 2$ dimensions is similar. For example
\begin{equation}
  \omega^{1 2 \ldots (q - 1)} = \frac{1}{2} \left( \bar e^0 \wedge \bar e^1
  \wedge \ldots \bar e^{q - 1} + \bar e^q \wedge \bar e^{q + 1} {\wedge \bar e^{2
  q - 1}}  \right)
\end{equation}
\begin{equation}
  \omega_{1 2 \ldots (q - 1)} = \frac{1}{2} \left( \bar e^0 \wedge \bar e^1
  \wedge \ldots \bar e^{q - 1} - \bar e^q \wedge \bar e^{q + 1} {\wedge \bar e^{2
  q - 1}}  \right)
\end{equation}
so that
\begin{equation}
  \sigma^{1 2 \ldots (q - 1)} = \frac{1}{2} \left( \Lambda^{1 2 \ldots (q -
  1)} \bar e^0 \wedge \bar e^1 \wedge \ldots \bar e^{q - 1} + \tilde{\Lambda}^{1
  2 \ldots (q - 1)} \bar e^q \wedge \bar e^{q + 1} {\wedge \bar e^{2 q - 1}} 
  \right)
\end{equation}
\begin{equation}
  \sigma_{1 2 \ldots (q - 1)} = \frac{1}{2} \left( \Lambda^{1 2 \ldots (q -
  1)} \bar e^0 \wedge \bar e^1 \wedge \ldots \bar e^{q - 1} - \tilde{\Lambda}^{1
  2 \ldots (q - 1)} \bar e^q \wedge \bar e^{q + 1} {\wedge \bar e^{2 q - 1}} 
  \right)
\end{equation}
where
\begin{equation}
  \Lambda^{1 2 \ldots (q - 1)} = \lambda^0 \lambda^1 \lambda^2 \ldots
  \lambda^{q - 1}, \quad \tilde{\Lambda}^{1 2 \ldots (q - 1)} = \lambda^q
  \lambda^{q + 1} \ldots \lambda^{2 q - 1}
\end{equation}
It is useful to introduce composite indices $A, B, \ldots$ corresponding to
ordered sets $\{ i_1 i_2 \ldots i_{q - 1} \}$ with $A \leftrightarrow i_1 i_2
\ldots i_{q - 1}$ with $i_1 < i_2 < \ldots < i_{q - 1}$, so that $\omega^A
\leftrightarrow \omega^{i_1 i_2 \ldots i_{q - 1}}$, $\omega_A  \leftrightarrow
\omega_{i_1 i_2 \ldots i_{q - 1}}$, and
\[ \Lambda^A \leftrightarrow \Lambda^{i_1 i_2 \ldots i_{q - 1}} \equiv
   \lambda^0 \lambda^{i_1} \lambda^{i_2} \ldots \lambda^{i_{q - 1}}, \quad
   \tilde{\Lambda}^A = \Lambda / \Lambda^A \]
with
\begin{equation}
  \Lambda = \det (f_{\mu}{}^{\nu}) = \det (f_{a}{}^{b}) =\prod_{\mu = 0}^{d - 1} \lambda^{a}
\end{equation}
With $E^A \leftrightarrow E^{0 i_1 \ldots i_{q - 1}}$, one has
\begin{equation}
  \omega^A = \bar{\Pi}_+ E^A, \quad \omega_A  = \bar{\Pi}_- E^A
\end{equation}
so that
\begin{equation}
  \sigma^A = \Phi (\omega^A) = \frac{1}{2} (\Lambda^A E^A + \tilde{\Lambda}^A
  \bar{\ast} E^A) = \frac{1}{2} (\Lambda^A + \tilde{\Lambda}^A) \omega^A +
  \frac{1}{2} (\Lambda^A - \tilde{\Lambda}^A) \omega_A
\end{equation}
\begin{equation}
  \sigma_A  = \Phi (\omega_A ) = \frac{1}{2} (\Lambda^A E^A -
  \tilde{\Lambda}^A \bar{\ast} E^A) = \frac{1}{2} (\Lambda^A -
  \tilde{\Lambda}^A) \omega^A + \frac{1}{2} (\Lambda^A + \tilde{\Lambda}^A)
  \omega_A
\end{equation}
Then  
\begin{equation}
  N^A{}_B = \frac{1}{2} (\Lambda^A + \tilde{\Lambda}^A) \delta^A_B, \quad K^{A
  B} = \frac{1}{2} (\Lambda^A - \tilde{\Lambda}^A) \delta^{A B}
\end{equation}
As $\frac{1}{2} (\Lambda^A + \tilde{\Lambda}^A) > 0$, $N$ is invertible, with
inverse
\begin{equation}
  \tilde{N}^A{}_B = \frac{2}{\Lambda^A + \tilde{\Lambda}^A} \delta^A_B
\end{equation}
Then
\begin{equation}
  M^{A B}  = \frac{\Lambda^A - \tilde{\Lambda}^A}{\Lambda^A +
  \tilde{\Lambda}^A} \delta^{A B}
\end{equation}

\subsection{Explicit forms for matrices in the non-diagonal case}
\subsubsection{Two Dimensions}

In the non-diagonalisable case, in the null basis with $a=+,-$ one has 
$$
\eta _{ab}=\left(\begin{array}{cc}
    \eta _{++}& \eta _{+-}\\
    \eta _{-+} & \eta _{--}
  \end{array}\right) =\left(\begin{array}{cc}
    0 & 1\\
    1 & 0
  \end{array}\right) 
$$
and
\begin{equation}
f_a {}^b = \left(\begin{array}{cc}
     \lambda & 0\\
     a / 2 \lambda & \lambda
   \end{array}\right) 
\end{equation}
Then
\begin{equation}
\sigma^a = \Phi (\omega^a)=\omega^b f_b{}^a
\end{equation}
gives
\begin{equation}
  \sigma^+ = \Phi (\omega^+) =  \left( \lambda \omega^+ +\frac a {2 \lambda} \omega^- \right)
\end{equation}
which is self-dual with respect to $g$ and
\begin{equation}
\sigma^- = \Phi (\omega^-) =  \lambda \omega^-  
\end{equation}
which is anti-self-dual with respect to $g$. Then  
\begin{equation}
  \sigma^+ = N \omega^+ + K \omega^-
\end{equation}
where
\begin{equation}
  N =  \lambda , \quad K = \frac a {2 \lambda} 
 \end{equation}
As $\lambda\ne  0 $, $N > 0$  is an invertible $1 \times 1$
matrix with inverse
\begin{equation}
  \tilde{N} = N ^{- 1} =  \lambda^{-1}
\end{equation}
Then $M $ is the $1 \times 1$ matrix
\begin{equation}
\label{mlamabc}
  M = \tilde{N}  K = \frac a {2 \lambda ^2} \end{equation}

%%%%%
\subsubsection{Six dimensions}

In $d = 6$ dimensions,  $E^{abc} = \bar e^a 
 \wedge \bar e^b \wedge \bar e^c$ where $a=(+,-,i,4)$ and $i,j=1,2,3$. It will be convenient to take
 $\epsilon_{-+1234}=1$.
 From subsection \ref{subs922}, a basis for the $3$-forms
that are self-dual with respect to $\bar{g}$ is given by
\begin{eqnarray}
\omega^{ +i  } &=& \bar{\Pi}_+ E^{+ i 4}
= \frac 1 2 \left( E^{+ i 4}+\frac 1 2 \epsilon ^{i}{}_{jk}E^{+ jk} \right)
\cr
\omega^{ -i  } &=&
 \bar{\Pi}_+ E^{- i 4}
= \frac 1 2 \left( E^{- i 4}-\frac 1 2 \epsilon ^{i}{}_{jk}E^{- jk} \right)
\cr
 \omega^{ i  } &=& \bar{\Pi}_+ E^{-+i }
= \frac 1 2 \left( E^{-+i }+\frac 1 2 \epsilon ^{i}{}_{jk}E^{j k4} \right)
\cr \omega^{ 4 } &=& \bar{\Pi}_+ E^{-+ 4 }
= \frac 1 2 \left( E^{-+ 4}+E^{123} \right)
\end{eqnarray}
while a basis for the $\eta$-anti-self-dual $3$-forms 
is
\begin{eqnarray}
\omega_{ +i  } &=& \bar{\Pi}_- E^{+ i 4}
= \frac 1 2 \left( E^{+ i 4}-\frac 1 2 \epsilon ^{i}{}_{jk}E^{+ jk} \right)
\cr
\omega_{ -i  } &=&
 \bar{\Pi}_- E^{- i 4}
= \frac 1 2 \left( E^{- i 4}+\frac 1 2 \epsilon ^{i}{}_{jk}E^{- jk} \right)
\cr
 \omega_{ i  } &=& \bar{\Pi}_- E^{-+ i }
= \frac 1 2 \left( E^{-+ i }-\frac 1 2 \epsilon ^{i}{}_{jk}E^{j k4} \right)
\cr \omega_{ 4 } &=& \bar{\Pi}_- E^{-+ 4 }
= \frac 1 2 \left( E^{-+ 4}-E^{123} \right)
\end{eqnarray}

Then
\begin{equation}
\Phi (\bar{e}_{\mu}{}^a)=  \bar{e}_{\mu}{}^b f_b{} ^a
\end{equation}
with
\begin{equation}
f_a {}^b = \left(\begin{array}{cc}
     \lambda &  0 \\
    a / 2 \lambda & \lambda
   \end{array}\right) \oplus  \rm{diag} (\lambda_1, \ldots, \lambda_{4})
\end{equation}
gives, for each $i$,
\begin{equation}
  \Phi (\bar e^{i}) = \lambda_{i} \bar e^{i},\qquad
   \Phi (\bar e^{4}) = \lambda_{4} \bar e^{4}
\end{equation}
with no summation over $i$
while
\begin{equation}
  \Phi (\bar e^+) =   \lambda \bar e^+ +\alpha \bar e^- 
 ,\qquad 
 \alpha\equiv  \frac a {2 \lambda}
\end{equation}
and
\begin{equation}
  \Phi (\bar e^-) =  \lambda \bar e^-  
\end{equation}

For example,
\begin{eqnarray}
\omega^{ +1  } &=&  \frac 1 2 \left( E^{+ 14}+E^{+ 23} \right)= \frac{1}{2} (\bar e^+ \wedge \bar e^1 \wedge \bar e^4 +\bar e^+ \wedge \bar e^2
  \wedge \bar e^3 )
\cr
\omega^{ -1} &=& \frac 1 2 \left( E^{- 14}-E^{-23} \right)= \frac{1}{2} (\bar e^- \wedge \bar e^1 \wedge \bar e^4 -\bar e^- \wedge \bar e^2
  \wedge \bar e^3 )
\cr \omega^{ 1  } &=& \frac 1 2 \left( E^{-+ 1}+E^{234} \right)=\frac{1}{2} (\bar e^- \wedge \bar e^+ \wedge \bar e^1 +\bar e^2 \wedge \bar e^3
  \wedge \bar e^4 )
\end{eqnarray}
and
\begin{eqnarray}
\omega_{ +1  } &=&  \frac 1 2 \left( E^{+ 14}-E^{+ 23} \right)= \frac{1}{2} (\bar e^+ \wedge \bar e^1 \wedge \bar e^4 -\bar e^+ \wedge \bar e^2
  \wedge \bar e^3 )
\cr
\omega_{ -1} &=& \frac 1 2 \left( E^{- 14}+E^{-23} \right)= \frac{1}{2} (\bar e^- \wedge \bar e^1 \wedge \bar e^4 +
\bar e^- \wedge \bar e^2
  \wedge \bar e^3 )
\cr 
\omega_{ 1  } &=& \frac 1 2 \left( E^{-+ 1}-E^{234} \right)=\frac{1}{2} (\bar e^- \wedge \bar e^+\wedge \bar e^1 
-
\bar e^2 \wedge \bar e^3
  \wedge \bar e^4 )
\end{eqnarray}
Then
\begin{eqnarray}
 \sigma^{ +1  } 
 =\Phi(\omega^{ +1  } )&=&  
 \frac{1}{2} 
  \left( \lambda \bar e^+ +\alpha \bar e^- \right)\wedge
  (   \lambda_1
  \lambda_4\,\bar e^1 \wedge \bar e^4 +  \lambda_2 \lambda_3\bar e^2
  \wedge \bar e^3)
\cr
 \sigma^{ -1}
 =\Phi(\omega^{ -1} )&=& 
 \frac{1}{2} 
  \lambda  \bar e^- \wedge
  (   \lambda_1
  \lambda_4\,\bar e^1 \wedge \bar e^4 -  \lambda_2 \lambda_3\bar e^2
  \wedge \bar e^3 )
  \cr
 \sigma^1
 = \Phi(\omega^{ 1  }) &=& 
 \frac{1}{2} (\lambda^2\lambda_1\bar e^-\wedge \bar e^+ \wedge \bar e^1 +\lambda_2\lambda_3 \lambda_4
 \bar e^2 \wedge \bar e^3
  \wedge \bar e^4 )
\end{eqnarray}
which can be rewritten as
\begin{eqnarray}
 \sigma^{ +1  } 
 &=&  
 \frac{1}{2} 
  \bigl[ 
  \lambda \lambda_1
  \lambda_4\,
   (\omega ^{+1}+\omega _{+1})
 + \lambda \lambda_2 \lambda_3
 (\omega ^{+1}-\omega _{+1}) 
  \cr
&  +&  \alpha \lambda_1
  \lambda_4\,
   (\omega ^{-1}+\omega _{-1})
  + \alpha \lambda_2 \lambda_3
     (-\omega ^{-1}+\omega _{-1})
  \bigr]
\cr
 \sigma^{ -1}
 &=& 
 \frac{1}{2} 
    \left( \lambda   \lambda_1
  \lambda_4\,
  (\omega ^{-1}+\omega _{-1})
   +
    \lambda  \lambda_2 \lambda_3
   (\omega ^{-1}-\omega _{-1}) \right)
  \cr
 \sigma^1
&=& 
 \frac{1}{2} (\lambda^2\lambda_1
 (\omega ^1+\omega _1)
  +\lambda_2\lambda_3 \lambda_4
(\omega ^1-\omega _1)
 )
\end{eqnarray}

Then expressing
$\sigma$ in the $\omega$ basis gives
\begin{eqnarray}
 \sigma^{ +1  } 
 &=&  
N^{ +1  } {}_{ +1  } \omega ^{+1}
+
N^{ +1  } {}_{ -1  } \omega ^{-1}
+ K^{ +1  \, +1  }\omega _{+1}
+ K^{ +1  \, -1  }\omega _{-1}
\cr
 \sigma^{ -1}
 &=& 
N^{ -1  } {}_{ -1  } \omega ^{-1}
+ K^{ -1  \, -1  }\omega _{-1}
  \cr
 \sigma^1
&=& 
N^{ 1  } {}_{ 1  } \omega ^{1}
+ K^{ 1  \, 1  }\omega _{1}
\end{eqnarray}
where
\begin{eqnarray}
N^{ +1  } {}_{ +1  }  
&=&  \frac{1}{2} \lambda( \lambda_1 \lambda_4 + \lambda_2
  \lambda_3 )
\qquad
N^{ +1  } {}_{ -1  }=\frac{1}{2} \alpha( \lambda_1 \lambda_4 - \lambda_2
  \lambda_3 )
\cr
K^{ +1  \, +1  }
&=&  \frac{1}{2} \lambda( \lambda_1 \lambda_4 - \lambda_2
  \lambda_3 )
\qquad
K^{ +1  \, -1  }=\frac{1}{2} \alpha( \lambda_1 \lambda_4 + \lambda_2
  \lambda_3 )
\cr
N^{ -1  } {}_{ -1  }
&=&   \frac{1}{2} \lambda( \lambda_1 \lambda_4 + \lambda_2
  \lambda_3 )
\qquad
K^{ -1  \, -1  }= \frac{1}{2} \lambda( \lambda_1 \lambda_4 - \lambda_2
  \lambda_3 )
  \cr
N^{ 1  } {}_{ 1  } 
&=&  \frac{1}{2}( \lambda^2 \lambda_1  + \lambda_2
  \lambda_3 \lambda_4)
\qquad
K^{ 1  \, 1  }=\frac{1}{2}( \lambda^2 \lambda_1  - \lambda_2
  \lambda_3 \lambda_4)
\end{eqnarray}

The general form of $N$ and $K$ is found similarly and is
\begin{eqnarray}
N^{+i }{}_{+j} &=& \frac{1}{2} \lambda \left(  \lambda_i \lambda_4 +
   \frac{\Lambda}{ \lambda_i } \right) \delta^{i }{}_{j} ,
   \qquad
   N^{+i }{}_{-j} = \frac{1}{2} \alpha \left(  \lambda_i \lambda_4 -
   \frac{\Lambda}{ \lambda_i } \right) \delta^{i }{}_{j} 
   \cr
   N^{-i }{}_{-j} &=& \frac{1}{2} \lambda \left(  \lambda_i \lambda_4 +
   \frac{\Lambda}{ \lambda_i } \right) \delta^{i }{}_{j} ,
   \qquad
   N^{-i }{}_{+j} =0  
   \cr
   N^{i }{}_{j} &=& \frac{1}{2}  \left( \lambda^2 \lambda_i +
   \frac{\Lambda}{ \lambda_i }  \lambda_4\right) \delta^{i }{}_{j} ,
    \end{eqnarray}
    
    \begin{eqnarray}
K^{+i \, +j} &=& \frac{1}{2} \lambda \left(  \lambda_i \lambda_4 -
   \frac{\Lambda}{ \lambda_i } \right) \delta^{i }{}_{j} ,
   \qquad
   K^{+i \, -j} = \frac{1}{2} \alpha \left(  \lambda_i \lambda_4 +
   \frac{\Lambda}{ \lambda_i } \right) \delta^{i }{}_{j} 
   \cr
   K^{-i \, -j} &=& \frac{1}{2} \lambda \left(  \lambda_i \lambda_4 -
   \frac{\Lambda}{ \lambda_i } \right) \delta^{i }{}_{j} ,
   \qquad
   K^{-i \, +j} =0  
   \cr
   K^{i \, j} &=& \frac{1}{2}  \left( \lambda^2 \lambda_i -
   \frac{\Lambda}{ \lambda_i }  \lambda_4\right) \delta^{i }{}_{j} ,
    \end{eqnarray}
 where there is no summation over repeated indices and
\[ \Lambda =     \lambda^1 \lambda^2 \lambda^3
     \]
Then
\begin{eqnarray}
N^{+i }{}_{+j} &=& \rho_i   \delta^{i }{}_{j} ,
   \qquad
   N^{+i }{}_{-j} =
   \mu_i 
     \delta^{i }{}_{j} 
   \cr
   N^{-i }{}_{-j} &=& \rho_i \delta^{i }{}_{j} ,
   \qquad
   N^{-i }{}_{+j} =0  
   \cr
   N^{i }{}_{j} &=&
   \tau _i 
    \delta^{i }{}_{j} ,
    \end{eqnarray}
(no summation over repeated indices) where
\begin{equation}
\rho_i =\frac{1}{2} \lambda \left(  \lambda_i \lambda_4 +
   \frac{\Lambda}{ \lambda_i } \right) ,\qquad
   \mu_i =\frac{1}{2} \alpha \left(  \lambda_i \lambda_4 -
   \frac{\Lambda}{ \lambda_i } \right)
   ,\qquad
   \tau _i =    \frac{1}{2}  \left( \lambda^2 \lambda_i +
   \frac{\Lambda}{ \lambda_i }  \lambda_4\right)
\end{equation}
As $\lambda _i>0$, $\rho_i>0$ and $\tau _i>0$  so that $N$ is invertible with
 (generalised) inverse $\tilde N$ is
\begin{eqnarray}
 \tilde N^{+i }{}_{+j} &=& \frac 1 {  \rho_i }  \delta^{i }{}_{j} ,
   \qquad
\tilde   N^{+i }{}_{-j} =
 - \frac 1 {  (\rho_i )^2}
   \mu_i 
     \delta^{i }{}_{j} 
   \cr
 \tilde  N^{-i }{}_{-j} &=& \frac 1 {  \rho_i }\delta^{i }{}_{j} ,
   \qquad
  \tilde N^{-i }{}_{+j} =0  
   \cr
 \tilde  N^{i }{}_{j} &=&
  \frac 1 {   \tau _i }
    \delta^{i }{}_{j} ,
    \end{eqnarray}
The matrix $M$ is then given by
\begin{equation}
  M^{AB}  = \tilde
N^{A}{}_{C} K^{CB}
  \end{equation}
where $A=(i+,i-,i)$ is a composite index.

\section{Coupling to Other Fields}
\label{secCoup}

\subsection{Coupling to Matter}

It has been  seen that the action
\begin{equation}
  \label{spq0} S_{P Q} = \int \left( \frac{1}{2} d P \wedge \bar{\ast} d P - 2
  Q \wedge d P - Q \wedge M (Q) \right)
\end{equation}
gives two field strengths $F, G$ satisfying
\begin{eqnarray*}
  F = \ast F, & \qquad & d F = 0\\
  G = \bar{\ast} G, &  \qquad & d G = 0
\end{eqnarray*}
so that $F$ couples to the metric $g$ and $G$ couples to the metric $\bar{g}$.
Moreover, $F$ can be viewed as having a kinetic term with the correct sign and
$G$ as having one with the wrong sign; this is confirmed by a Hamiltonian
analyis  \cite{Sen:2015nph,Sen:2019qit,Andriolo:2020ykk}.

To this can be added an action for matter fields denoted generically by $\psi$
and for the metric $g$ (possibly including an Einstein-Hilbert term or other
kinetic term for $g$)
\begin{equation}
  \label{spq+} S = S_{P Q} + S_{ \rm{mat}} [g, \psi]
\end{equation}
The metric $\bar{g}$ and the field strength $G$ completely decouple from the
physical fields: they do not couple to $\psi$, $A$  or $g$.

Note that there is a certain symmetry between the two sectors, $(g, F)$ and
$(\bar{g}, G)$, in the free theory. 
Instead of choosing $g,F$ to be the physical fields,
one could  choose $\bar{g}$ to be the
physical metric and $G$ to be the physical field strength, with $F$ the
auxiliary field, using the action $- S_{P Q}$ so that $G$ has positive energy. For this, consider an interacting action of the form
\begin{equation}
  \label{spq-} S = - S_{P Q} + S_{ \rm{mat}} [\bar{g}, \psi]
\end{equation}
so that now the  energy for $C$ has the right sign and that for $A$ is negative.

This  will now be generalised to include couplings between the matter fields and $F$
or $G$. As will be  seen, this introduces some interesting differences between
theories in which $(g, F)$ are physical and those in which $(\bar{g}, G)$ are
physical.

\subsection{Sen's Action in Minkowski Space with Chern-Simons Terms Revisited}

In section \ref{Sencs}, the following action for a self-dual gauge field with
Chern-Simons interactions in Minkowski space was considered:
\begin{equation}
  \label{actom2} S = \int \left( \frac{1}{2} d P \wedge \ast d P - 2 Q \wedge
  d P + 2 Q \wedge \Omega - \frac{1}{2} \Omega \wedge \ast \Omega
  \right)
\end{equation}
The field strengths (with (\ref{omplu}))
\begin{equation}
  \label{giss2} G \equiv \frac{1}{2}  (dP + \ast d P) + Q, \quad F \equiv Q +
  \Omega_+
\end{equation}
satisfied
\begin{equation}
  \ast G = G, \quad d G = 0, \quad \ast F = F, \quad d F = d \Omega
\end{equation}
so that there are potentials $A, C$ with
\begin{equation}
  F = d A + \Omega, \quad G = d C
\end{equation}
Under the transformations
\begin{equation}
  \label{symom2} \delta P = \Lambda, \quad \delta \Omega = d \Lambda, \quad
  \delta Q = - \frac{1}{2}  (1 + \ast) d \Lambda
\end{equation}
the field strengths (\ref{giss2}) are invariant, $\delta F = \delta G = 0,$
but the variation of the action under these transformations is
\begin{equation}
  \label{delss2} \delta S = - \int \Lambda \wedge d \Omega
\end{equation}
Thus the action is invariant under transformations for which this is
satisfied.

However, this variation of the action can be cancelled by adding the following
term to the action
\begin{equation}
  \label{splus} S' = \int P \wedge d \Omega
\end{equation}
Then $S + S'$ is invariant under the transformations (\ref{symom2}). However,
now the $P$ field equation is changed to
\begin{equation}
  \label{adpist2} d \left( \frac{1}{2} \ast d P + Q + \Omega \right) = 0
\end{equation}
while the $Q$ field equations remain unchanged as \
\begin{equation}
  \label{pcona} d P - \Omega = \ast (d P - \Omega)
\end{equation}
Now
\begin{equation}
  \label{giss3} G \equiv \frac{1}{2}  (dP + \ast d P) + Q
\end{equation}
satisfies
\begin{equation}
  \ast G = G, \quad d G = - d \Omega
\end{equation}
so that locally
\[ G = d C - \Omega \]
Taking the exterior derivative of (\ref{pcona}) gives
\[ - d \Omega = d \ast (d P - \Omega) \]
i.e.\
\[ - 2 d \Omega_- = d \ast (d P) \]
and using (\ref{adpist}) to eliminate $P$ gives
\begin{equation}
  d (Q + \Omega_+) = 0
\end{equation}
Then defining as before
\begin{equation}
  \label{fiss2} F \equiv Q + \Omega_+
\end{equation}
one has
\begin{equation}
  \ast F = F, \quad d F = 0
\end{equation}
Note that $F$ is invariant under (\ref{symom2}). Thus $G$ couples to $\bar{g}$
and to other fields through $\Omega$ while $F$ couples to $g$ but not to any
other fields.
Taking $S\to -S$ then gives  an action for a physical sector consisting of $\bar g, G,\psi$ and
an shadow sector $g,F$.

\subsection{Self-dual gauge fields with general couplings}
\label{SdgfGen}

In this subsection, \ the actions (\ref{spq+}) or (\ref{spq-}) with
(\ref{spq0}) will be generalised to include the coupling of a chiral $q -
1$-form gauge field to the matter fields $\psi$ through a $q$-form $\Omega$,
resulting in a
 self-dual field strength $$F = d A + \Omega$$
 Examples of such interactions were discussed in subsection \ref{Sencs}.
Such matter couplings were given in  \cite{Sen:2015nph,Sen:2019qit,Andriolo:2020ykk}
for the case $\bar{g} = \eta$ and these will be generalised to the case of
general $\bar{g}$. The Chern-Simons terms in $\Omega$ will typically not be
invariant under the gauge transformations of the gauge fields it is
constructed from and so gauge invariance of the resulting theory will be
non-trivial, and in particular the possibility of adding a term of the form
(\ref{splus}) will be discussed. Here  no assumptions will be made  about the
composition of $\Omega$, other than to allow the possibility of gauge
transformations under which it transforms as a total derivative.

The action with coupling to $\Omega$ for $\bar{g} = \eta$ of \cite{Sen:2015nph,Sen:2019qit}
generalises to
%\begin{equation}
%  S = \int \left( \frac{1}{2} d P \wedge \bar{\ast} d P - 2 Q \wedge (d P-\Omega)
%   - { (Q+\Omega_+)}x
 % \wedge M (Q+\Omega_+) + \frac{1}{2} \Omega
%  \wedge \ast \Omega 
%  +
%  \lambda  P \wedge d \Omega \right)
%   \label{actinst}
%\end{equation}
\begin{eqnarray}
S &=& \int \left( \frac{1}{2} d P \wedge \bar{\ast} d P - 2 Q \wedge (d P-\Omega) \right.
  \nonumber
  \\
   &-& 
   \left. { (Q+\Omega_+)}
  \wedge M (Q+\Omega_+) + \frac{1}{2} \Omega
  \wedge \ast \Omega 
  +
  \lambda  P \wedge d \Omega \right)
   \label{actinst}
\end{eqnarray}
where \[ \Omega_{\pm} = \bar{\Pi}_{\pm} \Omega \]
and $\lambda = 0$ or $\lambda = 1$. This gives the possibility of including the
term (\ref{splus}): if $\lambda = 1$, then it is included and if $\lambda = 0$
it is not.

The field equations for $P, Q$ (using symmetry and linearity of $M$), for
action $S_{P Q}$ are
\begin{equation}
  \label{pfom} d \left( \frac{1}{2}  \bar{\ast} d P + Q + \lambda \Omega
  \right) = 0
\end{equation}
and
\begin{equation}
  \label{qfom} 
  \frac 1 2 (d P-\bar{\ast} d P )+ M (Q+\Omega_+)-\Omega_-=0
\end{equation}

Then defining as before the field strength
\begin{equation}
  G \equiv \frac{1}{2}  (d P + \bar{\ast} d P) + Q
\end{equation}
which is self-dual
\begin{equation}
  \bar{\ast} G = G
\end{equation}
then from (\ref{pfom})
\begin{equation}
  dG = - \lambda d \Omega
\end{equation}
Taking the exterior derivative of (\ref{qfom}) and eliminating $P$ using
(\ref{pfom}) gives
\begin{equation}
  d [Q + M (Q + \Omega_+)] = d \Omega_- - \lambda d \Omega
\end{equation}
Let
\begin{equation}
  F \equiv Q + \Omega_+ + M (Q + \Omega_+)
\end{equation}
Then
\begin{equation}
  d F = (1 - \lambda) d \Omega
\end{equation}
and $F$ is \ $g$-self-dual
\begin{equation}
  \ast F = F
\end{equation}
Then potentials $A, C$ can be introduced so that
\begin{equation}
  F = d A + (1 - \lambda) \Omega, \quad G = d C - \lambda \Omega
\end{equation}

%%%%

The transformations
\begin{equation}
  \label{symom22} \delta P = \Lambda, \quad \delta \Omega = d \Lambda, \quad
  \delta Q = - \bar{\Pi}_+ d \Lambda
\end{equation}
leave the field strengths   invariant, $\delta F = \delta G = 0,$
but the variation of the action under these  is
\begin{equation}
  \label{delss2s} \delta S = - (1 - \lambda) \int \Lambda \wedge d \Omega
\end{equation}
In the case $\lambda =0$, $G=dC$ is a free field coupling only to $\bar g$ so that the shadow sector can be taken to be $\bar g, C$. The physical gauge field $A$ then couples to other physical fields through $F = d A + \Omega$.
The field equations are invariant under (\ref{symom22}) but the action
is invariant only under transformations for which (\ref{delss2s}) vanishes.
As was seen in subsection \ref{Sencs}, the variation (\ref{delss2s}) vanishes for all transformations in IIB supergravity but in 6 dimensions with a 2-form $A$ coupled to Yang-Mills Chern-Simons terms this gives a non-trivial constraint.

In the case $\lambda =1$, the action is invariant under  (\ref{symom22}) without further constraints. In this case,
$F=dA$ is a free field coupling only to $ g$ so that the shadow sector can be taken to be $g,A$.
Then $G=d C -  \Omega$ is self-dual with respect to $\bar g$ and includes couplings to other physical fields to that the physical sector can be taken to be $\bar g, C$ plus the other physical fields and will have positive energy if the sign of the action for $P,Q$ is flipped to give the action (\ref{spq-}).

\section{Non-Linear  Action in a General Spacetime}
\label{nonlasp}
 
The non-linear action of section \ref{sec22}
is  generalised to general spacetimes by replacing the flat metric $\eta$ with $\bar g$ to give
\begin{equation}
\label{raction}
  S = \int \left( \frac{1}{2} d P \wedge \bar \ast d P - 2 Q \wedge d P - {\mathcal F}(Q)  \right)
\end{equation}
where ${\mathcal F}(Q) $ is a top form constructed from $Q$ and as before $\bar \ast Q=Q$.
The variation of the interaction term 
\begin{equation}
\label{fvary}
\delta \int {\mathcal F}(Q) = 2 \int \delta Q \wedge {\mathcal R}(Q)
\end{equation}
defines a  $q$-form ${\mathcal R}(Q)$ which is anti-self-dual:
$$\bar \ast {\mathcal R}(Q)=-{\mathcal R}(Q)
$$
Then, as before, $G$ defined by (\ref{Giss}) satisfies $ \bar{\ast} G = G$ and $dG=0$ while
\begin{equation}
\label{FFist}
  F \equiv Q + {\mathcal R}(Q)
\end{equation}
satisfies
\begin{equation}
  d F = 0
\end{equation}

Let
\begin{equation}
\label{fhatis}
\hat F_\pm =\frac 1 2 (F+\ast F)
\end{equation}
The self-duality condition $\hat F_-=0$ arises from choosing ${\mathcal R}(Q)
$ to be the $M(Q)$ constructed in section  \ref{sec5}. The aim here is to find
an action giving the constraint
\begin{equation}
\hat F_-={\mathcal S} (\hat F_+)
\label{friss}
\end{equation}
for any  ${\mathcal S}$ that satisfies
\begin{equation}
\label{eryed}
\ast {\mathcal S} (\hat F_+)=-{\mathcal S} (\hat F_+)
\end{equation}
This means that for a given ${\mathcal S}$ the goal is first to find an
${\mathcal R}(Q)
$
 so that the $F$ given by (\ref{FFist}) satisfies (\ref{friss}). Then 
 the next step is to find 
 an
 ${\mathcal F}(Q)
$ with variation (\ref{fvary}) that gives this 
 ${\mathcal R}(Q)
$ so that the action (\ref{raction}) leads to (\ref{friss}).
This can be done following the arguments of section \ref{sec5}.

As   $\hat F_+$ is $g$-self-dual, acting with $\Phi^{- 1}$ on $\hat F_+$ gives a $q$-form
\begin{equation}
  \alpha \equiv \Phi^{- 1} (\hat F_+)
\end{equation}
which is $\bar{g}$-self-dual:
\begin{equation}
  \bar{\ast} \alpha = \alpha
\end{equation}
Then, if it is assumed that $F$ satisfies (\ref{friss}) for a given ${\mathcal S}$,
then
\begin{equation}
F=\hat F_++ \hat F_-=\hat F_++{\mathcal S} (\hat F_+)= \hat\Phi(\alpha)
\end{equation}
where
\begin{equation}
\hat\Phi\equiv (1+{\mathcal S} )\Phi
\end{equation}
i.e.\ $\hat\Phi(\alpha)\equiv \Phi(\alpha)+{\mathcal S} (\Phi(\alpha))$ with $\Phi(\alpha)=\hat F_+
$.
The  relation
\begin{equation}
  F = Q + {\mathcal R}  = \hat\Phi (\alpha)
\end{equation}
can be decomposed using the projectors ${\bar {\Pi}}_{{\pm}}$ to give
\begin{equation}
  F_+=Q = \hat N (\alpha)
\end{equation}
\begin{equation}
  F_-=
  {\mathcal R}
    = \hat K (\alpha)
\end{equation}
($F_\pm={\bar {\Pi}}_{{\pm}}F$)
where the maps $\hat N,\hat K$ are
\begin{equation}
 \hat N \equiv \bar{\Pi}_+ \hat \Phi \bar{\Pi}_+
\end{equation}
\begin{equation}
\hat  K \equiv \bar{\Pi}_- \hat\Phi \bar{\Pi}_+
\end{equation}
If $\hat N$ has a generalised inverse
 $\tilde{N}$ satisfying
\begin{equation}
  \label{genina} \tilde{N} \hat N = \bar{\Pi}_+
\end{equation}
then
\begin{equation}
  \alpha = \tilde{  N}  (Q)
\end{equation}
so that finally one can write ${\mathcal R}$ as a   function ${\mathcal R}(Q)$ of $Q$
\begin{equation}
\label{ewtafh}
  {\mathcal R}(Q) = \hat K \tilde{  N}  (Q)
\end{equation}
This is the desired function ${\mathcal R} (Q)$ and, by construction, satisfies
$\bar{\ast} {\mathcal R} = - {\mathcal R}$.
For any ${\mathcal S}$ that satisfies
(\ref{eryed}) 
this then gives the non-linear theory provided ${\mathcal R}$ can be derived from varying an interaction term ${\mathcal F}$, i.e.\ provided (\ref{fvary}) holds.

\section{Non-Linear  Action in a General Spacetime with Chern-Simons Coupling}

A non-linear version of the action for a self-dual gauge field with Chern-Simons coupling is given by combining the actions (\ref{raction}) and (\ref{actinst}) to give
\begin{equation}
  S = \int \left( \frac{1}{2} d P \wedge \bar{\ast} d P
   - 2 Q \wedge (d P-\Omega) - 
{\mathcal F} (Q+\Omega_+) + \frac{1}{2} \Omega
  \wedge \ast \Omega 
  +
  \lambda  P \wedge d \Omega \right)
   \label{actinstf}
\end{equation}

As in subsection \ref{SdgfGen},
the field equation for $P$ is 
\begin{equation}
  \label{pfomr} d \left( \frac{1}{2}  \bar{\ast} d P + Q + \lambda \Omega
  \right) = 0
\end{equation}
so that the field strength
\begin{equation}
  G \equiv \frac{1}{2}  (d P + \bar{\ast} d P) + Q
\end{equation}
 is self-dual
\begin{equation}
  \bar{\ast} G = G
\end{equation}
and satisfies
\begin{equation}
  dG = - \lambda d \Omega
\end{equation}

The field equation for $Q$ is
\begin{equation}
  \label{qfomr} 
  \frac 1 2 (d P-\bar{\ast} d P )+ {\mathcal R} (Q+\Omega_+)-\Omega_-=0
\end{equation}
where
${\mathcal R}$ is defined by (\ref{fvary}).
Taking the exterior derivative of (\ref{qfomr}) and eliminating $P$ using
(\ref{pfomr}) gives
\begin{equation}
  d [Q + {\mathcal R} (Q + \Omega_+)] = d \Omega_- - \lambda d \Omega
\end{equation}
Let
\begin{equation}
  F \equiv Q + \Omega_+ + {\mathcal R} (Q + \Omega_+)
\end{equation}
Then
\begin{equation}
  d F = (1 - \lambda) d \Omega
\end{equation}
and potentials $A, C$ can be introduced so that
\begin{equation}
  F = d A + (1 - \lambda) \Omega, \quad G = d C - \lambda \Omega
\end{equation}
If ${\mathcal R}$, ${\mathcal F}$ are chosen as in section \ref{nonlasp}, then
$F$ satisfies the self-duality equation
(\ref{friss}) with the definition (\ref{fhatis}).

\section{The Theory of Chiral Scalars  in Two dimensions}
\label{2dims}

\subsection{Coordinate Basis Formulation}

Consider the theory in two dimensions on $\mathbb{R}^2$ (or a cylinder) with
coordinates $x^{\pm} = \frac{1}{\sqrt{2}} (x^0 \pm x^1)$, and  $\bar{g} $ given by the Minkowski metric
$\bar{g} = \eta$, with line element $ds^2 = 2 d x^+ d x^-$, $\epsilon_{+ -} = 1$.
For a 1-form with components $V_{\mu}$
\begin{equation}
  (\bar{\ast} V)_{\pm} = \pm V_{\pm}
\end{equation}
\begin{equation}
  (\ast V)_{\mu} = \sqrt{g} \,\epsilon_{\mu \nu} \, g^{\nu \rho} V_{\rho}
  \label{vsdis}
\end{equation}
so that
\begin{eqnarray*}
  (\ast V)_+ & = & \quad \tilde{g}^{- +} V_+ + \tilde{g}^{- -} V_-\\
  (\ast V)_- & = & - \tilde{g}^{+ +} V_+ - \tilde{g}^{+ -} V_-
\end{eqnarray*}
where
\begin{equation}
  \tilde{g}^{\mu \nu} = \sqrt{g} g^{\mu \nu}
\end{equation}
For $g$-self-dual $F_{\mu}$, $F = \ast F$ gives
\begin{equation}
  \begin{array}{lll}
    F_- & = & - \tilde{g}^{+ +} F_+ - \tilde{g}^{+ -} F_-
  \end{array}
\end{equation}
so that
\begin{equation}
    F_-  =  - \frac{\tilde{g}^{+ +}}{1 + \tilde{g}^{+ -}} F_+
  \label{fsdis}
\end{equation}

The action (\ref{sensss}) for a scalar $P$ and 1-form $Q_{\mu}$ which satisfies $Q =
\bar{\ast} Q$ so that $Q_- = 0$ is given by
\begin{equation}
  S = \int d^2 x \sqrt{\bar{g}} \left(  \partial_+ P \partial_- P +
  2 Q_+ \partial_- P + M_{--} Q_+ Q_+ \right)
  \label{2dacts}
\end{equation}
with ${M(Q)} _ - = M_ {--}  Q_+$.
The closed 1-forms $F,G$ with $G=\bar\ast G$ so that $G_-=0$ and $F=\ast F$
are given by
\begin{equation}
G_+=\frac 1 2 \partial _+P+Q_+, \qquad
F_+=Q_+, \qquad F_-=M_{--}Q_+
\end{equation}
Then $F_- = M_{- -} F_+$ agrees with the $F=\ast F$ constraint (\ref{fsdis}) if
\begin{equation}
  M_{- -} = - \frac{\tilde{g}^{+ +}}{1 + \tilde{g}^{+ -}}
\end{equation}
The scalar fields  $A,C$ with $F=dA$, $G=dC$
then satisfy the self-duality conditions
\begin{equation}
      \partial _-C=0, \qquad \partial _-A=M_{--}\partial _+A
\end{equation}

If the metric $g$ satisfies the conformal gauge condition
$g_{\mu\nu} = e^\phi \eta_{\mu\nu}$ then $\ast V=\bar \ast V$, 
$\tilde g^{\mu\nu}=\eta^{\mu\nu}$ 
and $M=0$.
In the Polyakov gauge in which the metric $g$ is brought to the form
\begin{equation}
  g_{\mu \nu} = \left(\begin{array}{cc}
    0 & 1\\
    1 & g_{- -}
  \end{array}\right)
\end{equation}
one has $\tilde{g}^{+ +}=-g_{- -} $ so that
\begin{equation}
\label{werter}
  M_{- -} = \frac{1}{2}g_{- -} 
\end{equation}
and
\begin{equation}
  F_- = \frac{1}{2}g_{- -} F_+
\end{equation}
implying that the scalar field $A$ satisfies
\begin{equation}
        \partial _-A=\frac{1}{2}g_{--}\partial _+A
\end{equation}
In this case
\begin{equation}
f_\mu {}^\nu = \left(\begin{array}{cc}
     1 & 0\\
      \frac 1 2 g_{--}   & 1
   \end{array}\right) 
   \end{equation}
satisfies (\ref{fra}).
Then $\omega ^\pm =dx^\pm$ satisfy $\bar \ast \omega ^\pm=\pm \omega ^\pm$ and
\begin{equation}
\sigma^a = \Phi (\omega^a)=\omega^b f_b{}^a
\end{equation}
are $g$-(anti-)self-dual satisfying $  \ast \sigma ^\pm=\pm \sigma ^\pm$, with 
\begin{equation}
  \sigma^+ = dx^+
  +
  \frac 1 2 g_{--}
  dx^-, \qquad \sigma^- = dx^-
  \end{equation}
   Then  
\begin{equation}
  \sigma^+ = N \omega^+ + K \omega^-
\end{equation}
where
\begin{equation}
  N = \tilde{N} = 1 , \quad K =   \frac 1 2 g_{--}
 \end{equation}
Then $M $ is the $1 \times 1$ matrix
\begin{equation}
  M = \tilde{N}  K =   \frac{1}{2}g_{- -} 
    \end{equation}
in agreement with (\ref{werter}).

\subsection{Frame Formulation}

Consider now the two dimensional case on a general 2-dimensional manifold with Lorentzian signature metrics $g,\bar g$.
Introducing frames that are orthonormal with respect to $\bar{g}$ using a 
zweibein $\bar{e}_{\mu}^a$, (\ref{vsdis}) has frame components
\begin{equation}
  (\ast V)_a = \epsilon_{a b} \tilde{G}^{b c} V_c
\end{equation}
where $G_{a b} = \bar{e}_a^{\mu} \bar{e}_b^{\nu} g_{\mu \nu}$ and
\begin{equation}
  \tilde{G}^{a b} = \sqrt{G} G^{a b}, \quad G = | \det G_{a b} |
\end{equation}
In a null basis with $a, b = +, -$, $\bar{e}^{\pm} = \frac{1}{\sqrt{2}}
(\bar{e}^0 \pm \bar{e}^1)$ so that $\bar{g}_{\mu \nu} = \bar{e}_{\mu}^+
\bar{e}_{\nu}^- + \bar{e}_{\mu}^- \bar{e}_{\nu}^+$ and $\epsilon_{a b}$
satisfies $\epsilon_{+ -} = 1$, the $\bar{g}$ dual is $(\bar{\ast} V)_{\pm} =
\pm V_{\pm}$.

For $g$-self-dual $F_a$, $F = \ast F$ gives
\begin{equation}
\label{ertye}
    F_-  =  - \frac{\tilde{G}^{+ +}}{1 + \tilde{G}^{+ -}} F_+
  \end{equation}
Then the action (\ref{act}),(\ref{act3}) for a scalar $P$ and 1-form $Q_a$ which satisfies $Q =
\bar{\ast} Q$ so that $Q_- = 0$ is given by
\begin{equation}
  S = \int d^2 x \sqrt{\bar{g}} \left(   \partial_+ P \partial_- P +
  2 Q_+ \partial_- P +
   M_{- -} Q_+ Q_+ \right)
\end{equation}
where $\partial_a = \bar{e}_a^{\mu} \partial_{\mu}$. (This is of course of the
same form as (\ref{2dacts}), but with $+, -$ now frame indices rather than coordinate
indices.) 
The closed 1-forms $F_a,G_a$ with $G=\bar\ast G$ so that $G_-=0$ and $F=\ast F$
are again given by
\begin{equation}
G_+=\frac 1 2 \partial _+P+Q_+, \qquad
F_+=Q_+, \qquad F_-=M_{--}Q_+
\end{equation} 
This then agrees with (\ref{ertye}) if
\begin{equation}
  M_{- -} = - \frac{\tilde{G}^{+ +}}{1 + \tilde{G}^{+ -}}
\end{equation}

This can now be checked against the results of section \ref{sec8}.
In the case in which $G$ is brought to the diagonal form
\begin{equation}
  G_{a b} = \left(\begin{array}{cc}
    - \lambda_0^2 & 0\\
    0 & \lambda_1^2
  \end{array}\right)
\end{equation}
in the basis in which $a = 0, 1$, then converting to the $a = +, -$ basis gives
\begin{equation}
  G_{a b} = \frac{1}{2} \left(\begin{array}{cc}
    \lambda_0^2 - \lambda_1^2 & \lambda_0^2 + \lambda_1^2\\
    \lambda_0^2 + \lambda_1^2 & \lambda_0^2 - \lambda_1^2
  \end{array}\right)
\end{equation}
Then it is readily checked that
\begin{equation}
  M_{- -} = - \frac{\tilde{G}^{+ +}}{1 + \tilde{G}^{+ -}} = \frac{\lambda_0 -
  \lambda_1}{\lambda_0 + \lambda_1}
\end{equation}
in agreement with (\ref{mlam}).

If instead $G$ is brought to the form in the null $a = +, -$ basis given by
\begin{equation}
  G_{a b} = \left(\begin{array}{cc}
    0 & \lambda^2\\
    \lambda^2 & a
  \end{array}\right)
\end{equation}
then
\begin{equation}
  \tilde{G}^{a b}  = \left(\begin{array}{cc}
    - a / \lambda^2  & 1\\
    1 & 0
  \end{array}\right)
\end{equation}
so that
\begin{equation}
  M_{- -} = - \frac{\tilde{G}^{+ +}}{1 + \tilde{G}^{+ -}} = \frac{a}{2
  \lambda ^2}
\end{equation}
in agreement with (\ref{mlamabc}) and the self-duality condition  is
\begin{equation}
  F_- = \frac{a}{2 \lambda^2 } F_+
\end{equation}
in agreement with (\ref{ertye}).

Despite appearances, the result for $M$ is fully covariant.
First, $\det (G_{ab})$ is a scalar
as
\begin{equation}
{\cal D}\equiv \det (G_{ab})=\frac 1 2
\epsilon ^{ac}\epsilon ^{bd}G_{ab}G_{cd}
=
\frac 1 2
(\eta ^{ab}\eta ^{cd}-\eta ^{ac}\eta ^{bd}
)
G_{ab}G_{cd}
\end{equation}
so that 
\begin{equation}
{\cal D}
= \frac 1 2[ (G_a{}^a)^2-
G_a{}^bG_b{}^a]=
\frac 1 2[ (
\bar g^{\mu\nu}g_{\mu\nu}
)^2-
\bar g^{\mu\nu}g_{\nu\rho}
\bar g^{\rho\sigma}g_{\sigma\mu}
]
\end{equation}
Then
\begin{equation}
\tilde{G}^{+ -}=  \frac 1 2{\cal D}{G}^{ab}\eta_{ab}=\frac 1 2{\cal D}
 g^{\mu\nu}\bar g_{\mu\nu}
\end{equation}
is also a scalar.
Then
$M_{--}Q_+Q_+=M^{\mu\nu}Q_\mu Q_\nu$
where
\begin{equation}
M^{\mu\nu}=
\frac {\cal D}
{1+\frac 1 2{\cal D}g^{
\lambda \tau
}\bar g_{\lambda \tau}
}
g^{\rho\sigma}
\bar \Pi_+{}_\rho{}^\mu
\bar \Pi_+{}_\sigma{}^\nu
\end{equation}
and is manifestly covariant.
The operator $M(Q)$ is
\begin{equation}
M(Q)_\mu= M_\mu {}^\nu Q_\nu, \qquad M_\mu {}^\nu=\eta_{\mu\rho}M^{\rho \nu}
\end{equation}

\section{Coupling to Branes}

A $q-1$ form gauge field $A$ couples to $q-2$ branes with a coupling of the form
\begin{equation}
\label{brane}
S_{brane} =\mu \int _\Sigma A
\end{equation}
where $\Sigma$ is the $q-1$ dimensional submanifold on which the
brane is located and $\mu$ is the charge of the  brane.

Of particular interest here is the case in which 
 the gauge field $A$ has self-dual field strength. 
 For  the 4-form gauge field in IIB supergravity  (with $q=5$) the  coupling would be to a D3 brane in 10 dimensions while for $q=2$  a 2-form gauge field  would couple to a self-dual string in 6 dimensions.
 The self-dual gauge field $A$ 
  can be formulated in the way discussed in this paper, but then the coupling $\mu \int A$ is problematic as $A$ is not a fundamental field in the action.
  However, the coupling (\ref{brane}) can be rewritten as
  \begin{equation}
\label{branes}
S_{brane} =\mu \int _X F
\end{equation}
  where $X$ is a $q$-dimensional subspace with boundary $\Sigma$.
  The functional integral is then independent of the choice of $X$ provided the flux of $\mu F$ through any closed $q$ surface is quantised  appropriately, as is the case in string theory. 
  
The action (\ref{act})  for $Q,P$  can be coupled to a $q-1$ brane by adding the interaction term
   \begin{equation}
\label{branest}
S_{brane} =\mu \int _X Q+M(Q)
\end{equation}
which then agrees with (\ref{branes}) using the definition of $F$ from section
\ref{twometrics}, which was $F=Q+M(Q)$.

\section{Conclusion}

In this paper, Sen's action for  $p$-form gauge fields with self-dual field strengths has been generalised to one that can be used to formulate such theories in any spacetime.
It features two metrics on spacetime, the physical dynamical metric $g$ and an auxiliary metric $\bar g$.
It can be coupled to other physical fields, in which case the action takes the form
\begin{equation}
S=S_c+S_m[\phi,g]
\end{equation}
where $S_c$ is the action for the chiral gauge field $A$ given by (\ref{act}) or an interacting generalisation such as (\ref{actinst}) or (\ref{raction}), while $S_m$ is an action for the other physical fields denoted collectively by $\phi$ that may contain an Einstein-Hilbert action or other kinetic term for the metric $g$. The action then gives the dynamics of the physical sector consisting of $g,A,\phi$ with field strength $F$ that is self-dual with respect to the physical metric $g$ but the theory also has a shadow sector consisting of a gauge field $C$ with
with field strength $G$ that is self-dual with respect to the auxiliary metric $\bar g$.
The key point is that the shadow sector $\bar g,C$ decouples from the physical sector.
This action can then be used to give a complete action for IIB supergravity, for example.

The theory is unusual in that the physical field $A$ does not appear directly in the action.
The degrees of freedom governed by the action are formulated in terms of the fields $Q,P$
which both couple directly to the auxiliary metric $\bar g$ while $Q$ couples also to the physical metric $g$ through the interaction term involving $M(Q)$ or $\mathcal{F}(Q)$.
The complicated coupled field equations for $Q,P$ are then disentangled by writing them in terms of $F$ which only couples to $g$ and $G$ which only couples to $\bar g$. 

For the study of a chiral gauge field in a  spacetime background with {\it fixed} non-dynamical metric $g$, for most purposes one could simply take $g=\bar g$ so that $M(Q)=0$, giving the minimal coupling of the Minkowski-space action (\ref{senact}) to $g$. This then describes two gauge fields with field strengths $F,G$ that are both self-dual with respect to $g$. However, for a theory in which gravity is dynamical it is necessary to have $\bar g$ distinct from the physical metric $g$ which gives the gravitational field.
For many purposes, the auxiliary metric $\bar g$ can be taken to be an  arbitrary fixed background metric on the spacetime that plays no role in the physics. 
However, it could instead be taken to be a dynamical field if a kinetic term for $\bar g$ is added.
Indeed, one could in principle add another term $S_{aux}[\bar g,C, \psi]$ to the action involving $\bar g$, the auxiliary gauge field $C$ and an arbitrary set of further auxiliary fields $\psi$. The auxiliary fields $\bar g,C, \psi$ only couple to each other and not to the physical sector so that they would constitute a truly hidden sector.  

For example, for an anti-de Sitter spacetime, $\bar g$ can be taken to be the anti-de Sitter metric (treated as a background metric) while $g$ is taken to be the dynamical metric satisfying asymptotically AdS boundary conditions. For example, such an action could  be used for IIB supergravity in $AdS_5\times S^5$.
For chiral gauge fields in a {\it fixed} AdS background, one could take $g=\bar g$ to be the anti-de Sitter metric.

It is convenient to take the field $\bar g$ to be a tensor field, so that it is a second metric on the spacetime. This leads to a covariant theory which is clearly  independent of the choice of coordinates and is the line that has been taken in most of this paper.
However, it was seen in section \ref{sec7} that there also is an intriguing possibility of taking the field $\bar g$ not to  be a tensor field but to be instead  a gauge field whose transition functions consist of diffeomorphisms combined with gauge transformations. This in turn led to the possibility of taking the field $\bar g$ to be the same in all coordinate patches. Choosing $\bar g=\eta$ in all patches then recovers Sen's action (\ref{sensss}) with transition functions given simply by the $\zeta$-transformations discussed in section \ref{symsec}. 

A manifold with two metrics has  interesting geometry, some of which has been explored here. In particular, the interpolating structure $f$ and the map $\Phi$ play an important role in the construction of the action here.
Spacetimes with two metrics also play a role in massive gravity \cite{deRham:2010kj,deRham:2010ik,deRham:2011rn},  where the graviton interacts with a fixed background metric, and in bi-gravity \cite{Hassan:2011tf,Hassan:2011zd} in which there are two dynamical gravitons with non-trivial interactions. Some of the structure used here should be of relevance for these theories.

\acknowledgments
It is a pleasure to acknowledge discussions with Neil Lambert, who suggested this problem, and with Dan Waldram.
This work was supported by the STFC Consolidated Grant ST/T000791/1.

\appendix
\section{Appendix : Hodge duals and bi-metric geometry}

\noindent
It will be assumed that spacetime is an orientable manifold.
The epsilon symbol is
\begin{equation}
\epsilon_{\mu_1\mu_2 \ldots \mu_d} =
\begin{cases}
+1 \qquad \mathrm{for} \quad (\mu_1, \mu_2,  \ldots \mu_d) \quad \mathrm{an \; even \; permutation \; of} \quad (1,2,\ldots,d) 
\\
-1 \qquad \mathrm{for} \quad (\mu_1, \mu_2,  \ldots \mu_d) \quad \mathrm{an \; odd \; permutation \; of} \quad (1,2,\ldots,d)
\\
  0 \qquad \mathrm{otherwise}
  \label{epon}
\end{cases}
\end{equation}
Given a metric $g_{\mu \nu}$, there is a corresponding pseudo-tensor
\begin{equation}
\varepsilon_{\mu_1\mu_2 \ldots \mu_d} =\sqrt{g}\,
\epsilon_{\mu_1\mu_2 \ldots \mu_d} 
\label{epg}
\end{equation}
where
\begin{equation}
g=| \det {g_{\mu \nu}} |
\end{equation}
and the positive square root, $\sqrt{g}>0$ will be taken.
 For a general $r$-form $\omega = \frac{1}{r!} \omega_{\mu_1 \ldots \mu_r} dx^{\mu_1} \wedge \ldots \wedge dx^{\mu_r}$, the Hodge dual is
\begin{eqnarray}
\star \; \omega = \frac 1{ r! (m-r)! } \omega_{\mu_1 \ldots \mu_r} \varepsilon^{\mu_1 \ldots \mu_r}_{~~~~~~\mu_{r+1} \ldots \mu_d} dx^{\mu_{r+1}} \wedge \ldots \wedge dx^{\mu_d}
\end{eqnarray}
with components
\begin{eqnarray}
(\star \; \omega) _{\mu_{r+1} \ldots \mu_d}  = \frac{1}{ r!  } \omega_{\mu_1 \ldots \mu_r} \varepsilon^{\mu_1 \ldots \mu_r}_{~~~~~~\mu_{r+1} \ldots \mu_d} 
\end{eqnarray}
with indices raised and lowered using $ g_{\mu\nu}$.
In Lorentzian signature
\begin{equation}
\star \star \; \omega  =  - (-1)^{r (d-r)} \omega
\end{equation}
In particular, for $r=q=d/2 =2n+1$, $\star \star \; \omega=\omega$, so that  a real $q$-form can be  decomposed into a  real self-dual  part and an anti-self-dual one.
The volume form can be written as 
$$\star 1 = \Omega_g=
\frac{ 1 }{ d!}  \varepsilon_{\mu_1 \ldots \mu_d} dx^{\mu_{1}} \wedge \ldots \wedge dx^{\mu_d}
= \sqrt{g}\,
dx^1\wedge dx^2\wedge\ldots \wedge dx^d
$$
This determines an orientation on the spacetime. If the negative square root had been chosen in (\ref{epg}), this would take $\varepsilon_{\mu_1 \ldots \mu_d} \to -\varepsilon_{\mu_1 \ldots \mu_d} $ and give the opposite orientation.

A vielbein $e_\mu{}^a$ satisfying
\begin{equation}
g_{\mu\nu}=e_\mu{}^a e_\nu {}^b \eta _{ab}
\end{equation}
 defines an orthonormal frame and converts coordinate indices $\mu ,\nu,\dots $ to tangent space indices $a,b,\dots$. 
For an oriented spacetime, one can choose $e_\mu{}^a$ to satisfy
\begin{equation}
\det( e_\mu{}^a)>0
\end{equation}
so that the one-form bases $\{ dx^\mu \}$ and $\{ e_\mu{}^a dx^\mu \}$ have the same orientation.
Then
\begin{equation}
\varepsilon _{a_1\dots a_d}=
e_{a_1}{}^{\mu_1}e_{a_2}{}^{\mu_2}\dots e_{a_d}{}^{\mu_d}
\varepsilon _{\mu_1\mu_2 \ldots \mu_d}
\label{dthyrt}
\end{equation}
giving
$$\varepsilon _{a_1\dots a_d}
=
e_{a_1}{}^{\mu_1}e_{a_2}{}^{\mu_2}\dots e_{a_d}{}^{\mu_d}
\sqrt {\det (g_{\mu\nu})} \,  \epsilon_{\mu_1\mu_2 \ldots \mu_d} $$
But
$$
e_{a_1}{}^{\mu_1}e_{a_2}{}^{\mu_2}\dots e_{a_d}{}^{\mu_d}
  \,  \epsilon_{\mu_1\mu_2 \ldots \mu_d} 
  =det(e_{a}{}^{\mu}) \epsilon_{a_1\dots a_d}
  $$
 giving
  \begin{equation}
\varepsilon _{a_1\dots a_d}=\epsilon_{a_1\dots a_d}
\end{equation}
As a result,
 \begin{eqnarray}
(\star \; \omega) _
{\mu_{r+1} \ldots \mu_d} 
 = \frac{ 1}{ r!  } 
 \omega_{\mu_1 \ldots \mu_r} 
 \varepsilon^
 {\mu_1 \ldots \mu_r}{}_{\mu_{r+1} \ldots \mu_d} 
\end{eqnarray}
has frame components
 \begin{eqnarray}
(\star \; \omega) _{a_{r+1} \ldots a_d}  = \frac{ 1}{ r!  } \omega_{a_1 \ldots a_r} \epsilon^{a_1 \ldots a_r}{}_{a_{r+1} \ldots a_d} 
\end{eqnarray}

A second metric $\bar g_{\mu\nu}$ can be used to  define 
similar structures  by replacing  $ g_{\mu\nu}$  with $\bar g_{\mu\nu}$  in the above formulae.
There is  then a second pseudo-tensor
\begin{equation}
\bar \varepsilon_{\mu_1\mu_2 \ldots \mu_d} =\sqrt{\bar g}\,
\epsilon_{\mu_1\mu_2 \ldots \mu_d} 
\end{equation}
where
\begin{equation}
\bar g=| \det {\bar g_{\mu \nu}} |
\end{equation}
and again  the positive square root is taken, $\sqrt{\bar g}>0$.
This gives
\begin{equation}
\bar \varepsilon_{\mu_1\mu_2 \ldots \mu_d} =\frac {\sqrt{\bar g}    } {\sqrt{ g}    }\,
\varepsilon_{\mu_1\mu_2 \ldots \mu_d} 
\end{equation}
 For a general $r$-form $\omega = \frac{1}{r!} \omega_{\mu_1 \ldots \mu_r} dx^{\mu_1} \wedge \ldots \wedge dx^{\mu_r}$, the $\bar g$ Hodge dual has  components
\begin{eqnarray}
(\bar \star \; \omega) _{\mu_{r+1} \ldots \mu_d}  = \frac{1}{ r!  } \omega_{\mu_1 \ldots \mu_r}\bar  \varepsilon^{\mu_1 \ldots \mu_r}{}_{\mu_{r+1} \ldots \mu_d} 
\end{eqnarray}
with indices raised and lowered using $\bar g_{\mu\nu}$.
The new volume form can be written as 
$$\bar \star 1 = \Omega_{\bar g}=
\frac{ 1 }{ d!} \bar  \varepsilon_{\mu_1 \ldots \mu_d} dx^{\mu_{1}} \wedge \ldots \wedge dx^{\mu_d}
= \sqrt{\bar g}\,
dx^1\wedge dx^2\wedge\ldots \wedge dx^d
$$
With $ \sqrt{ g}>0, \sqrt{\bar g}>0$, $\bar \star 1 $ determines the same orientation as $\star 1$.

A vielbein $\bar e_\mu{}^a$ satisfying
\begin{equation}
\bar g_{\mu\nu}=\bar e_\mu{}^a \bar e_\nu {}^b \eta _{ab}
\end{equation}
gives a second way to convert coordinate indices $\mu ,\nu,\dots $ to tangent space indices $a,b,\dots$. 
Then, as before,
\begin{equation}
\bar \varepsilon _{a_1\dots a_d}=
\bar e_{a_1}{}^{\mu_1}\bar e_{a_2}{}^{\mu_2}\dots \bar e_{a_d}{}^{\mu_d}
\bar \varepsilon _{\mu_1\mu_2 \ldots \mu_d} 
=\epsilon_{a_1\dots a_d}
\label{sdfgt}
\end{equation}
Now (\ref{dthyrt}),(\ref{sdfgt}) give
\begin{equation}
e_{a_1}{}^{\mu_1}e_{a_2}{}^{\mu_2}\dots e_{a_d}{}^{\mu_d}
\varepsilon _{\mu_1\mu_2 \ldots \mu_d}
=
\bar e_{a_1}{}^{\mu_1}\bar e_{a_2}{}^{\mu_2}\dots \bar e_{a_d}{}^{\mu_d}
\bar \varepsilon _{\mu_1\mu_2 \ldots \mu_d} 
\end{equation}
Then
\begin{equation}
\varepsilon _{\mu_1\mu_2 \ldots \mu_m} = f_{\mu_1}{}^{\nu_1}\dots f_{\mu_m}{}^{\nu_m}\bar
\varepsilon _{\nu_1\nu_2 \ldots \nu_m}
\label{ftrans}
\end{equation}
where
 \begin{equation}
f_\mu {}^\nu=e_\mu{}^a \bar e_a{}^\nu
\end{equation}
The map $\Phi$ on $r$-forms  
$$\Phi: \omega\to \omega' =\Phi(\omega)$$
acts on the components  in a coordinate basis as
 \begin{equation}
\omega'   _{\mu_1\mu_2 \ldots \mu_r} = f_{\mu_1}{}^{\nu_1}\dots f_{\mu_r}{}^{\nu_r}\omega
 _{\nu_1\nu_2 \ldots \nu_r} 
 \label{sdfgs}
 \end{equation}
 Then (\ref{ftrans}) can be written as
 $$\Phi(\bar \varepsilon) = \varepsilon$$
 
 It will now be shown that the Hodge star operator  $*$ for the metric $g$ and the Hodge star operator  $\bar *$ for the metric $\bar g$ are related by
 $$ \Phi(\bar * \omega)=*\Phi(\omega)
$$
The $\bar g$-dual of an $r$-form $\omega$ has components
\begin{eqnarray}
(\bar \star \; \omega) _{\mu_{r+1} \ldots \mu_m}  = \frac{ 1}{ r!  } \omega_{\mu_1 \ldots \mu_r} \bar \varepsilon^{\mu_1 \ldots \mu_r}_{~~~~~~\mu_{r+1} \ldots \mu_m} 
\end{eqnarray}
while $\omega' =\Phi(\omega)$ 
has $g$-dual
\begin{eqnarray}
(\star \; \omega') _{\mu_{r+1} \ldots \mu_m}  =
 \frac{ 1}{ r!  } \omega'_{\nu_1 \ldots \nu_r} 
 g^{\mu_1\nu_1}\dots g^{\mu_r\nu_r}
 \varepsilon_{\mu_1 \ldots \mu_r\mu_{r+1} \ldots \mu_m} 
\end{eqnarray}
which, using (\ref{ftrans}), gives
\begin{eqnarray}
(\star \; \omega') _{\mu_{r+1} \ldots \mu_m}  =
 \frac{ 1}{ r!  } \omega'_{\nu_1 \ldots \nu_r} 
 g^{\mu_1\nu_1}\dots g^{\mu_r\nu_r}
f_{\mu_1}{}^{\sigma_1}\dots f_{\mu_m}{}^{\sigma_m}\bar
\varepsilon _{\sigma_1\sigma_2 \ldots \sigma_m}
\end{eqnarray}
Now $\omega' =\Phi(\omega)$ 
has components (\ref{sdfgs}) so that \begin{eqnarray}
(\star \; \omega') _{\mu_{r+1} \ldots \mu_m}  =
 \frac{ 1}{ r!  } 
 [f_{\nu_1}{}^{\rho_1}\dots f_{\nu_r}{}^{\rho_r} g^{\mu_1\nu_1}\dots g^{\mu_r\nu_r}
f_{\mu_1}{}^{\sigma_1}\dots f_{\mu_m}{}^{\sigma_m}]
 \omega
 _{\rho_1\rho_2 \ldots \rho_m} 
\bar
\varepsilon _{\sigma_1\sigma_2 \ldots \sigma_m}\end{eqnarray}
Using
$$ f_{\nu}{}^{\rho}g^{\mu\nu}f_{\mu}{}^{\sigma}=\bar g^{\rho\sigma}
$$
gives
\begin{eqnarray}
(\star \; \omega') _{\mu_{r+1} \ldots \mu_m}  =
f_{\mu_{r+1}}{}^{\sigma_{r+1}}\dots f_{\mu_m}{}^{\sigma_m}
\left[ \frac{ 1}{ r!  } \omega_{\nu_1 \ldots \nu_r} 
 \bar g^{\mu_1\nu_1}\dots \bar g^{\mu_r\nu_r}
\bar \varepsilon_{\mu_1 \ldots \mu_r\sigma_{r+1} \ldots \sigma_m} \right]
\end{eqnarray}
The RHS is the component form of $\Phi(\bar * \omega)$, 
so that $*\Phi (\omega)=\Phi(\bar * \omega)$
as required.

\begin{eqnarray}
(\star \; \omega') _{\mu_{r+1} \ldots \mu_m}  = \frac{ 1}{ r!  } \omega'_{\mu_1 \ldots \mu_r} \varepsilon^{\mu_1 \ldots \mu_r}_{~~~~~~\mu_{r+1} \ldots \mu_m} 
\end{eqnarray}

where
$$\omega'   _{\mu_1\mu_2 \ldots \mu_r} = f_{\mu_1}{}^{\nu_1}\dots f_{\mu_r}{}^{\nu_r}\omega
 _{\nu_1\nu_2 \ldots \nu_m} $$
and 
$$\varepsilon _{\mu_1\mu_2 \ldots \mu_m} = f_{\mu_1}{}^{\nu_1}\dots f_{\mu_m}{}^{\nu_m}\bar
\varepsilon _{\nu_1\nu_2 \ldots \nu_m} $$
Then the components  of $*\Phi(\omega)$ are
\begin{eqnarray}
(\star \; \omega') _{\mu_{r+1} \ldots \mu_m}  =
 \frac{ 1}{ r!  } \omega'_{\nu_1 \ldots \nu_r} 
 g^{\mu_1\nu_1}\dots g^{\mu_r\nu_r}
 \varepsilon_{\mu_1 \ldots \mu_r\mu_{r+1} \ldots \mu_m} 
\end{eqnarray}
giving

Then using
$$\omega'   _{\nu_1\nu_2 \ldots \nu_r} = f_{\nu_1}{}^{\rho_1}\dots f_{\nu_r}{}^{\rho_r}\omega
 _{\rho_1\rho_2 \ldots \rho_m} $$
 gives
 \begin{eqnarray}
(\star \; \omega') _{\mu_{r+1} \ldots \mu_m}  =
 \frac{ 1}{ r!  } 
 f_{\nu_1}{}^{\rho_1}\dots f_{\nu_r}{}^{\rho_r}\omega
 _{\rho_1\rho_2 \ldots \rho_m} 
 g^{\mu_1\nu_1}\dots g^{\mu_r\nu_r}
f_{\mu_1}{}^{\sigma_1}\dots f_{\mu_m}{}^{\sigma_m}\bar
\varepsilon _{\sigma_1\sigma_2 \ldots \sigma_m}\end{eqnarray}
which can be written as
\begin{eqnarray}
(\star \; \omega') _{\mu_{r+1} \ldots \mu_m}  =
 \frac{ 1}{ r!  } 
 [f_{\nu_1}{}^{\rho_1}\dots f_{\nu_r}{}^{\rho_r} g^{\mu_1\nu_1}\dots g^{\mu_r\nu_r}
f_{\mu_1}{}^{\sigma_1}\dots f_{\mu_m}{}^{\sigma_m}]
 \omega
 _{\rho_1\rho_2 \ldots \rho_m} 
\bar
\varepsilon _{\sigma_1\sigma_2 \ldots \sigma_m}\end{eqnarray}

\section{Appendix : Perturbative Construction}

Explicit perturbative expressions for the maps in  section \ref{sec5} can be
found following  \cite{Sen:2015nph,Sen:2019qit}. The relation
\begin{equation}
  Q = \bar{\Pi}_+ \Phi \alpha = [1 + \bar{\Pi}_+ (\Phi - 1)] \alpha
\end{equation}
(where $\Phi$ is the map taking $\alpha$ to $\Phi \alpha = \Phi (\alpha)$) can be formally
inverted to give
\begin{equation}
  \label{uiss} \alpha = [1 + \bar{\Pi}_+ (\Phi - 1)]^{- 1} Q
\end{equation}
which can be formally expanded as a power series. Then with
\begin{equation}
  N = \bar{\Pi}_+ \Phi \bar{\Pi}_+
\end{equation}
the generalised inverse is
\begin{equation}
  \label{nti} \tilde{N} = \bar{\Pi}_+ [1 + \bar{\Pi}_+ (\Phi - 1)]^{- 1}
  \bar{\Pi}_+
\end{equation}
which satisfies (\ref{genin}).

Next
\begin{equation}
  M = K (\alpha) = \bar{\Pi}_- \Phi \bar{\Pi}_+ \alpha
\end{equation}
with $\alpha$ given by (\ref{uiss}) gives
\[ M Q = \bar{\Pi}_- \Phi \bar{\Pi}_+ [1 + \bar{\Pi}_+ (\Phi - 1)]^{- 1} Q \]
which can be rewritten as
\begin{equation}
  M Q = \bar{\Pi}_-  (\Phi - 1) \bar{\Pi}_+ [1 + \bar{\Pi}_+ (\Phi - 1)]^{- 1}
  Q
\end{equation}
from which the map $M$ is
\begin{equation}
  \label{mist} M = \bar{\Pi}_-  (\Phi - 1) \bar{\Pi}_+ [1 + \bar{\Pi}_+ (\Phi
  - 1)]^{- 1} \bar{\Pi}_+
\end{equation}

The formal expressions for $\tilde{N}$ and $M$ given in
(\ref{nti}),(\ref{mist}) can be expanded as formal power series in $(\Phi -
1)$, so that e.g.\
\begin{equation}
  \label{npow} \tilde{N} = \bar{\Pi}_+ \: \sum_{k = 0}^{\infty} [\bar{\Pi}_+
  (\Phi - 1)]^k\, \bar{\Pi}_+
\end{equation}
If $g$ is sufficiently close to $\bar{g}$ with
\begin{equation}
  e^a_{\mu} = \bar{e}_{\mu}^a + k^a_{\mu}
\end{equation}
with $k^a_{\mu}$ small, then $f_{\mu}{}^{\nu} =
\delta_{\mu}{}^{ \nu} + k^a_{\mu} \bar{e}_a{}^{ \nu}$
and $(\Phi - 1)$ will also be small, so that such power series will be the sum
of decreasing terms.

\section{Appendix : Symmetry of the Action}

Consider the variation of the action
\begin{equation}
  S = \int \left( \frac{1}{2} d P \wedge \bar{\ast} d P - 2 Q \wedge d P - Q
  \wedge M (Q) \right)
\end{equation}
under the transformations
\begin{equation}
  \quad \delta \bar{g} = 0, \quad \delta g =\mathcal{L}_{\zeta} g, \quad
  \label{Qtransa} \delta Q = - \frac{1}{2}  (1 + \bar{\ast}) d \delta P
\end{equation}
under which $\delta G = 0$.

Since $M$ is a symmetric linear operator (which can be written in components
as (\ref{Mcomps}))
\begin{equation}
  \delta (M (Q)) = M (\delta Q) + (\delta M) (Q)
\end{equation}
where $M (Q)$ changes to $(M + \delta M) (Q)$ as $g \rightarrow g + \delta g$
and
\begin{equation}
  Q \wedge M (\delta Q) = \delta Q \wedge M (Q)
\end{equation}
Then
\begin{equation}
  \delta \int Q \wedge M (Q) = \int 2 \delta Q \wedge M (Q) + Q \wedge (\delta
  M) (Q)
\end{equation}
where $\delta M$ is the change in $M$ due to the change in $g$. The variation
of the action gives
\begin{equation}
  \delta S = \label{varS} \int  \left\{ - 2 \left( \frac{1}{2}  \bar{\ast} d P
  + Q \right) \wedge d \delta P - 2 \delta Q \wedge (d P + M) - Q \wedge
  [\delta M] (Q) \right\}
\end{equation}
Using (\ref{Qtransa}),
\[ - 2 \delta Q \wedge (d P + M) = - 2 d \delta P \wedge \bar{\Pi}_- (d
   P + M) \]
Then using $\bar{\Pi}_- M = M$ and, dropping a total derivative term, the
variation (\ref{varS}) becomes
\begin{equation}
  \delta S = \label{varS22} \int  \{ - 2 F \wedge d \delta P - Q \wedge
  [\delta M] (Q) \}
\end{equation}
As $\bar{\Pi}_- M = M$ and $\delta \overline{g} = 0$, the variation should
also be anti-self-dual, $\bar{\Pi}_- \delta M = \delta M$. Then $M \wedge
\delta M = 0$ so $Q \wedge \delta M = F \wedge \delta M$ and the
variation becomes
\begin{equation}
  \delta S = \label{varS33} \int  \left\{ - 2 F \wedge \left( d \delta P -
  \frac{1}{2} Q \wedge [\delta M] (Q) \right) \right\}
\end{equation}

Under the transformations $\delta \bar{g} = 0, \delta g =\mathcal{L}_{\zeta}
g$ with parameter a vector field $\zeta^{\mu}$
\[ \delta f_{\mu}{}^{\nu} = \zeta_{\mu}{}^{\rho}
   f_{\rho}{}^{\nu} \]
where
\[ \zeta_{\mu}{}^{ \nu} = (\nabla_{\mu} \zeta^{\nu}) \]
and $\nabla_{\mu}$ is the Levi-Civita connection for the metric $g$. To see
this, note that the one-form $e^a = e^a_{\mu} d x^{\mu}$ transforms under a
diffeomorphism with parameter $\zeta^{\mu}$ and a frame rotation (local
Lorentz transformation) with parameter $L^a {}_b$ as
\begin{eqnarray}
\delta e ^a &=&\mathcal{L}_{\zeta} e ^a + L^a{} _b e^b = ( {di}_{\zeta} +
   i_{\zeta} d) e^a + L^a {}_b e^b
   \cr 
   &=& d \zeta^a - i_{\zeta} \left(
   \omega^a{}_{ b} \wedge e^b \right) + L^a {}_b e^b = D \zeta^a
   + \left( L^a {}_b - i_{\zeta} \omega^a{}_{ b} \right) e^b 
\end{eqnarray}
where $\zeta^a = e^{a }_{\mu} \zeta^{\mu}$ and
\[ D \zeta^a = d \zeta^a + \omega^a{}_{ b} \zeta^b \]
The term $- \left( i_{\zeta} \omega^a{}_{ b} \right) e^b$ can be
absorbed into the frame rotation, $L^a _b {\rightarrow L'}^a _b - L^a _b -
i_{\zeta} \omega^a{}_{b}$, giving
\[ \delta e ^a = D \zeta^a {+ L'}^a _b e^b \]
Then from $f_{\mu}  ^{\nu} = e_{\mu}^a  \bar{e}_a^{\nu}$
\[ \delta f_{\mu}{}^{ \nu} = \delta e_{\mu}^a  \bar{e}_a^{\nu} =
   (D_{\mu} \zeta^a) \bar{e}_a^{\nu} = (\nabla_{\mu} \zeta^{\rho}) e_{\rho }^a
   \bar{e}_a^{\nu} = (\nabla_{\mu} \zeta^{\rho}) f_{\rho}^{\nu}
\]
The map $R_{\zeta}$ on $r$-forms $X$ is given by
\[ R_{\zeta} (X)_{\mu_1 \ldots \mu_r} = r! (\nabla_{[  \mu_1}
   \zeta^{\rho}) X_{| \rho | \mu_2 \ldots   \mu_r]} \]
This can be rewritten as
\[ R_{\zeta} (X)_{\mu_1 \ldots \mu_r} = r \nabla_{[  \mu_1}
   (\zeta^{\rho} X_{| \rho | \mu_2 \ldots   \mu_r]}) - r \zeta^{\rho}
   \nabla_{[  \mu_1} X_{| \rho | \mu_2 \ldots   \mu_r]} \]
giving
\[ R_{\zeta} (X)_{\mu_1 \ldots \mu_r} = r \nabla_{[  \mu_1}
   (\zeta^{\rho} X_{| \rho | \mu_2 \ldots   \mu_r]}) - (r + 1)
   \zeta^{\rho} \nabla_{[  \mu_1} X_{\rho \mu_2 \ldots  
   \mu_r]} + \zeta^{\rho} \nabla_{\rho} X_{\mu_1 \ldots \mu_r} \]
For an $r$-form
\[ X = \frac{1}{r!} X_{\mu_1 \ldots \mu_r} d x^{\mu_1} \wedge \cdots
   \wedge d x^{\mu_r} \]
the exterior derivative is
\[ d X = \frac{1}{(r + 1) !} (d X )_{\mu_1 \ldots \mu_{r + 1}} d x^{\mu_1}
   \wedge \cdots \wedge d x^{\mu_{r + 1}} \]
with components
\[ (d X )_{\mu_1 \ldots \mu_{r + 1}} = (r + 1) \partial_{[\mu_1  }
   X_{\mu_1 \ldots \mu_{r + 1}  ]} \]
so
\[ R_{\zeta} (X) = (i_{\zeta} d X + d i_{\zeta} X) + \zeta^{\rho}
   \nabla_{\rho} X \]
giving
\begin{equation}
R_{\zeta} (X) =\mathcal{L}_{\zeta} {X + \nabla_{\zeta}}  X 
\label{dergee}
\end{equation}
where ${\nabla_{\zeta}}  X = \zeta^{\rho} \nabla_{\rho} X$.

The variation of $M$ given in  (\ref{delm}) is
\begin{equation}
  (\delta M) (Q) = R_{\zeta} (F) - \{ \Xi + M (\Xi) \}
\end{equation}
where
\begin{equation}
  \Xi \equiv \bar{\Pi}_+ R_{\zeta} (F)
\end{equation}
and this can now be used in the variation of the action (\ref{varS33}).
Since $\Psi (\Xi) = \{ \Xi + M (\Xi) \}$ is $g$-self-dual, $\Psi (\Xi) = \ast
\Psi (\Xi)$,
\begin{equation}
  \text{} F \wedge \{ \Xi + M (\Xi) \} = 0
\end{equation}
and
\begin{equation}
  \delta S = \label{varS3b} \int  \left\{ - 2 F \wedge \left( d \delta P -
  \frac{1}{2} R_{\zeta} (F) \right) \right\}
\end{equation}
Then using $R_{\zeta} (X) =\mathcal{L}_{\zeta} {X + \nabla_{\zeta}}  X$
\begin{equation}
  \int F \wedge R_{\zeta} (F) = \int F \wedge \left( \mathcal{L}_{\zeta}
  {F + \nabla_{\zeta}}  F \right) = \int F \wedge \mathcal{L}_{\zeta} F =
  \int F \wedge (i_{\zeta} d F + d i_{\zeta} F)
\end{equation}
where ${\ast \nabla_{\zeta}}  {F = \nabla_{\zeta}}  F$ has been used.

Using $F = \ast F$ and dropping a total derivative
\begin{equation}
  \int F \wedge R_{\zeta} (F) = 2 \int F \wedge d i_{\zeta} F
\end{equation}
Then
\begin{equation}
  \delta S = \label{varS3c} \int  \{ - 2 F \wedge (d \delta P - d i_{\zeta}
  F) \}
\end{equation}
and is invariant if
\begin{equation}
  \delta P = i_{\zeta} F
\end{equation}
(More generally, it is invariant if $\delta P = i_{\zeta} F + d \alpha$ for any
$\alpha$.)

 Turn now to the variation of $F = Q + M (Q)$, so that
\[ \delta F = \delta Q + M (\delta Q) + (\delta M) (Q) \]
gives, using (\ref{siis}),(\ref{delm}),(\ref{xis}),
\begin{equation}
  \delta F = \Psi (\delta Q - \Xi) + R_{\zeta} (F)
\end{equation}
Now
\begin{equation}
  \delta Q - \Xi = \bar{\Pi}_+ (d i_{\zeta} F - R_{\zeta} (F))
\end{equation}
with
\begin{equation}
  d i_{\zeta} F - R_{\zeta} (F) = d i_{\zeta} F -\mathcal{L}_{\zeta} {F -
  \nabla_{\zeta}}  F = - i_{\zeta} {d F - \nabla_{\zeta}}  F
\end{equation}
Then
\begin{equation}
  \delta F = - \Psi (\bar{\Pi}_+ i_{\zeta} d F) - \Psi \left( \bar{\Pi}_+
  {\nabla_{\zeta}}  F \right) + R_{\zeta} (F)
\end{equation}
As ${\nabla_{\zeta}}  {F = \ast \nabla_{\zeta}}  F$, (\ref{psiiden}) implies
\begin{equation}
  \Psi \left( \bar{\Pi}_+ {\nabla_{\zeta}}  F \right) {= \nabla_{\zeta}}  F
\end{equation}
Then using (\ref{dergee})
\begin{equation}
  \delta F =\mathcal{L}_{\zeta} F - \Psi (\bar{\Pi}_+ i_{\zeta} d F)
\end{equation}
The field equations imply $d F = 0$, so that on-shell
\begin{equation}
  \delta F \approx \mathcal{L}_{\zeta} F
\end{equation}

\end{document}